\documentclass[a4paper,11pt]{article}
\pdfoutput=1
\usepackage{jheppub}

\usepackage{amsmath}
\usepackage{amssymb}
\usepackage{array}
\usepackage{calc}
\usepackage{orcidlink}
\usepackage{multirow}
\usepackage{nicefrac}
\usepackage[colorlinks=true,
            linkcolor=blue,
            citecolor=blue,
            urlcolor=blue]{hyperref}
\usepackage{cleveref}
\usepackage{pstricks}
\usepackage{graphicx}
\graphicspath{{figures/}}
\usepackage{xspace}
\usepackage{listings}
\usepackage{
  pgf,
  tikz}
\usetikzlibrary{
  patterns,
  shapes.multipart,
  arrows,
  trees,
  scopes,
  decorations.pathreplacing,
  decorations.pathmorphing,
  decorations.markings,
  decorations.text,
  positioning,
  calc
}
\usepackage{mciteplus}

\usepackage{subfig}
\usepackage{relsize}
\usepackage{todonotes}
\setuptodonotes{fancyline, inline, color=blue!30}

\usepackage{newunicodechar}
\newunicodechar{−}{-}
\newcommand{\myfigure}[5][h]{
  \begin{figure}[#1]
    \begin{center}
      \includegraphics[width=#2\textwidth]{#3}\\[0.5em]
      \parbox[t]{#2\textwidth}{
        \caption{#4}
        \ifx&#5&\else\label{#5}\fi
      }
    \end{center}
  \end{figure}
}

\numberwithin{equation}{section}

\newcommand{\ddone}{{\mathrm d}}

\newcommand{\bea}{\begin{eqnarray}}
\newcommand{\eea}{\end{eqnarray}}
\newcommand{\beq}{\begin{equation}}
\newcommand{\eeq}{\end{equation}}

\newcommand{\bs}{\begin{split}}
\newcommand{\es}{\end{split}}

\newcommand{\bi}{\begin{itemize}}
\newcommand{\ei}{\end{itemize}}
\newcommand{\bc}{\begin{center}}
\newcommand{\ec}{\end{center}}
\newcommand{\bac}{\begin{array}{c}}
\newcommand{\bacc}{\begin{array}{cc}}
\newcommand{\ea}{\end{array}}

\def\spa#1.#2{\langle#1\,#2\rangle}
\def\spb#1.#2{[#1\,#2]}

\renewcommand\Re{\operatorname{\mathfrak{Re}}}

\newcommand{\sqrts}{\ensuremath{\sqrt{s}}\xspace}

\newcommand{\hsqrts}{\ensuremath{\frac{\sqrt{s}}{2}}\xspace}

\newcommand{\sla}[1]{\ensuremath{{#1\kern-0.45em/}}}

\newcommand{\eik}[1]{\ensuremath{\tilde{S}\left(k_{#1}\right)}}

\newcommand{\CMD}{CMD\xspace}
\newcommand{\KLOEI}{KLOE-I\xspace}
\newcommand{\KLOEII}{KLOE-II\xspace}
\newcommand{\BES}{BESIII\xspace}
\newcommand{\Bscen}{B\xspace}
\newcommand{\Babar}{BaBar\xspace}
\newcommand{\CMDthree}{CMD-3\xspace}
\newcommand{\KLOE}{KLOE\xspace}
\newcommand{\GVMD}{GVMD\xspace}
\newcommand{\FsQED}{FsQED\xspace}
\newcommand{\FxsQED}{F$\times$sQED\xspace}
\newcommand{\sQED}{sQED\xspace}

\newcommand{\kkmc}{KK\protect\scalebox{0.8}{MC}\xspace}

\newcommand{\ACRAIC}{A\protect\scalebox{0.8}{CRAIC}\xspace}

\newcommand{\Collier}{C\protect\scalebox{0.8}{OLLIER}\xspace}

\newcommand{\OpenLoops}{O\protect\scalebox{0.8}{PEN}L\protect\scalebox{0.8}{OOPS}\xspace}

\newcommand{\LoopTools}{L\protect\scalebox{0.8}{OOP}T\protect\scalebox{0.8}{OOLS}\xspace}

\newcommand{\Sherpa}{S\protect\scalebox{0.8}{HERPA}\xspace}
\newcommand{\Comix}{C\protect\scalebox{0.8}{OMIX}\xspace}

\newcommand{\FeynArts}{F\protect\scalebox{0.8}{EYN}A\protect\scalebox{0.8}{RTS}\xspace}
\newcommand{\FeynCalc}{F\protect\scalebox{0.8}{EYN}C\protect\scalebox{0.8}{ALC}\xspace}

\newcommand{\McMule}{McMule\xspace}
\newcommand{\BabaYaga}{BabaYaga@NLO\xspace}
\newcommand{\Phokhara}{PHOKHARA\xspace}

\DeclareOldFontCommand{\rm}{\normalfont\rmfamily}{\mathrm}
\DeclareOldFontCommand{\sf}{\normalfont\sffamily}{\mathsf}
\DeclareOldFontCommand{\tt}{\normalfont\ttfamily}{\mathtt}
\DeclareOldFontCommand{\bf}{\normalfont\bfseries}{\mathbf}
\DeclareOldFontCommand{\it}{\normalfont\itshape}{\mathit}
\DeclareOldFontCommand{\sl}{\normalfont\slshape}{\@nomath\sl}
\DeclareOldFontCommand{\sc}{\normalfont\scshape}{\@nomath\sc}

\usepackage{slashed}

\usepackage[T1]{fontenc}
\usepackage{placeins}
\usepackage{changes}
\usepackage{lineno}

\begin{document}

\title{QED corrections to $e^+e^-\to\pi^+\pi^-(\gamma)$  with amplitude-level resummation}
\preprint{
  \begin{tabular}{r}
    MCNET-26-36\\
    IPPP/26/81
  \end{tabular}
}

\author[a]{St\'ephane~Delorme\orcidlink{0000-0002-2447-5989},}
\author[b]
{Charles Earnshaw\orcidlink{0009-0004-7653-3943},}
\author[b]
{Frank Krauss\orcidlink{0000-0001-5043-3099},}
\author[c]{Alan~Price\orcidlink{0000-0002-0372-1060},}
\author[a]{Nora~Salone\orcidlink{0000-0003-2365-8916},}
\author[c]{and Andrzej Si\'{o}dmok\orcidlink{0000-0001-9614-7856}}
\affiliation[a]{Institute of Physics, University of Silesia, Katowice, Poland}
\affiliation[b]{IPPP, Department of Physics, Durham University, Durham DH1 3LE, United Kingdom}
\affiliation[c]{Jagiellonian University, ul.\ prof.\ Stanis\l{}awa \L{}ojasiewicza 11, 30-348 Krak\'{o}w, Poland}
\emailAdd{stephane.delorme@us.edu.pl}
\emailAdd{charlie.earnshaw@durham.ac.uk}
\emailAdd{frank.krauss@durham.ac.uk}
\emailAdd{alan.price@uj.edu.pl}
\emailAdd{nora.salone@us.edu.pl}
\emailAdd{andrzej.siodmok@uj.edu.pl}

\date{\today}

\abstract{
We present QED corrections to $e^+e^-\to\pi^+\pi^-(\gamma)$ in two independent Monte Carlo event generators, \kkmc and \Sherpa, for point-like scalar pions dressed with the pion form factor.
All-order soft-photon exponentiation is combined with exact one- and two-photon tree amplitudes, the complete one-loop correction including the initial-final boxes with coherent interference, and at second order the photonic two-loop vertices and the exact one-loop correction to one-photon emission.
At first order their fiducial cross sections agree to $\lesssim0.04\%$ and their distributions within statistical precision.
The second-order band is $\sim0.005\%$ for \CMD{}, $\sim0.4\%$ at $4$~GeV and $\sim0.9\%$ at $10$~GeV.
Theory nuisance parameters estimate missing higher orders, and we construct a new rule for treating radiative return, where first-order corrections are of order one.
}

\maketitle

\section{\label{sec:intro} Introduction} For decades, a persistent discrepancy existed between theoretical predictions~\cite{Aoyama:2020ynm} and experimental measurements~\cite{Muong-2:2021ojo,Muong-2:2023cdq,Muong-2:2025xyk} of the muon anomalous magnetic moment $a_\mu$~\cite{Jegerlehner:2009ry}, motivating extensive searches for new physics~\cite{Athron:2025ets}.
The Muon $g-2$ experiment (E989) at Fermilab~\cite{Muong-2:2015xgu}, combining its full dataset with the earlier Brookhaven (BNL) measurement~\cite{Muong-2:2006rrc}, recently reached a precision of $124~\text{ppb}$~\cite{Muong-2:2025xyk,Muong-2:2026qnz}.

While the 2020 White Paper~\cite{Aoyama:2020ynm} yielded $a_\mu(\text{EXP}) - a_\mu(\text{SM}) = (251 \pm 59) \times 10^{-11}$ -- a $4.2\sigma$ discrepancy, recent updates report $a_\mu(\text{EXP}) - a_\mu(\text{SM}) = (38 \pm 63) \times 10^{-11}$~\cite{Aliberti:2025beg}, in agreement within current
uncertainties.
This shift is driven by lattice QCD calculations of the leading-order hadronic vacuum polarisation (HVP) contribution $a_\mu^{\text{HVP, LO}}$ by the Budapest-Marseille-Wuppertal (BMW) Collaboration~\cite{Borsanyi:2020mff,Boccaletti:2024guq}, since confirmed by other lattice collaborations~\cite{Djukanovic:2024cmq,RBC:2024fic,FermilabLatticeHPQCD:2024ppc}.

A substantial tension remains within the HVP sector itself.
Lattice calculations disagree with data-driven estimates based on $e^+e^- \to \text{hadrons}$ cross sections, and the $e^+e^-$ datasets disagree among themselves~\cite{Davier:2023fpl}, with \CMDthree{}~\cite{CMD-3:2023alj,CMD-3:2023rfe} yielding a higher $e^+e^-\to\pi^+\pi^-$ cross section than \KLOE{}~\cite{KLOE:2008fmq,KLOE:2010qei,KLOE:2012anl,KLOE-2:2017fda} and \Babar~\cite{BaBar:2012bdw}.
\Babar has since presented a new measurement with its full initial-state-radiation data set, which separates the pion and muon pairs through their angular distributions~\cite{BaBar:2026pic,BaBar:2026jzp}.
The \CMDthree{} result is therefore typically excluded from global data-driven averages of $a_\mu^{\text{HVP, LO}}$.
For comprehensive reviews of these HVP tensions, see~\cite{Aliberti:2025beg, Jegerlehner:2026zwk, Hertzog:2025ssc}.
HVP dominates the error budget of $a_\mu$, and the two-pion final state accounts for the largest part of the HVP contribution~\cite{Aoyama:2020ynm,Aliberti:2025beg}.
Pion channels are equally crucial for the hadronic contribution to the running coupling $\alpha(s)$, obtained from a dispersion integral over $e^+e^-$ annihilation data, and for $\alpha(M_Z^2)$, a key input for precision studies of electroweak pseudo-observables at future colliders such as FCC-ee~\cite{Gluza:2024sop,FCC:2025lpp}.

At low-energy flavour factories, the $e^+e^- \to \pi^+\pi^-$ cross section is measured either by an energy scan, varying the centre-of-mass energy, or by radiative return~\cite{Arbuzov:1998te,Binner:1999bt}, where an initial-state (ISR) photon gives access to a range of energies at fixed beam energy.
Both methods require careful treatment of radiative corrections, which can significantly affect the measured cross section and thus $a_\mu^{\text{HVP}}$.
Radiative corrections to $e^+e^-$ processes have a long history~\cite{Kuraev:1985hb,Berends:1987ab,Nicrosini:1987ch,Nicrosini:1988hi,Berends:1973jb,Berends:1983mi,Jadach:1984hwn,Jadach:1988ec,Jadach:1996is,Jadach:1993hs,Yennie:1961ad,Baier:1965jz,Baier:1966jf,Kuraev:1977rp,Baier:1980kx,Berends:1981rb,Chen:1974wv,Budnev:1975poe,Passarino:1979jh,Bardin:1989tq,Bardin:1999ak}, including the structure-function and parton-shower methods for QED corrections.
Today such corrections are implemented in Monte Carlo (MC) event generators, essential tools for experimental analyses, whose precision is often limited by the perturbative order of the radiative corrections.
A reliable extraction of $a_\mu^{\text{HVP}}$ from $e^+e^- \to \pi^+\pi^-$ requires higher orders.
The standard generator for the pion channel, \Phokhara~\cite{Rodrigo:2001kf,Czyz:2000wh,Czyz:2002np,Czyz:2003ue,Czyz:2004ua,Czyz:2005as,Czyz:2007wi,Czyz:2010hj}, includes the next-to-leading-order (NLO) radiative corrections for both the $\pi^+\pi^-\gamma$ and $\mu^+\mu^-\gamma$ channels~\cite{Campanario:2013uea,Campanario:2019mjh}, but as a fixed-order prediction it does not include multiple photon emission~\cite{WorkingGrouponRadiativeCorrections:2010bjp,BaBar:2023xiy}.
For pions, multiple photon emission matched to NLO was first implemented in the \BabaYaga event generator~\cite{Balossini:2006wc,Balossini:2008xr} with a parton shower~\cite{Budassi:2024whw}, recently extended to the $2\rightarrow3$ radiative channel~\cite{Budassi:2026lmr}.
Fixed-order corrections to $e^+e^- \to \pi^+\pi^-(\gamma)$ are also available from the MC integrator \McMule~\cite{Banerjee:2020rww}, which provides the $\pi^+\pi^-$ NLO corrections with the pion form factor inside the loops and the hadronic vacuum polarisation in the NNLO initial-state corrections~\cite{Fang:2025mhn}.
Its NLO initial-state corrections to $\pi^+\pi^-\gamma$ were presented together with those of other codes in~\cite{Aliberti:2024fpq}.
The disperon QED method~\cite{Fang:2025mhn} brings dispersive input, such as the pion vector form factor and the hadronic vacuum polarisation, into the loop amplitudes of Monte Carlo calculations, with the form factor entering the loops in form-factor scalar QED (\FsQED{}).

A different treatment of the form factor in the loops, attaching it directly to the photon propagators via generalised vector-meson dominance (\GVMD{}), was implemented for this process in Ref.~\cite{Budassi:2024whw} and extended to the NLO corrections to $\pi^+\pi^-\gamma$~\cite{CarloniCalame:2026vfc,CarloniCalame:2026gdm}.
Towards full NNLO~\cite{Dave:2026pvq,PetitRosas:2026iuq,Pozzoli:2026eiu,Fadin:2023phc,Badger:2023xtl,PetitRosas:2025xhm}, tensor decompositions within the \GVMD{} framework for radiative return have provided the higher-order-in-$\epsilon$ real-virtual building blocks needed to add two-loop virtual and double-real photon corrections to Monte Carlo event generators.

In this work, we present a new calculation of the radiative corrections to $e^+e^- \to \pi^+\pi^-(\gamma)$ in the Yennie-Frautschi-Suura (YFS) formalism~\cite{Yennie:1961ad}, which resums the soft photons that accompany every event to all orders, unlike fixed-order calculations.
The calculation was first implemented in \Sherpa~\cite{Sherpa:2024mfk} with the methods of~\cite{Krauss:2022ajk,Price:2025fiu}.
We present its implementation in the well-established $e^+e^-$ generator \kkmc~\cite{Jadach:1999vf,Jadach:2022mbe}, with the higher-order corrections constructed in the coherent exclusive exponentiation (CEEX) method~\cite{Jadach:2000ir}, and in \Sherpa with the newly implemented \ACRAIC~\cite{Price:2026qop} method.
The two independent implementations provide a powerful cross-check of each other.
We estimate the missing higher orders with theory nuisance parameters~\cite{Tackmann:2024kci,Lim:2024nsk}, as introduced for YFS resummation in Ref.~\cite{Price:2026qop}, which give an uncertainty for every bin of every distribution
tied to identified missing terms (\cref{sec:tnp}).

\section{\label{sec:theory} Theory} The charged pions are treated as point-like scalars, and all corrections are computed in scalar QED (\sQED{}) dressed with the pion form factor.
We model the $\pi^\pm$ as dynamical scalar particles which couple to the photon through the \sQED{} Lagrangian,
\begin{equation}
  \mathcal{L}_{\rm \sQED{}} = (D_\mu\pi)^\dagger (D^\mu\pi) - m_\pi^2\,\pi^\dagger\pi\,,
  \qquad D_\mu = \partial_\mu + i e A_\mu\,.
  \label{eq:sqed}
\end{equation}
With the momenta labelled $e^-(p_1)\,e^+(p_2)\to\pi^-(p_3)\,\pi^+(p_4)\,\gamma(k_1)\cdots\gamma(k_{n_\gamma})$, the Feynman rules are
\begin{align}
  \gamma^*_\mu(q)\,\pi^+\pi^-:      & \quad i e\,(p_4-p_3)_\mu\,F_\pi(q^2)\,,
  \label{eq:fr_ssv}                                                            \\
  \gamma_\mu\gamma_\nu\,\pi^+\pi^-: & \quad 2 i e^2\,g_{\mu\nu}\,F_\pi(q^2)\,,
  \label{eq:fr_vvss}
\end{align}
with the scalar propagator $i/(p^2-m_\pi^2+i0)$ and the usual QED rules for the electrons.
The form factor is evaluated in every amplitude, with resolved or unresolved photons, and $F_\pi\equiv1$ recovers point-like \sQED{}.

By the YFS theorem the all-order differential cross-section is,
\begin{align}\label{eq:masterYFS}
  \ddone\sigma^{\left(\infty\right)} & = \sum_{n_\gamma=0}^\infty
  \frac{e^{Y(\Omega)}}{n_\gamma!}\,
  \ddone{\Phi_n}
  \left[\prod_{i=1}^{n_\gamma}\ddone{\Phi_i^\gamma}\,\eik{i}\,\Theta(k_i,\Omega)\right]
  \nonumber                                                                                \\
                                     & \quad\times\left(\tilde{\beta}_0\left(\Phi_n\right)
  + \sum_{j=1}^{n_\gamma}\frac{\tilde{\beta}_{1}(\Phi_{n+1})}{\eik{j}}
  +\sum_{\substack{j,k=1\\ j<k}}^{n_\gamma}
  \frac{\tilde{\beta}_{2}(\Phi_{n+2})}{\eik{j}\eik{k}}
  + \cdots\right)\, ,
\end{align}
where $\eik{i}$ is the eikonal factor of photon $i$, $Y(\Omega)$ the YFS form factor, the sum of the virtual and real soft-photon integrals of every pair of charged legs over the soft-photon domain $\Omega$, outside which $\Theta(k_i,\Omega)$ keeps the resolved photons, and $\tilde\beta_{n_\gamma}$ the infrared-finite residuals, each summed over any number of virtual
photons.
These are defined as in Refs.~\cite{Jadach:2000ir,Schonherr:2008av,Price:2026qop}.
Both generators implement the amplitude-level form of \cref{eq:masterYFS}.

\subsection{Pion Form Factor}
\label{app:pion_ff}\label{sec:ff_amplitudes}
  The pion electromagnetic form factor $F_\pi(q^2)$ is parameterised in
  vector meson dominance (VMD) \cite{PhysRev.124.953, RevModPhys.50.261} as a normalised sum of resonance
  propagators $P_v$ with mixing factors $M_v$,
  \begin{equation}
    F_\pi(q^2)=\frac{\sum_v c_v\,P_v(q^2)\,M_v(q^2)}{\sum_v c_v}\,,\qquad
    M_v(q^2)=1+\sum_x c_x\,\frac{q^2}{m_x^2}\,b_x(q^2)\,,
    \label{eq:pion_ff}
  \end{equation}
  with complex amplitudes $c=|c|\,e^{i\varphi}$, free masses and widths, and $P_v(0)=1$, so that
  $F_\pi(0)=1$. We use the STRONG2020
  parametrisation~\cite{Aliberti:2024fpq} in both generators, with $v\in\{\rho,\rho',\rho'',\rho'''\}$
  in the Gounaris--Sakurai form, $\rho$--$\omega$ and $\rho$--$\phi$ mixing on
  the $\rho$ only, $M_{\rho'}=M_{\rho''}=M_{\rho'''}=1$, and its auxiliary functions and resonance parameters.
The narrow $\omega$ and $\phi$ of the mixing factors are constant-width Breit--Wigner propagators,
\begin{equation}
  b_v(q^2) = \frac{m_v^2}{m_v^2 - q^2 - i m_v \Gamma_v}\,,
  \qquad v \in \{\omega,\phi\}\,.
  \label{eq:bw}
\end{equation}

The broad $\rho$ resonances, with their energy-dependent width, take the Gounaris--Sakurai
(GS) form~\cite{Gounaris:1968mw},
\begin{equation}
  P_v(q^2) = \frac{m_v^2 + d(m_v)\,m_v\,\Gamma_v}
  {m_v^2 - q^2 + f(q^2,m_v,\Gamma_v) - i\,m_v\,\Gamma(q^2,m_v,\Gamma_v)}\,,
  \qquad v \in \{\rho,\rho',\rho'',\rho'''\}\,,
  \label{eq:gs}
\end{equation}
where $\Gamma(q^2,m_v,\Gamma_v)$ is the energy-dependent width of the
$\pi^+\pi^-$ channel and $f$ and $d$ are the Gounaris--Sakurai corrections
that keep the propagator analytic and normalise it at $q^2=0$.
In \Comix the form factor is a vertex factor at the virtuality of the photon current
joining the pion pair, so real photons couple point-like and the same currents serve every multiplicity.

The measured form factor fixes the coupling of one virtual photon to the pion pair, and how it enters amplitudes with more photons, in particular inside loops, is a matter of modelling,
with three treatments in use. The factorised approach,
\FxsQED{}~\cite{Campanario:2019mjh,Aliberti:2024fpq}, multiplies the point-like
scalar-QED amplitudes by $F_\pi$ at the virtuality of the $s$-channel
photon~\cite{Hoefer:2001mx,Bystritskiy:2005ib,Arbuzov:2020foj}. In the
generalised vector-meson dominance (\GVMD{})
approach~\cite{Sakurai:1972wk,Ignatov:2022iou} the form factor is a
sum of vector-meson poles attached to every photon coupling to the
pions, loop photons included, which turns the loops into point-like
integrals with massive photons. \FsQED{} dresses every pion vertex with
form factors and evaluates the loops through dispersion
relations~\cite{Colangelo:2014dfa,Colangelo:2022lzg}. Disperon
QED~\cite{Fang:2025mhn} brings \FsQED{} to loop-level Monte Carlo by writing
the form factor on a photon propagator as a spectral integral over photons of variable mass, the
disperons, whose loops are standard while the spectral
integral is sampled with the phase space. \GVMD{} and \FsQED{} agree well
for the charge asymmetry, and both describe the \CMDthree{}
measurement~\cite{CMD-3:2023alj,Ignatov:2022iou,Budassi:2024whw}, which the
factorised approach does not. The diagrams and the form-factor arguments of the
three approaches at next-to-leading order are collected in~\cite{Budassi:2024whw}.

\begin{figure}[htb]
  \centering
  \includegraphics[width=\textwidth,height=0.57\textheight,keepaspectratio]{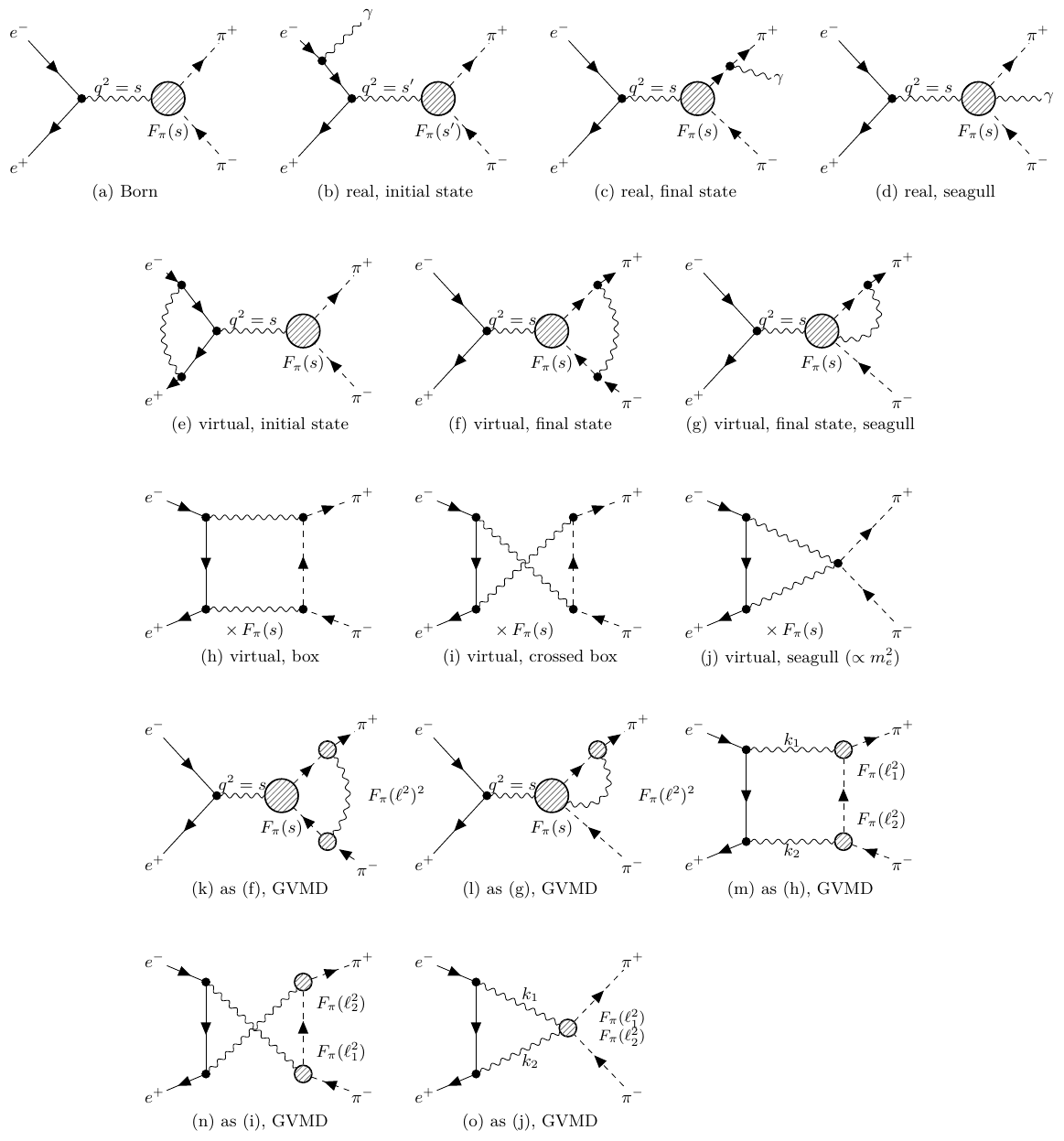}
  \caption{Where the pion form factor enters the amplitudes, see
    \cref{sec:ff_amplitudes}. Large hatched circles are $F_\pi(q^2)$ at the
    virtuality of the $s$-channel photon, small hatched circles are
    $F_\pi(\ell^2)$ of a loop photon, and dots are point-like vertices. Tree-level
    and real-emission amplitudes (a)--(d), one-loop corrections for point-like
    pions in the factorised approach (e)--(j), and the same loops in \GVMD{}
    (k)--(o). Mirror diagrams and counterterms are not shown.}
  \label{fig:ff_insertion}
\end{figure}

\ACRAIC and \kkmc implement the factorised approach,
\Cref{fig:ff_insertion}~(a)--(j), and optionally \GVMD{} in the loops,
panels (k)--(o). In both generators the
form factor sits on the amplitude, separately for each partition of the
photons. It is evaluated at $q^2=s$ in the Born amplitude (a), at the
virtuality $s'=(p_1+p_2-k_1)^2$ of the $s$-channel photon of the partition for
a photon from the beams (b), and at $q^2=s$ for a photon from the pions (c) and in
the seagull (d), where the same $F_\pi(s)$ on both vertices keeps the sum
gauge invariant.
The one-loop corrections (e)--(j), computed for point-like pions,
multiply the Born amplitude with $F_\pi(s)$ and comprise the initial- and
final-state vertices (e, f), the final-state seagull triangle
(g), the initial-final box and crossed box (h, i) and the two-photon seagull
(j), suppressed by $m_e^2$.

In \GVMD{} the initial-state correction is unchanged, the
form factor sitting on the $s$-channel vertex. In the final and initial-final
loops $F_\pi(\ell^2)$ sits on every $\gamma\pi\pi$ and
$\gamma\gamma\pi\pi$ vertex of a loop photon of virtuality $\ell^2$, so a final-state loop photon (k, l)
carries $F_\pi(\ell^2)^2$ and the two photons of the
boxes and the two-photon seagull (m)--(o) carry
$F_\pi(\ell_1^2)F_\pi(\ell_2^2)$. Inside the loops we
use the five-pole Breit--Wigner fit to the \CMDthree{} data of~\cite{Budassi:2024whw},
and the tree amplitudes keep \cref{eq:pion_ff}. The loop kernels with massive photons, rederived and
checked in \cref{sec:virtual-2to2}, enter as
relative corrections to the Born amplitude.

\subsection{Real Corrections}
\label{sec:real}
The real corrections enter through the residuals $\tilde\beta_{n_\gamma}$ of \cref{eq:masterYFS}, the $n_\gamma$-photon amplitudes with their soft limits removed.
Both generators work with amplitudes for fermion helicities $\{\lambda\}$ and photon helicity $\sigma$, in the conventions of CEEX~\cite{Jadach:2000ir} and \ACRAIC~\cite{Price:2026qop}.
For a photon $(k,\sigma)$ the one-photon residual is the exact amplitude minus the eikonal current times the Born amplitude,
\begin{equation}\label{eq:RealSub}
  \hat\beta_1(k,\sigma;\{\lambda\})=\mathfrak{M}_1(k,\sigma;\{\lambda\})
  -\mathfrak{s}^{(\sigma)}(k)\,\mathfrak{M}_0(\{\lambda\})\,,\qquad
  \mathfrak{s}^{(\sigma)}(k)=e\sum_iQ_i\,Z_i\,
  \frac{\varepsilon^{(\sigma)*}\cdot p_i}{k\cdot p_i}\,,
\end{equation}
summed over the charged legs $i$ with charge $Q_i$ and $Z_i=1\,(-1)$ for incoming (outgoing) legs, so that $\sum_\sigma|\mathfrak{s}^{(\sigma)}(k)|^2=16\pi^3\,\tilde S(k)$.
Both $\mathfrak{s}$ and $\mathfrak{M}_1$ use the polarisation vectors $\varepsilon^{(\sigma)}(k)$ of the external photon currents of \Comix.
The residual is finite as $k\to0$, and the residuals of all photons and photon pairs are summed coherently before squaring, which keeps the initial-final interference at every order.

In \Sherpa every tree-level amplitude in the residuals is generated by \Comix~\cite{Gleisberg:2008fv}, from the charged scalar of the model with its $\gamma\pi\pi$ and $\gamma\gamma\pi\pi$ vertices and the form factor as a vertex factor.
Both electrons and pions are kept massive and the residuals are evaluated at the physical event momenta.
Assigning a photon to the initial or final stage fixes the virtuality of the $s$-channel propagator and the form factor, which \Comix applies as a shift of the propagator momentum without rebuilding the currents.
For a new final state \ACRAIC therefore needs only the model with its form factor and the virtual correction.

In \kkmc the pion amplitudes are coded analytically in the CEEX spinor formalism~\cite{Jadach:1998wp}. With the lepton current $J^\mu=\bar v(p_2)\gamma^\mu u(p_1)$, the pion current $K=p_4-p_3$ and $s'=P_X^2$ for the $s$-channel momentum $P_X$ of the partition, the Born amplitude and the initial- and final-state eikonal currents of \cref{eq:RealSub}, with $\varepsilon\equiv\varepsilon^{(\sigma)*}(k)$, are
\begin{align}
  \mathfrak{M}_0 & =e^2Q_eQ_\pi\,F_\pi(s')\,\frac{J\cdot K}{s'}\,, \nonumber                                            \\
  \mathfrak{s}_I & =e\,\Big(\frac{\varepsilon\cdot p_2}{k\cdot p_2}-\frac{\varepsilon\cdot p_1}{k\cdot p_1}\Big),\qquad
  \mathfrak{s}_F=e\,\Big(\frac{\varepsilon\cdot p_3}{k\cdot p_3}-\frac{\varepsilon\cdot p_4}{k\cdot p_4}\Big).
  \label{eq:kkmc_born}
\end{align}
The one-photon residuals $h=\mathfrak{M}_1-\mathfrak{s}\,\mathfrak{M}_0$ are, for initial-state emission,
\begin{equation}
  h_I=e^3Q_e^2Q_\pi\,\frac{F_\pi(s')}{s'}\,\bar v(p_2)\Big[
    \frac{\slashed{K}(\slashed{p}_1-\slashed{k}+m_e)\slashed{\varepsilon}}{-2p_1\cdot k}
    +\frac{\slashed{\varepsilon}(\slashed{k}-\slashed{p}_2+m_e)\slashed{K}}{-2p_2\cdot k}\Big]u(p_1)
  -\mathfrak{s}_I\,\mathfrak{M}_0\,,
  \label{eq:kkmc_hI}
\end{equation}
and for final-state emission the two pion diagrams and the seagull, contracted with $\varepsilon$, reduce to
\begin{equation}
  h_F=e^3Q_eQ_\pi^2\,\frac{F_\pi(s')}{s'}\,\big(2\,\varepsilon\cdot J-D\,k\cdot J\big)\,,\qquad
  D=\frac{\varepsilon\cdot p_3}{k\cdot p_3}+\frac{\varepsilon\cdot p_4}{k\cdot p_4}\,,
  \label{eq:kkmc_hF}
\end{equation}
which is finite as $k\to0$ and vanishes for $\varepsilon\to k$.
The two-photon residuals follow likewise from the exact amplitudes $\mathfrak{M}_{12}$, with all photon orderings and seagulls,
\begin{equation}
  \hat\beta_2^{II,FF}=\mathfrak{M}_{12}-\mathfrak{s}_1\mathfrak{s}_2\,\mathfrak{M}_0-\mathfrak{s}_1h_2-\mathfrak{s}_2h_1\,,\qquad
  \hat\beta_2^{IF}=\frac{L_1\cdot h_2}{s'}\,,
  \label{eq:kkmc_beta2}
\end{equation}
where for one initial- and one final-state photon the amplitude factorises through the virtual photon as $(\mathfrak{s}_1J+L_1)\cdot(\mathfrak{s}_2H_0+h_2)/s'$, with $H_0$ the photonless pion current and $L_1$, $h_2$ the hard parts of the beam and pion currents.
In the CEEX sum each residual carries the eikonal factors of the other photons of its partition, and at second order the one-photon residuals also the one-loop initial- and final-state vertex factors.
For a momentum $p=(E,\mathbf p)$ with helicity $\lambda=\pm1$ and
\[
  |\mathbf p|=\sqrt{p_x^2+p_y^2+p_z^2}\,,\qquad
  \theta=\arccos\!\bigl(p_z/|\mathbf p|\bigr)\,,\qquad
  \phi=\operatorname{atan2}(p_y,p_x)\,,
\]
the spinors in the Dirac basis are
\begin{gather}
  u(p,\lambda)=\begin{pmatrix}\sqrt{E+m}\,\chi_\lambda(\mathbf p) \\[0.4em]
    \frac{\boldsymbol\sigma\cdot\mathbf p}{\sqrt{E+m}}\,\chi_\lambda(\mathbf p)\end{pmatrix},\qquad
  v(p,\lambda)=\begin{pmatrix}\frac{\boldsymbol\sigma\cdot \mathbf p}{\sqrt{E+m}}\,\chi_\lambda(\mathbf p) \\[0.4em]
    \sqrt{E+m}\,\chi_\lambda(\mathbf p)\end{pmatrix}, \nonumber \\
  \chi_+=\begin{pmatrix}\cos\frac{\theta}{2} \\[0.2em] e^{i\phi}\sin\frac{\theta}{2}\end{pmatrix},\qquad
  \chi_-=\begin{pmatrix}-\sin\frac{\theta}{2} \\[0.2em] e^{i\phi}\cos\frac{\theta}{2}\end{pmatrix}.
  \label{eq:kkmc_spinors}
\end{gather}

\subsection{Virtual Corrections}
\label{sec:virtual-2to2}

We compute the complete one-loop QED corrections to
$e^+e^-\to\pi^+\pi^-$ (momenta as in \cref{sec:theory}), also given in~\cite{Budassi:2024whw} and
available in tools such as \OpenLoops~\cite{Buccioni:2019sur}, in scalar QED, where point-like
pions couple to the photon through the three-point and the two-photon seagull
vertices, keeping the full electron-mass dependence. The amplitudes were generated with
\FeynArts~\cite{Hahn:2000kx}, the Dirac and Lorentz algebra was done
with \FeynCalc~\cite{Mertig:1990an,Shtabovenko:2023idz}, and the tensor integrals
were reduced to $A_0$, $B_0$, $C_0$ and
$D_0$~\cite{Passarino:1978jh,Denner:1991kt}, evaluated with \LoopTools~\cite{Hahn:1998yk}.
The poles, the on-shell renormalisation constants and the analytic finite
parts that validate the numerics came from Package-X~\cite{Patel:2015tea,Patel:2016fam}
through FeynHelpers~\cite{Shtabovenko:2016whf}.

Independent electron and pion charges $Q_e$ and $Q_\pi$ split the
interference $2\Re\big(\mathcal{M}_1\mathcal{M}_0^*\big)$ into initial-state (ISR),
final-state (FSR) and initial-final interference (IFI) corrections, proportional to
$Q_e^4Q_\pi^2$, $Q_e^2Q_\pi^4$ and $Q_e^3Q_\pi^3$ respectively. Since gauge invariance
holds for any two charges, each subset is separately gauge invariant and
UV finite after renormalisation. ISR is the electron vertex correction, FSR the
pion vertex correction and the two diagrams with a virtual photon between the
seagull vertex and an outgoing pion, and IFI the direct and crossed boxes and
the diagram with both photons between the electron line and the seagull vertex.
We renormalise on shell~\cite{Denner:1991kt} with $\alpha=\alpha(0)$. As in the
STRONG2020 comparisons, vacuum polarisation in the $s$-channel photon is not
included anywhere in this paper, so the charge and photon-field
renormalisations cancel and only the field renormalisation of the external legs remains. The wave-function
renormalisation constants are
\begin{equation}
    \delta Z_e = -\Re \left[\Sigma_e^V (m_e^2)\right] -  2m_e^2 \frac{\partial}{\partial p^2}\Re \left.\left[ \Sigma_e^V (p^2) + \Sigma_e^S (p^2)\right]\right|_{p^2 = m_e^2},
\end{equation}
and
\begin{equation}
    \delta Z_\pi = -\Re\left.\left[\frac{\partial \Sigma_\pi (p^2)}{\partial p^2}\right]\right|_{p^2 = m_\pi^2},
\end{equation}
with $\Sigma_e(p)=\slashed{p}\,\Sigma_e^V(p^2)+m_e\Sigma_e^S(p^2)$ split into the vector and
scalar parts of the electron self-energy and $\Sigma_{\pi}$ the pion self-energy. Reduced to
Passarino-Veltman master integrals they read
\begin{align}
  \delta Z_e   & = \frac{\alpha}{4\pi}\left[1-\frac{A_0(m_e^2)}{m_e^2}
  +4m_e^2\,B_0'(m_e^2,0,m_e^2)\right],               \\
  \delta Z_\pi & = \frac{\alpha}{4\pi}\left[2B_0(m_\pi^2,0,m_\pi^2)
  +4m_\pi^2\,B_0'(m_\pi^2,0,m_\pi^2)\right],
\end{align}
with $B_0'$ the derivative in the external momentum
squared, the electron constant reproducing the familiar
$\delta Z_e=-\frac{\alpha}{4\pi}\big(\Delta_{\rm UV}+2\Delta_{\rm IR}
+3\ln(\mu^2/m_e^2)+4\big)$. Mass counterterms, which sit on
internal electron and pion propagators, do not appear. The only internal line of
the tree is the $s$-channel photon, the loop propagators already carry the physical
mass, a counterterm there being of order $\alpha^2$, and on the external legs the
on-shell mass counterterms cancel the self-energies at $p^2=m^2$, leaving the
residues $\delta Z_e$ and $\delta Z_\pi$, which are needed for the real-virtual
correction of \cref{sec:realvirtual}. UV and IR singularities are regulated
dimensionally in $D=4-2\epsilon$, and an infrared photon
mass changes only the master integrals, not the coefficients below.

We write the virtual correction as
\begin{equation}
  \mathrm{d}\sigma_V = \mathrm{d}\sigma_0\left(\delta_{\rm ISR}+\delta_{\rm FSR}
  +\delta_{\rm IFI}\right),
  \qquad
  \delta_X = \frac{\alpha}{2\pi}\sum_k r^X_k\,\Re\,I_k + \delta^X_{\rm CT},
\end{equation}
with $\delta^{\rm ISR}_{\rm CT}=2\delta Z_e$, $\delta^{\rm FSR}_{\rm CT}=2\delta
Z_\pi$ and $\delta^{\rm IFI}_{\rm CT}=0$. The scalar integrals $I_k$ are
normalised as in \LoopTools, $I=\mu^{4-D}/(i\pi^{D/2}r_\Gamma)\int\mathrm{d}^Dq\,
(\ldots)$. The results are given at $D=4$, including the rational terms left
by the $\mathcal{O}(\epsilon)$ parts of $D$-dependent coefficients of UV poles, in terms of
$s=(p_1+p_2)^2$, $t=(p_1-p_3)^2$ and $u=(p_1-p_4)^2=2m_e^2+2m_\pi^2-s-t$. With $M^2=m_e^2+m_\pi^2$
the Born matrix element is proportional to
\begin{equation}
  K = (t-M^2)(u-M^2)-s\,m_\pi^2\,.
\end{equation}

\paragraph{Initial-state corrections.}

  Relative to the Born, the loop gives coefficients of
  $A_0(m_e^2)$, $B_0(s,m_e^2,m_e^2)$, $B_0(m_e^2,0,m_e^2)$ and the vertex triangle,
  with the rational term $m_e^2\rho-2$ at $D=4$. Combining
  with $2\delta Z_e$ through $A_0(m_e^2)=m_e^2\big[B_0(m_e^2,0,m_e^2)-1\big]$,
  all constant terms cancel and
  \begin{align}
    \delta_{\rm ISR} = \frac{\alpha}{2\pi}\Big\{
     & \left(m_e^2\rho-3\right)\Big[B_0(s,m_e^2,m_e^2)-B_0(m_e^2,0,m_e^2)\Big]
    \nonumber \\
     & - 2\left(s-2m_e^2\right)C_0^{ee\gamma}(s)
    + 4m_e^2\,B_0'(m_e^2,0,m_e^2)\Big\},
    \label{eq:virt-isr}
  \end{align}
  where $C_0^{ee\gamma}(s)$ is the vertex triangle with two electron
  and one photon propagator, and the UV poles cancel between the two $B_0$
  functions.

\paragraph{Final-state corrections.}
For point-like pions the FSR correction does not depend on the scattering angle,
\begin{align}
  \delta_{\rm FSR} = \frac{\alpha}{2\pi}\Bigg\{
   & \frac{4\left(s-2m_\pi^2\right)}{s-4m_\pi^2}
    \Big[B_0(m_\pi^2,0,m_\pi^2)-B_0(s,m_\pi^2,m_\pi^2)\Big]
  - 2\left(s-2m_\pi^2\right)C_0^{\pi\pi\gamma}(s)
  \nonumber \\
   & + 4m_\pi^2\,B_0'(m_\pi^2,0,m_\pi^2)\Bigg\},
  \label{eq:virt-fsr}
\end{align}
with $2\delta Z_\pi$ combined with the loop and
$C_0^{\pi\pi\gamma}(s)$ the vertex triangle with two pion and one
photon propagator. The UV poles again cancel between the two $B_0$ functions, and the
structure is that of~\cite{Budassi:2024whw}.

\paragraph{Initial-final interference.}
The IFI is odd under $t\leftrightarrow u$ and reads, at $D=4$,
\begin{align}
  \delta_{\rm IFI} = \frac{\alpha}{2\pi}\,\frac{s}{K}\Bigg\{
   & \frac{4\,m_e^2\,(t-u)}{s-4m_e^2}
  \Big[B_0(m_e^2,0,m_e^2)-B_0(s,0,0)\Big]
  \nonumber                                                                     \\
   & + (t-u)\left[\frac{s^2-8\,s\,m_e^2+8m_e^4}{s-4m_e^2}\,
    C_0^{\gamma\gamma e}(s) + \left(s-2m_\pi^2\right)C_0^{\gamma\gamma\pi}(s)\right]
  \nonumber                                                                     \\
   & + 2\Big[(t-m_e^2)(t-M^2)\,C_0(t)-(u-m_e^2)(u-M^2)\,C_0(u)\Big]
  \nonumber                                                                     \\
   & + (t-M^2)\Big[2(t-M^2)^2+s\,(t-m_e^2)\Big]D_0(t)
  \nonumber \\
   & - (u-M^2)\Big[2(u-M^2)^2+s\,(u-m_e^2)\Big]D_0(u)\Bigg\}.
  \label{eq:virt-ifi}
\end{align}
Here $D_0(t)$ and $D_0(u)$ are the direct and crossed boxes, with two
photon, one electron and one pion propagator, and the triangles
arise from their reduction. $C_0^{\gamma\gamma e}(s)$ and $C_0^{\gamma\gamma\pi}(s)$
have two photon and one electron or pion propagator, with external
invariants $(s,m_e^2,m_e^2)$ and $(s,m_\pi^2,m_\pi^2)$, and $C_0(t)$ and $C_0(u)$ have one
photon, one electron and one pion propagator, with $(t,m_e^2,m_\pi^2)$ and
$(u,m_e^2,m_\pi^2)$.
The $B_0$ term is UV finite and vanishes for $m_e\to0$, and
$C_0^{\gamma\gamma e}$ is IR finite since neither photon connects two external legs,
so their coefficients leave no rational term at $D=4$. The IR singularity sits in
the boxes and in $C_0(t)$ and $C_0(u)$, and the IR poles of $\delta_{\rm ISR}$,
$\delta_{\rm FSR}$ and $\delta_{\rm IFI}$ cancel against the YFS form factor. The
finite parts and the $1/\epsilon$ poles of each subset agree with~\cite{Budassi:2024whw}
to $10^{-16}$ at every phase-space point tested, and the expressions are
algebraically identical.
In the \GVMD{} treatment of the loops (\cref{sec:ff_amplitudes}) the final-state and
initial-final corrections combine the same kernels with loop
photons of complex squared mass, Eqs.~(44)--(46) of~\cite{Budassi:2024whw}.

\subsection{Real-Virtual Corrections}
\label{sec:realvirtual}

\begin{figure}[htb]
  \centering
  \includegraphics[width=\textwidth]{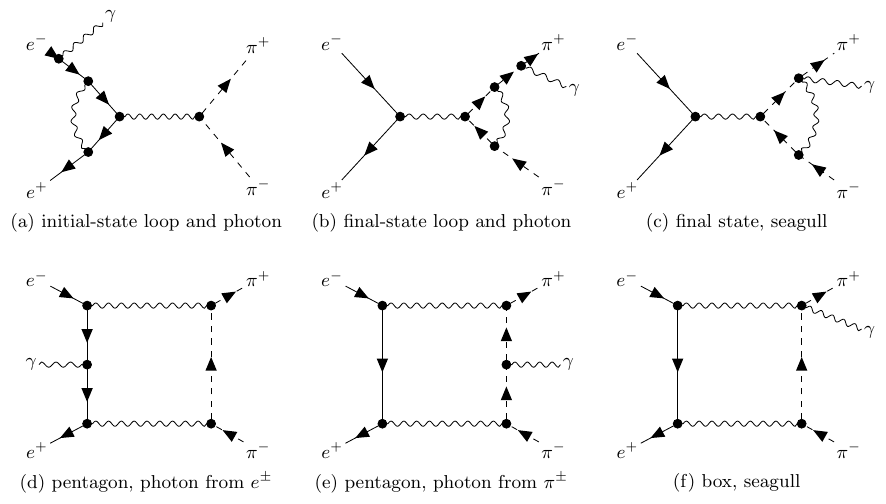}
  \caption{One-loop diagrams of $e^+e^-\to\pi^+\pi^-\gamma$
    with point-like pions, without mirror diagrams, crossed boxes and pentagons,
    other placements of the real photon and counterterms.
    Interfered with the tree diagrams with initial-state (final-state)
    emission, (a) contributes to $\mathcal{V}_{62}$ ($\mathcal{V}_{53}$), (b)
    and (c) to $\mathcal{V}_{35}$ ($\mathcal{V}_{26}$), (d) to
    $\mathcal{V}_{53}$ ($\mathcal{V}_{44}$), and (e) and (f) to
    $\mathcal{V}_{44}$ ($\mathcal{V}_{35}$).}
  \label{fig:rv_diagrams}
\end{figure}

The complete one-loop QED correction to
$e^+e^-\to\pi^+\pi^-\gamma$, with momenta $e^-(p_1)$, $e^+(p_2)$,
$\pi^-(p_3)$, $\pi^+(p_4)$ and $\gamma(k_1)$, is computed in the same way, in scalar QED
with point-like pions and the full electron mass. Its initial-state~\cite{Rodrigo:2001jr,Kuhn:2002xg}
and final-state~\cite{Czyz:2003ue} parts and the complete \sQED{}
correction~\cite{Campanario:2019mjh} are known, and it has been matched to a parton
shower~\cite{Budassi:2026lmr}. It provides the exact real-virtual correction to
$\tilde\beta_1$ at second order and the virtual correction for processes with a
photon in the Born, such as the tagged-photon \BES{} and \Bscen{} scenarios.
\FeynArts~\cite{Hahn:2000kx} and \FeynCalc~\cite{Mertig:1990an,Shtabovenko:2023idz}
give the amplitudes as tensor coefficients of one- to five-point functions, evaluated
with \LoopTools~\cite{Hahn:1998yk} or \Collier~\cite{Denner:2016kdg}. The tree
amplitude has five diagrams, two with the photon from the electrons, two from
the pions and one seagull (loop examples in \cref{fig:rv_diagrams}).
The on-shell counterterms in the $\alpha(0)$ scheme comprise the field counterterms
$\delta Z_e$ and $\delta Z_\pi$ of \cref{sec:virtual-2to2} and, since the real photon
puts an electron or pion line off shell, mass counterterms on these internal propagators,
\begin{align}
  \delta m_e     & = m_e\,\frac{\alpha}{4\pi}\Big[3\Delta_{\rm UV}
  +3\ln\frac{\mu^2}{m_e^2}+4\Big],                                            \\
  \delta m_\pi^2 & = \frac{\alpha}{4\pi}\Big[4m_\pi^2\,B_0(m_\pi^2,0,m_\pi^2)
    -A_0(m_\pi^2)\Big]
\end{align}
with $m_0=m+\delta m_e$ and $m_{\pi,0}^2=m_\pi^2+\delta m_\pi^2$, entering the
subsets with initial- and final-state emission respectively. As for the $2\to2$
correction, closed charged loops, and thus vacuum polarisation, are dropped.

Tagging every photon coupling with the charge of its line gives,
\begin{equation}
  2\,\mathrm{Re}\big(\mathcal{M}_1\mathcal{M}_0^*\big)
  = \sum_{m+n=8} Q_e^m Q_\pi^n\, \mathcal{V}_{mn}\,.
\end{equation}
With three photon couplings in the tree and five in the
loop, every term is of eighth order in the charges, and since both amplitudes couple at
least once to each line, $2\le m,n\le6$, leaving five separately gauge-invariant
subsets. $\mathcal{V}_{62}$ holds the initial-state loop on
initial-state emission, $\mathcal{V}_{26}$ the final-state loop on
final-state emission, $\mathcal{V}_{44}$ the mixed terms, including the
boxes and pentagons times the Born with final-state emission, while
$\mathcal{V}_{53}$ and $\mathcal{V}_{35}$, odd under $\pi^+\leftrightarrow\pi^-$,
generate the charge asymmetry. The
coefficients are kept as functions of $D=4-2\epsilon$, so the rational terms
are exact, and the finite part can be given in conventional dimensional
regularisation, with a $D$-dimensional photon polarisation sum, or in the
'tHooft--Veltman scheme, which differ by the infrared pole times the
$\mathcal{O}(\epsilon)$ part of the Born. The YFS subtraction uses the
four-dimensional eikonal factor times the Born.

Each charge subset is checked at hard points at $\sqrts=0.7$ and $1.02$~GeV. The
ultraviolet poles cancel to $10^{-19}$ relative to the Born. A fit of one coefficient per
renormalisation constant over all points and subsets returns $\pm1$, with the on-shell constants
computed independently. The $1/\epsilon$ infrared pole equals the eikonal sum over the four
charged legs times the tree $|\mathcal{M}_0|^2$ to $10^{-13}$. In the soft limit the finite part
of every subset tends, linearly in the photon energy, to the eikonal factor times the $2\to2$
correction of \cref{sec:virtual-2to2}. This validates it against the $2\to2$ result and shows
that $\tilde\beta_1^{(2)}$ is infrared finite. Replacing the photon polarisation by its momentum
gives zero to $\sim10^{-8}$ or better for the initial-state loops, the mixed subset and the
final-state subsets. Unlike the ultraviolet and infrared checks, this one catches missing
diagrams such as the seagulls.

\Cref{tab:rv_benchmark} gives the correction at two hard points
at $\sqrts=0.7$~GeV and one at $1.02$~GeV, built in the centre-of-mass frame with the
electron along $+z$. The photon has energy $k_1^0 = x\sqrts/2$ and angles
$\theta_\gamma$ and $\phi_\gamma$, and the pions are back to back in their rest frame,
the $\pi^-$ along $(\theta^*,\phi^*)$ in the same axes, boosted with the photon recoil.
With $m_e=0.51099895$~MeV, $m_\pi=139.57039$~MeV,
$\alpha=1/137.035999139$, the integrals normalised as in
\cref{sec:virtual-2to2}, \LoopTools and \Collier agree
to $10^{-10}$. The finite parts are a sizeable fraction of the tree because, before
the YFS subtraction, they contain the electron-mass collinear logarithms multiplying
the infrared pole. The ancillary material gives the
four-momenta to twelve digits and the finite parts in conventional
dimensional regularisation.
\begin{table}[htb]
  \centering
  \renewcommand{\arraystretch}{1.2}
  \begin{tabular}{llrrr}
    \hline\hline
    & & $P_1$ & $P_2$ & $P_3$ \\
    \hline
    \multicolumn{2}{l}{$\sqrts$ [GeV]} & $0.7$ & $0.7$ & $1.02$ \\
    \multicolumn{2}{l}{$x=2k^0/\sqrts$} & $0.3$ & $0.15$ & $0.45$ \\
    \multicolumn{2}{l}{$\cos\theta_\gamma$} & $0.4$ & $-0.7$ & $0.1$ \\
    \multicolumn{2}{l}{$\phi_\gamma$} & $0.3$ & $2$ & $4$ \\
    \multicolumn{2}{l}{$\cos\theta^*$} & $-0.5$ & $0.2$ & $0.6$ \\
    \multicolumn{2}{l}{$\phi^*$} & $1.1$ & $0.4$ & $2.5$ \\
    \multicolumn{2}{l}{$|\mathcal{M}_0|^2$ [GeV$^{-2}$]} & $0.26481039$ & $1.05830509$ & $0.01286222$ \\
    \hline
    $\delta_{62}$ & $1/\epsilon$ & $+1.192287$ & $+2.551765$ & $+4.789373$ \\
     & finite & $+11.196206$ & $+23.956939$ & $+43.043677$ \\
    $\delta_{53}$ & $1/\epsilon$ & $+1.456243$ & $-0.517547$ & $-4.809762$ \\
     & finite & $+12.351991$ & $-3.756933$ & $-36.845039$ \\
    $\delta_{44}$ & $1/\epsilon$ & $+0.980337$ & $+1.346518$ & $+3.968569$ \\
     & finite & $+7.473811$ & $+11.596277$ & $+28.083788$ \\
    $\delta_{35}$ & $1/\epsilon$ & $+0.259695$ & $-0.107268$ & $-1.075739$ \\
     & finite & $+1.413656$ & $-0.459939$ & $-4.457636$ \\
    $\delta_{26}$ & $1/\epsilon$ & $+0.084391$ & $+0.139825$ & $+0.410427$ \\
     & finite & $+0.637996$ & $+0.989850$ & $+2.672504$ \\
    sum & $1/\epsilon$ & $+3.972952$ & $+3.413294$ & $+3.282867$ \\
     & finite & $+33.073661$ & $+32.326194$ & $+32.497293$ \\
    \hline\hline
  \end{tabular}
  \caption{Benchmark values of the one-loop correction to $e^+e^-\to\pi^+\pi^-\gamma$ for point-like
    pions at three phase-space points, constructed as described in the text. For each charge subset,
    $\delta_{mn}=Q_e^mQ_\pi^n\mathcal{V}_{mn}/|\mathcal{M}_0|^2$ is given in units of $10^{-2}$: the
    coefficient of the infrared pole $1/\epsilon$ and the renormalised finite part in the
    't~Hooft--Veltman scheme at $\mu=1$~GeV. $|\mathcal{M}_0|^2$ is the spin-averaged point-like tree,
    summed over its three charge subsets.}
  \label{tab:rv_benchmark}
\end{table}

As in the $2\to2$ case, in the factorised approach the correction to the
point-like diagrams of \cref{fig:rv_diagrams}, taken relative to the point-like tree, multiplies the tree with the form factor placed as in
\cref{fig:ff_insertion}~(b)--(d), so all diagrams, the boxes and pentagons (d)--(f)
included, carry the form factor of the tree they interfere with. In the \GVMD{} option
the loop photons carry $F_\pi(\ell^2)$ as in \cref{fig:ff_insertion}~(k)--(o). With
$F_\pi(\ell^2)=\sum_vc_v\,L_v/(L_v-\ell^2)$ and $\sum_vc_v=1$, the photon
propagator times the form factor splits into
  \begin{equation}
    \frac{F_\pi(\ell^2)}{\ell^2}=\sum_vc_v\Big[\frac{1}{\ell^2}-\frac{1}{\ell^2-L_v}\Big]\,,
    \label{eq:gvmd_pf}
  \end{equation}
and a loop photon between the two pions, carrying $F_\pi^2$, adds terms with
two masses and, for coinciding poles, their derivative. The resulting massive-photon
loops are again taken relative to the tree with the same pole form factor and multiply
the tree with the Gounaris--Sakurai form factor. These tables pass the ultraviolet and
infrared checks, with the infrared pole equal to the eikonal factor times the
form-factor tree, and reduce to the point-like ones for massless photons.
Disperon QED~\cite{Fang:2025mhn,Fang:2026naz} gives the correction with the
form factor inside the loops, with $\mathcal{V}_{62}$ complete, $\mathcal{V}_{53}$ in
\FsQED{}, $\mathcal{V}_{44}$ estimated and the final-state subsets left out, and is the
reference for the form-factor dependence of $\mathcal{V}_{53}$, absent from our
point-like loops. It has also
been computed in \GVMD{}~\cite{CarloniCalame:2026vfc} and with a dispersive form
factor~\cite{CarloniCalame:2026gdm}.

\subsubsection{Numerical stability}
\label{sec:rv_numerics}
  The exact correction matters for hard initial-state photons
  away from the beams, its logarithmic expansion describing it well close to them. As the photon
  approaches a beam the box and pentagon coefficients cancel against each other, with
  both \LoopTools and \Collier, to a degree controlled by
  \begin{equation}
    r=\frac{\min(p_1\cdot k_1,\,p_2\cdot k_1)}{p_1\cdot p_2}\,,
    \label{eq:rv_r}
  \end{equation}
  which combines the photon energy with its angle to the nearer beam, so
  the failure region is not set by the angle alone. In
  double precision the finite part loses all significance for $r\lesssim10^{-6}$,
  reached by a photon with $x\sim10^{-2}$ already at $\sim10^{-2}$~rad. The precision
  is lost in the coefficient arithmetic, not the loop functions. Quadruple precision
  inside \LoopTools or a different loop-momentum routing through the pentagons does not
  help, while quadruple-precision coefficients with double-precision loop functions
  restore the amplitude down to $r\sim10^{-7}$. An independent calculation evaluates
  pentagons, coefficients and loop functions alike in quadruple precision close to the
  beams~\cite{Budassi:2026lmr}. All predictions use our
  own amplitude, which also provides the form-factor variants.
  The predictions of \cref{sec:results} use the exact correction except
  for initial-state photons with $r<r_c=10^{-3}$, where the exact remainder is replaced
  by its leading and next-to-leading logarithmic limit $V_2$~\cite{Jadach:2001jx}. This
  changes the fiducial cross sections by $\sim0.01\%$ at most, checked with both the
  independent evaluation and the quadruple-precision coefficients. The remainder is taken relative to the one-loop
  correction that \ACRAIC applies to $\hat\beta_1$, with the initial-state vertex evaluated
  at $s$.
  In \kkmc the initial-state loops on a beam photon, $\mathcal{V}_{62}$,
  multiply the one-photon initial-state CEEX amplitude for $r\ge r_c$, while the other
  subsets, without amplitude-level counterpart in CEEX, enter as an extra event weight,
  the same remainder at the hardest photon. At \KLOEI{} the
  remainder comes from $\mathcal{V}_{44}$, $\mathcal{V}_{35}$ and $\mathcal{V}_{26}$,
  each a few tenths of a per cent and largely cancelling, while $\mathcal{V}_{62}$ and
  $\mathcal{V}_{53}$ give $\sim0.01\%$, and the two generators agree on the correction.
  The exact remainder dominates the generation cost, by about an order of
  magnitude with hard photons. The correction therefore comes from a
  smaller sample whose events carry a second weight with $V_2$ in place of the exact
  remainder, and the ratio of the two distributions, far more precise than either as both
  weights belong to the same events, is applied bin by bin to the $V_2$ samples at
  under a tenth of the cost of new samples.

\subsection{Fixed order and resummation}
\label{sec:fo_theory}
To test the convergence of the all-order resummation of soft and soft-collinear logarithms in \ACRAIC and \kkmc, its expansion in $\alpha$ is truncated at second and third order.
Writing the bracket of \cref{eq:masterYFS} for $n_\gamma$ photons as $R_{n_\gamma}$ and counting $Y(\Omega)$, the virtual correction $v$ and each photon as one power of $\alpha$, schematically
\begin{equation}\label{eq:fo_truncation}
  \sigma^{(k)} = \sum_{n_\gamma=0}^{k}\frac{1}{n_\gamma!}\int\ddone{\Phi_n}
  \left[\prod_{i=1}^{n_\gamma}\ddone{\Phi_i^\gamma}\,\eik{i}\right]
  \Big[e^{Y(\Omega)}R_{n_\gamma}\Big]_{\alpha^{k}}\,,
\end{equation}
where $[\ldots]_{\alpha^{k}}$ keeps the terms up to $\alpha^k$. Without real photons, for example,
$[e^{Y}(1+v/2+\delta^{(2)})^2]_{\alpha^2}=1+Y+v+Y^2/2+Yv+v^2/4+2\delta^{(2)}$, with
$\delta^{(2)}$ the two-loop vertex.
The strict fixed-order $\mathcal{O}(\alpha)$ prediction is
$\sigma^{(1)}$ of \cref{eq:fo_truncation} with the exact one-photon matrix
element and the one-loop virtual correction, the next-to-leading order of
$e^+e^-\to\pi^+\pi^-$ computed by the fixed-order generators. In the
tagged-photon scenarios its leading term is the tree-level
$e^+e^-\to\pi^+\pi^-\gamma$, a leading-order prediction for the observed final
state. The truncations $\sigma^{(2)}$ and $\sigma^{(3)}$ of the
$\mathcal{O}(\alpha^2L)$ prediction measure how much of the resummed result
lies beyond each order. They test only
the higher orders the resummation contains, the powers of the soft and
collinear logarithms from $e^{Y(\Omega)}$, the eikonal factors and the virtual
correction. Corrections it does not contain, such as the hard three-photon
matrix element or the three-loop virtual correction, appear in no truncation
and are covered only by the theory nuisance parameters.

\subsection{Theoretical Uncertainties}
\label{sec:tnp}
The uncertainty of the generators used in radiative-return and
scan measurements has so far been judged from the spread between generators
or from the size of the last correction
included~\cite{WorkingGrouponRadiativeCorrections:2010bjp,Aliberti:2024fpq},
without a systematic estimate of their missing higher orders that we know of.
Neither measure is reliable here, as generators sharing an approximation
agree without being right, and in radiative return the first-order
correction is of order one, so the last correction no longer estimates the next.
We therefore use theory nuisance
parameters~\cite{Tackmann:2024kci,Lim:2024nsk}, applied to YFS resummation as
in~\cite{Price:2026qop}, each standing for an identified missing term. The
amplitude of an event is a sum of corrections, $\mathcal{C}=\sum_c\mathcal{C}_c$,
the $\tilde\beta_0$ term, the $\tilde\beta_1$ of initial- and final-state
photons and, at second order, the $\tilde\beta_2$ of
photon pairs. The template of a correction is the response of the event weight
to rescaling $\mathcal{C}_c$ alone, and each $\theta_t$ multiplies one template
with a normalisation $N_t$, the size of the first term the class omits,
\begin{equation}
  T_c=\frac{2\,\mathrm{Re}\,\langle\mathcal{C}_c,\mathcal{C}\rangle}{\langle\mathcal{C},\mathcal{C}\rangle}\,,
  \qquad \sum_cT_c=2\,,\qquad
  \mathrm{d}\sigma(\theta)=\mathrm{d}\sigma\,\Big[1+\sum_t\theta_tN_tT_{c(t)}\Big]\,,
  \label{eq:tnp_template}
\end{equation}
with $\langle A,B\rangle=\overline{\sum}_{\{\lambda\},\sigma}A^*B$ the helicity-averaged inner product and $\mathcal{C}$ the coherent amplitude of~\cite{Price:2026qop}. Observables
are linear in the parameters, and for independent parameters of unit variance
the band adds the $\theta_t=\pm1$ variations in quadrature,
\begin{equation}
  \delta\sigma=\Big[\sum_t\Big(N_t\int\mathrm{d}\sigma\,T_{c(t)}\Big)^2\Big]^{1/2}.
  \label{eq:tnp_band}
\end{equation}
With
$g_e=(\alpha/\pi)\,[\ln(s/m_e^2)-1]$ and
$g_\pi=(\alpha/\pi)\,[\ln(s/m_\pi^2)-1]$, the normalisations
of~\cite{Price:2026qop} are $N=g^k/(2\,k!)$ for a one-photon class of
a prediction truncated at order $k$, with $g=g_e$ ($g_\pi$)
for initial-state (final-state) photons, and the $\tilde\beta_0$
classes have $N=g^m/(2\,m!)$ for the first order $m$ they miss, separately
for the initial and final state and their products. The
factor $1/2$ converts a relative cross-section correction to the amplitude. The $\tilde\beta_2$ classes, absent from the
first-order prediction, enter its band with $N=1$ and one parameter for the
three pair classes, on templates from the second-order prediction on the same events.

Two features of pion production need further parameters. First, the second
order is incomplete for pions, the two-loop pion vertex being approximated and the
two-loop initial-final boxes absent, so the $\mathcal{O}(\alpha^2L)$ band
carries one $\tilde\beta_0$ parameter with $N=(\alpha/\pi)^2\pi^2/2$, again halved to
the amplitude. Second, in radiative return a series whose next term is $g$ times the
last no longer describes the order-one first-order correction. There a radiative-return parameter takes as template in each
bin the relative size $r_k=\sigma_k/\sigma_{k-1}-1$ of the last
correction included, applied to the current prediction, $r_k\,\sigma_k$, and as
normalisation the YFS exponent $\gamma=2(\alpha/\pi)(L-1)$, so the next
relative correction is $\gamma\,r_k$, one further power of the logarithm rather
than $r_k$ itself. Its use and the size of every parameter are discussed in
\cref{sec:tnp_breakdown}. The templates are additional weights on the same
events, so every histogram carries its band, correlated across bins
and observables through the shared parameters. In \kkmc the classes are
the corresponding $\hat\beta$ terms of the CEEX amplitudes, so the bands of
the two generators differ only where they assign a correction to different classes.

\section{Numerical Results}\label{sec:results}
\subsection{Input parameters}\label{sec:inputs}
Both generators use on-shell masses, the fine-structure constant at
zero momentum transfer, no vacuum polarisation and the STRONG2020 pion form
factor of \cref{eq:pion_ff} with the resonance parameters of
Ref.~\cite{Aliberti:2024fpq} (\cref{tab:inputs}). The soft cut-offs of the YFS
integrals and the thresholds below which a photon enters the hard corrections
only through the eikonal factor are technical parameters, and the fiducial
cross sections do not depend on the soft cut-off.

\begin{table}[!htb]
  \centering
  \renewcommand{\arraystretch}{1.2}
  \begin{tabular}{lc}
    \hline\hline
    Parameter                                          & Value       \\
    \hline
    $\alpha^{-1}(0)$                                   & $137.03599976$ \\
    $m_e$ [MeV]                                        & $0.51099895$\\
    $m_\pi$ [MeV]                                      & $139.57039$ \\
    $F_\pi(q^2)$                                       & STRONG2020~\cite{Aliberti:2024fpq} \\
    vacuum polarisation                                & off\\
    soft cut-off                                       & $E_{\rm min}=1.75$~keV \\
    \hline\hline
  \end{tabular}
  \caption{Input parameters, common to both generators and all scenarios.}
  \label{tab:inputs}
\end{table}

\subsection{Fiducial selection}
\Cref{tab:fiducial_cuts} summarises the fiducial selections.

\begin{table}[!htb]
  \centering
  \footnotesize
  \setlength{\tabcolsep}{4pt}
  \renewcommand{\arraystretch}{1.2}
  \resizebox{\textwidth}{!}{
    \begin{tabular}{l c l l l}
      \hline\hline
      Scenario  & $\sqrts$ [GeV] & Pion selection                                                 & Photon selection                                                              & Further cuts                                     \\
      \hline
      \CMD{}    & 0.7            & $\theta_{\rm avg}\in[1,\pi-1]$~rad                             & --                                                                            & $\delta\phi<0.15$~rad                            \\
                &                & $p^\pm>0.45\hsqrts$                                            &                                                                               & $\xi<0.25$~rad                                   \\[3pt]
      \KLOEI{}  & 1.02           & $\theta^\pm\in[50^\circ,130^\circ]$                            & $\theta_\gamma\in[50^\circ,130^\circ]\land E_\gamma>20\,\mathrm{MeV}$         & $M_{\pi^+\pi^-}^2\in[0.1,0.85]\,\mathrm{GeV}^2$  \\
                &                & $|p_z^\pm|>90\,\mathrm{MeV}\lor p_\perp^\pm>160\,\mathrm{MeV}$ &                                                                               &                                                  \\[3pt]
      \KLOEII{} & 1.02           & as \KLOEI{}                                                    & $\theta_{\widetilde\gamma}<15^\circ \lor \theta_{\widetilde\gamma}>165^\circ$ & $M_{\pi^+\pi^-}^2\in[0.35,0.95]\,\mathrm{GeV}^2$ \\[3pt]
      \BES{}    & 4              & $|\cos\theta^\pm|<0.93$                                        & $|\cos\theta_\gamma|<0.8\land E_\gamma>25\,\mathrm{MeV}$                      & $\exists!\,\gamma: E_\gamma>400\,\mathrm{MeV}$   \\
                &                & $p_\perp^\pm>300$~MeV                                          & $\lor\;0.86<|\cos\theta_\gamma|<0.92\land E_\gamma>50\,\mathrm{MeV}$          &                                                  \\[3pt]
      \Bscen{}  & 10             & $\theta^\pm\in[0.65,2.75]$~rad                                 & $\theta_\gamma\in[0.6,2.7]\,\mathrm{rad}\land E_\gamma>3\,\mathrm{GeV}$       & $\theta_{\gamma^{(h)},\widetilde\gamma}<0.3$~rad \\
                &                & $p^\pm>1$~GeV                                                  &                                                                               & $M_{\pi^+\pi^-\gamma}>8$~GeV                     \\
      \hline\hline
    \end{tabular}}
  \caption{Fiducial selection criteria for the five scenarios.
    $\delta\phi=||\phi^+-\phi^-|-\pi|$ and $\xi=|\theta^++\theta^--\pi|$ are
    the acoplanarity and acollinearity of the pion pair, and
    $\theta_{\rm avg}=(\theta^--\theta^++\pi)/2$. $p^+=|\vec p_4|$, $p^-=|\vec p_3|$ and $\theta^\pm$ are the momentum and the polar angle of the
    $\pi^\pm$ in the centre-of-mass frame of the beams, the angle measured
    from the $e^-$ beam. $\widetilde\gamma$ denotes
    the direction opposite the di-pion momentum, $\gamma^{(h)}$ the hardest
    selected photon. Definitions follow
    ~\cite{radiomontecarlow2-results}, and the scenarios correspond to the measurements of
    ~\cite{CMD-3:2023alj} (\CMD{}), \cite{KLOE:2010qei} (\KLOEI{}),
    \cite{KLOE:2008fmq,KLOE:2012anl} (\KLOEII{}), \cite{BESIII:2015equ} (\BES{}) and
    \cite{BaBar:2012bdw} (\Bscen{})}.
  \label{tab:fiducial_cuts}
\end{table}

Both generators couple the pions to photons in point-like scalar QED with
$F_\pi$ of \Cref{app:pion_ff} on the $s$-channel photon, exponentiate the soft
photons to all orders through the YFS form factor and the eikonal factors of all
dipoles, and include the initial-final interference at amplitude level through
the sum over partitions~\cite{Jadach:2000ir}.

At $\mathcal{O}(\alpha)$:
\begin{itemize}
  \item $\tilde\beta_0$ with the one-loop virtual correction of
        \Cref{sec:virtual-2to2}, the electron vertex, the pion vertex and the
        initial-final boxes, computed for point-like pions and supplemented with $F_\pi(s)$
        either in \FxsQED or \GVMD. The pion vertex of \kkmc, $F_{\rm fin}$, equals
        $\delta_{\rm FSR}$ of \Cref{eq:virt-fsr} to $10^{-16}$, and the boxes
        are evaluated with the same code. Each generator removes
        the infrared singularity with its own YFS form factor, which its
        exponent puts back. \kkmc evaluates the boxes once per event at the Born point
        and applies the two vertices as $(1+F_{\rm ini})(1+F_{\rm fin})$, so
        its $\mathcal{O}(\alpha)$ also contains the $\mathcal{O}(\alpha^2)$
        product $F_{\rm ini}F_{\rm fin}$, which raises the \CMD{}
        cross section by $\sim0.01\%$ and is optional in \ACRAIC.
  \item $\tilde\beta_1$ from the exact tree amplitudes for one real photon,
        emitted from the electrons or from the pions, with $F_\pi$
        evaluated at the virtuality of the $s$-channel photon of each
        partition. \ACRAIC takes them from \Comix, \kkmc from its own
        scalar-QED helicity amplitudes.
\end{itemize}

At $\mathcal{O}(\alpha^2L)$, in addition:
\begin{itemize}
  \item $\tilde\beta_2$ from the exact tree amplitudes for two real
        photons, both from the electrons, both from the pions or one from each, coherently.
  \item The exact one-loop correction to the emission of one photon from the
        electrons or the pions, for every charge subset.
  \item The square of the one-loop amplitude, in both.
  \item The photonic two-loop vertices in $\tilde\beta_0$. \kkmc adds the two-loop
        electron vertex at leading and next-to-leading logarithmic accuracy and, for
        the pions, $F_{\rm fin}+F_{\rm fin}^2/2$. \ACRAIC adds the leading logarithms
        of every pair of charged legs, including the product of the initial- and
        final-state vertices, the next-to-leading logarithms and the constant of the
        massive two-loop vertex~\cite{Penin:2005eh}.
\end{itemize}
Neither generator includes the two-loop boxes, the genuine two-loop
scalar-QED pion vertex, or the photonic corrections of third order.
The second order therefore contains the $\alpha^2L^2$ and the dominant
$\alpha^2L$ terms, with $L=\ln(s/m_e^2)$, but not the non-logarithmic
$\alpha^2$ terms, hence the label $\mathcal{O}(\alpha^2L)$, while
$\mathcal{O}(\alpha^2)$ refers to the perturbative order itself.
\Cref{sec:ff_amplitudes} describes where $F_\pi$
enters the amplitudes of both generators.

\begin{table}[htb]
  \centering
  \renewcommand{\arraystretch}{1.3}
  \resizebox{\textwidth}{!}{
    \begin{tabular}{llccc}
      \hline\hline
      Scenario  & Order                                        & \kkmc CEEX $\sigma_{\rm fid}$ [nb] & \Sherpa \ACRAIC $\sigma_{\rm fid}$ [nb] & Ratio \ACRAIC/CEEX       \\
      \hline
      \CMD{}    & $\mathcal{O}(\alpha)$                        & $570.078(9)$           & $570.066(5)$                            & $0.99998(2)$  \\
                & $\mathcal{O}(\alpha^2L)$                     & $570.242(9)$           & $570.287(5)$                            & $1.00008(2)$  \\
                & $\mathcal{O}(\alpha^2L)/\mathcal{O}(\alpha)$ & $1.00029$                          & $1.00039$                               & --                        \\
      \KLOEI{}  & $\mathcal{O}(\alpha)$                        & $3.2512(3)$            & $3.2513(2)$                 & $1.00005(11)$ \\
                & $\mathcal{O}(\alpha^2L)$                     & $3.2012(3)$            & $3.2033(2)$                 & $1.00067(11)$ \\
                & $\mathcal{O}(\alpha^2L)/\mathcal{O}(\alpha)$ & $0.98462$                          & $0.98524$                   & --                        \\
      \KLOEII{} & $\mathcal{O}(\alpha)$                        & $15.966(1)$            & $15.965(1)$                 & $0.99995(11)$ \\
                & $\mathcal{O}(\alpha^2L)$                     & $16.020(1)$            & $16.017(1)$                 & $0.99979(11)$ \\
                & $\mathcal{O}(\alpha^2L)/\mathcal{O}(\alpha)$ & $1.00337$              & $1.00322$                   & --                        \\
      \BES{}    & $\mathcal{O}(\alpha)$                        & $0.088510(11)$         & $0.088507(10)$              & $0.99997(16)$ \\
                & $\mathcal{O}(\alpha^2L)$                     & $0.083889(13)$         & $0.083911(9)$               & $1.00026(19)$ \\
                & $\mathcal{O}(\alpha^2L)/\mathcal{O}(\alpha)$ & $0.94780$              & $0.94807$                   & --                        \\
                & $\pi^+\pi^-\gamma$ resummed NLO              & --                                 & $0.083677(13)$                          & --                        \\
      \Bscen{}  & $\mathcal{O}(\alpha)$                        & $0.011160(1)$          & $0.011157(2)$               & $0.99971(17)$ \\
                & $\mathcal{O}(\alpha^2L)$                     & $0.010056(3)$          & $0.010058(1)$               & $1.00015(28)$ \\
                & $\mathcal{O}(\alpha^2L)/\mathcal{O}(\alpha)$ & $0.90104$              & $0.90143$                   & --                        \\
                & $\pi^+\pi^-\gamma$ resummed NLO              & --                                 & $0.010052(1)$                           & --                        \\
      \hline\hline
    \end{tabular}}
  \caption{Fiducial cross sections of \kkmc CEEX and \Sherpa \ACRAIC. The $\pi^+\pi^-\gamma$ rows are the explicit-photon calculation of \ACRAIC at next-to-leading order (NLO), resummed with the matched photons.}
  \label{tab:fid_xsec}
\end{table}
\FloatBarrier
\subsection{\CMD{}}

\begin{figure}[htb]
  \centering
  \subfloat[]{\includegraphics[width=0.49\textwidth]{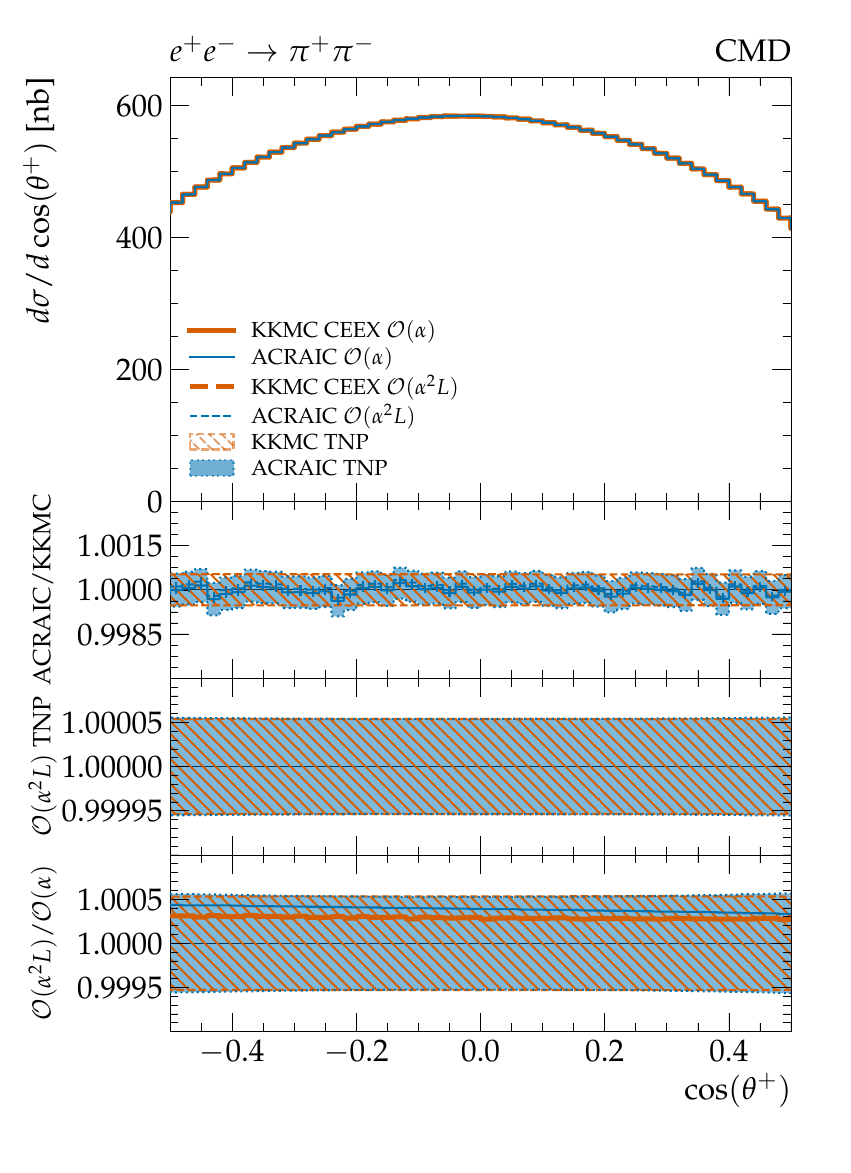}}\hfill
  \subfloat[]{\includegraphics[width=0.49\textwidth]{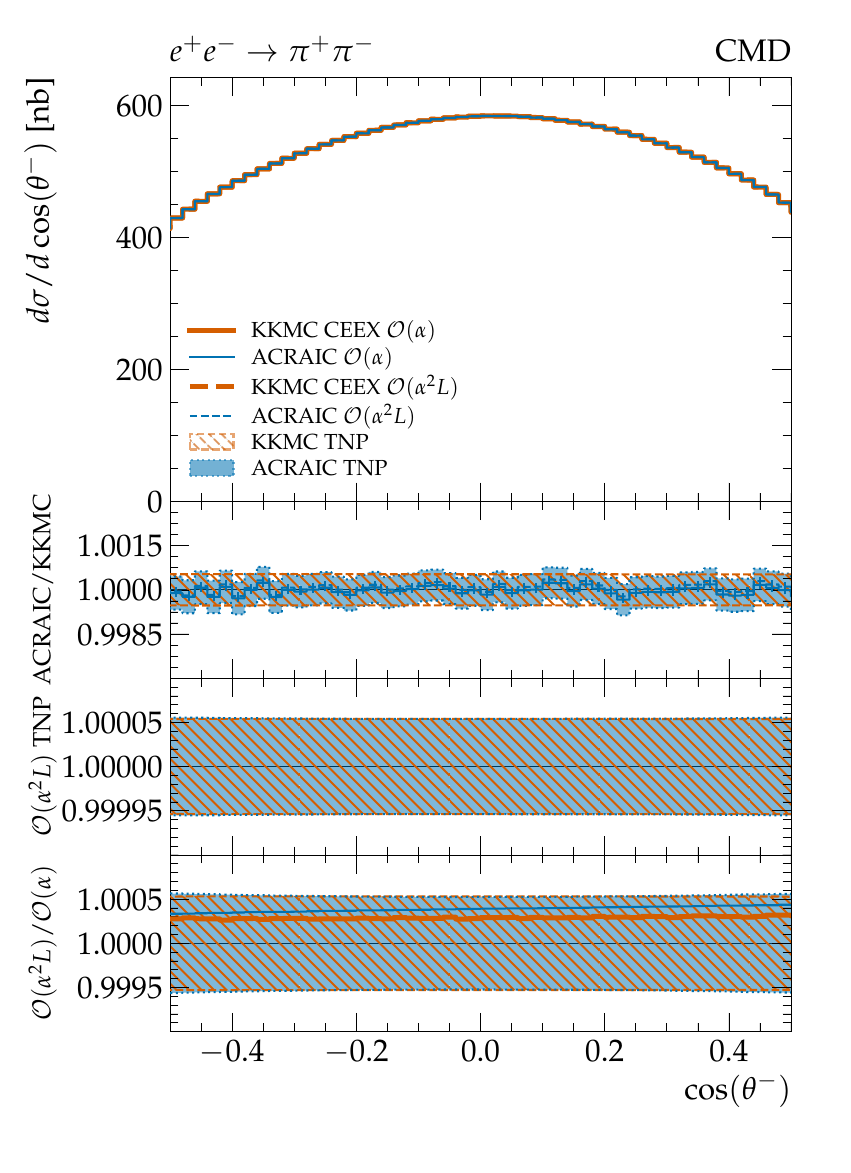}}\hfill
  \caption{The distributions of $\cos\theta^+$ (left) and $\cos\theta^-$
    (right) in the \CMD{} fiducial region. \kkmc CEEX and \Sherpa \ACRAIC  at $\mathcal{O}(\alpha)$ (solid) and $\mathcal{O}(\alpha^2L)$
    (dashed). The second panel shows \ACRAIC/\kkmc at each order with the
    $\mathcal{O}(\alpha)$ theory-nuisance-parameter bands of both generators, the
    third panel their $\mathcal{O}(\alpha^2L)$ bands, each relative to its own
    prediction and on its own scale, and the bottom panel
    $\mathcal{O}(\alpha^2L)/\mathcal{O}(\alpha)$ for each generator
    with the $\mathcal{O}(\alpha)$ bands around one.}
  \label{fig:cmd_costheta}
\end{figure}

At $\mathcal{O}(\alpha)$ the pion angles of \kkmc CEEX and \Sherpa
\ACRAIC agree to better than $0.1\%$ in every bin with $|\cos\theta^\pm|<0.5$
(\cref{fig:cmd_costheta}), and the fiducial cross sections
(\cref{tab:fid_xsec}) to $\sim0.002\%$, within the statistical uncertainty.
This is a first non-trivial cross-check between two independent
amplitude-level implementations of these corrections for
$e^+e^-\to\pi^+\pi^-$.

\begin{figure}[htb]
  \centering
  \subfloat[]{\includegraphics[width=0.5\textwidth]{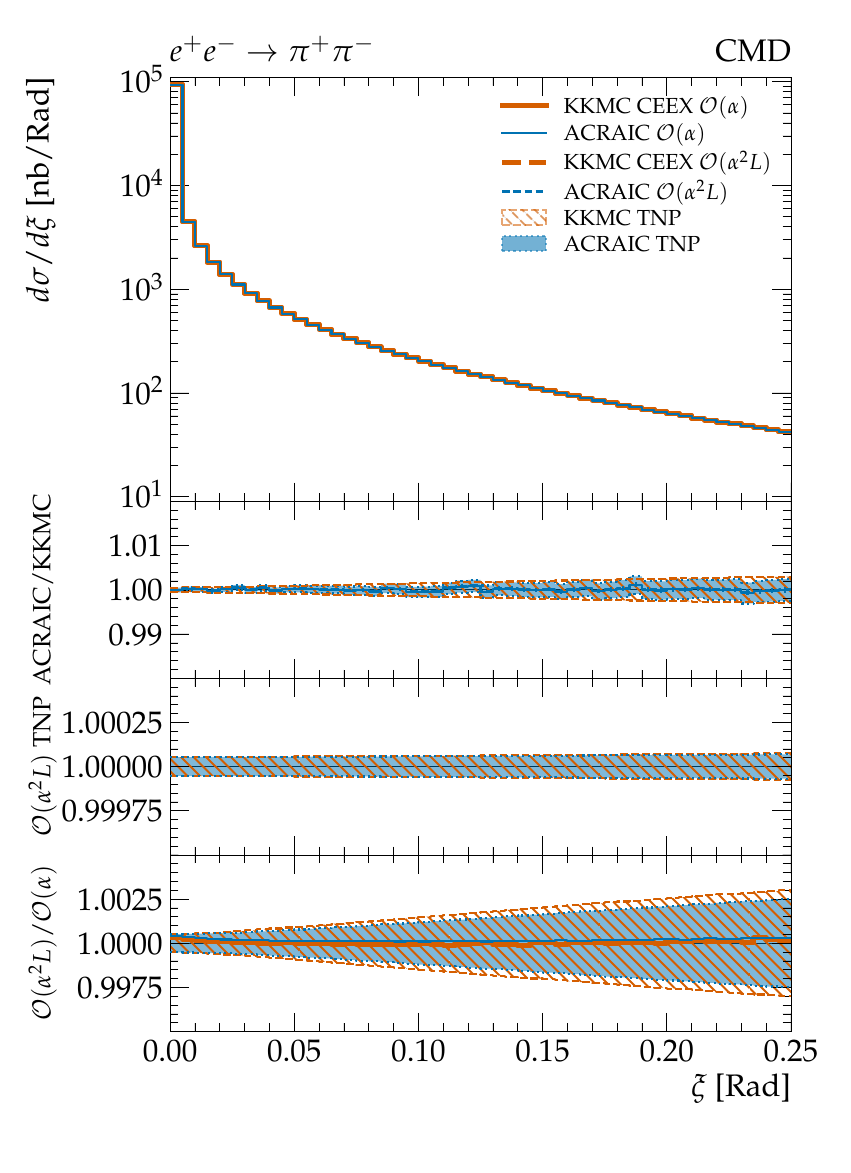}}\hfill
  \subfloat[]{\includegraphics[width=0.5\textwidth]{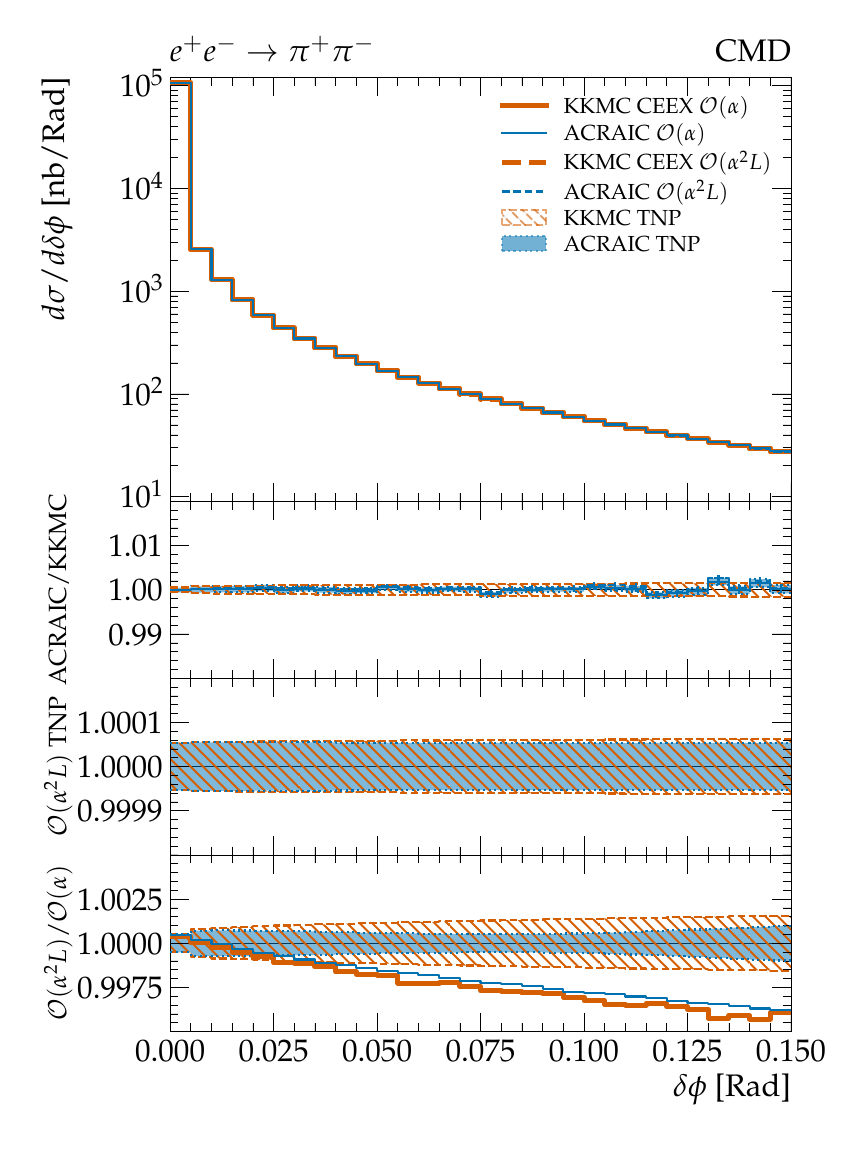}}
  \caption{The acollinearity $\xi$ (a) and the acoplanarity $\delta\phi$ (b)
    in the \CMD{} fiducial region. Curves, bands and panels as in~\Cref{fig:cmd_costheta}.}
  \label{fig:cmd_angles}
\end{figure}

In the acollinearity and acoplanarity (\cref{fig:cmd_angles}) the
two calculations agree within the statistical fluctuations even in the tails,
which fall by several orders of magnitude. There the
$\mathcal{O}(\alpha^2L)$ correction, negative and a few tenths of a per cent in
both generators, exceeds the $\mathcal{O}(\alpha)$ band several times. It is
the one-loop correction to the one-photon events, which also changes the soft
part of the one-photon amplitude, since the virtual correction differs between
the radiative and the Born point, whereas the $\tilde\beta_1$ parameters
rescale only the hard part, small here.

\begin{figure}[htb]
  \centering
  \subfloat[]{\includegraphics[width=0.48\textwidth]{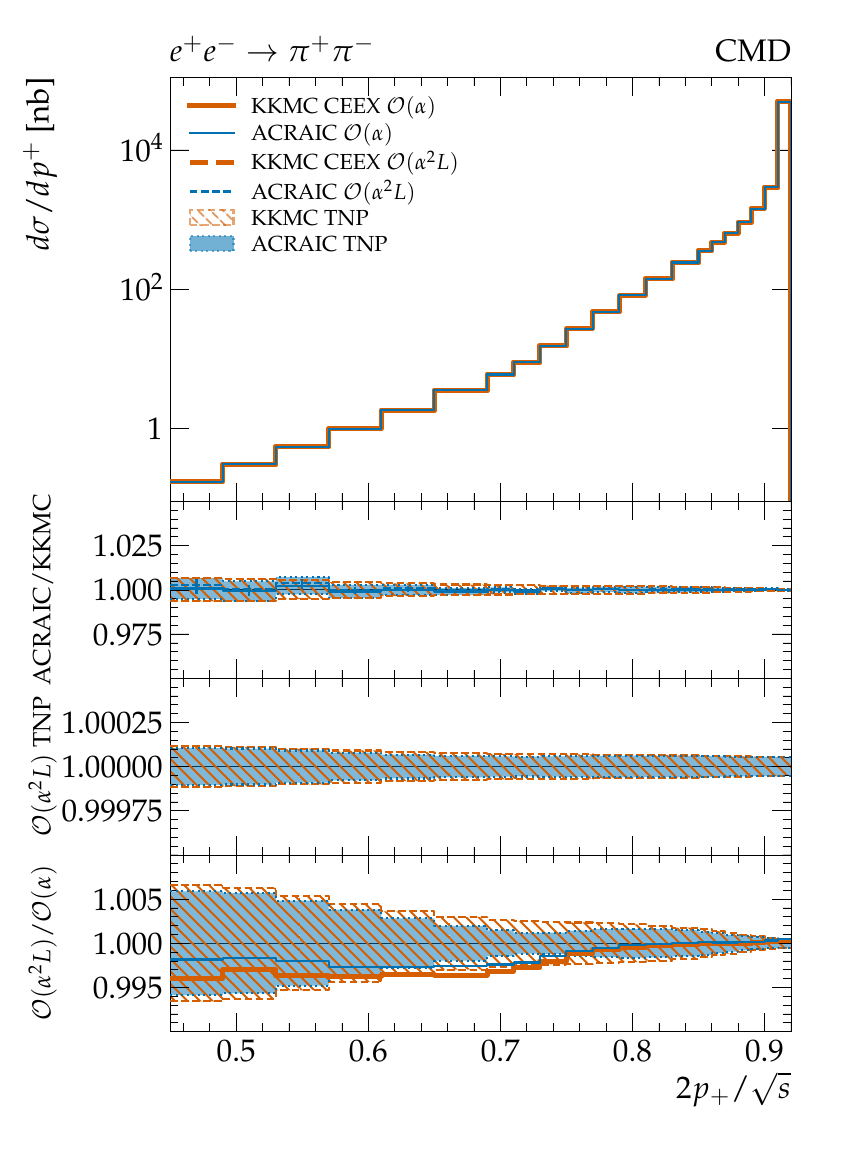}}\hfill
  \subfloat[]{\includegraphics[width=0.48\textwidth]{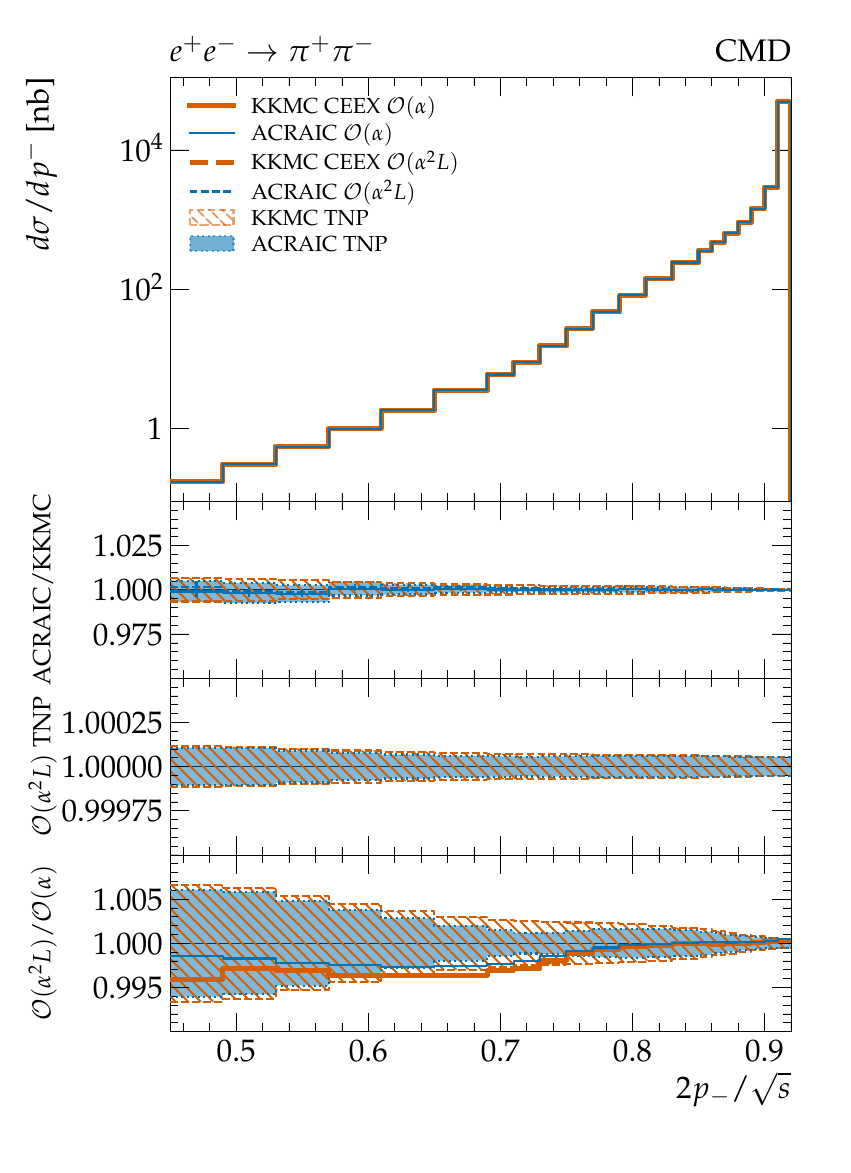}}
  \caption{Scaled pion momenta $2p^+/\sqrts$ (left) and
    $2p^-/\sqrts$ (right) for the \CMD{} fiducial region. Curves, bands and panels as in~\Cref{fig:cmd_costheta}.}
  \label{fig:cmd_momenta}
\end{figure}

The charge-symmetric $\pi^\pm$ momentum spectra (\cref{fig:cmd_momenta})
agree from the radiative tail to the elastic peak.
\begin{figure}[htb]
  \centering
  \subfloat[]{\includegraphics[width=0.45\textwidth]{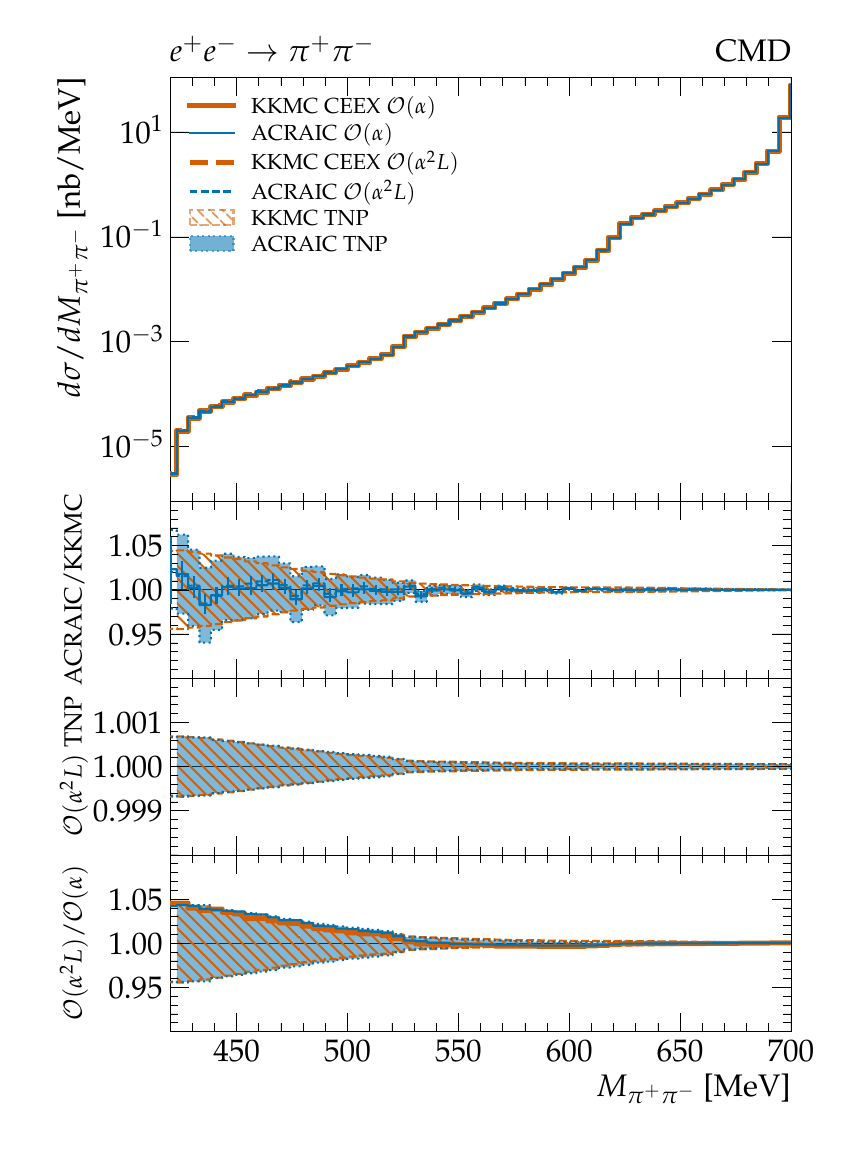}}\hfill
  \subfloat[]{\includegraphics[width=0.45\textwidth]{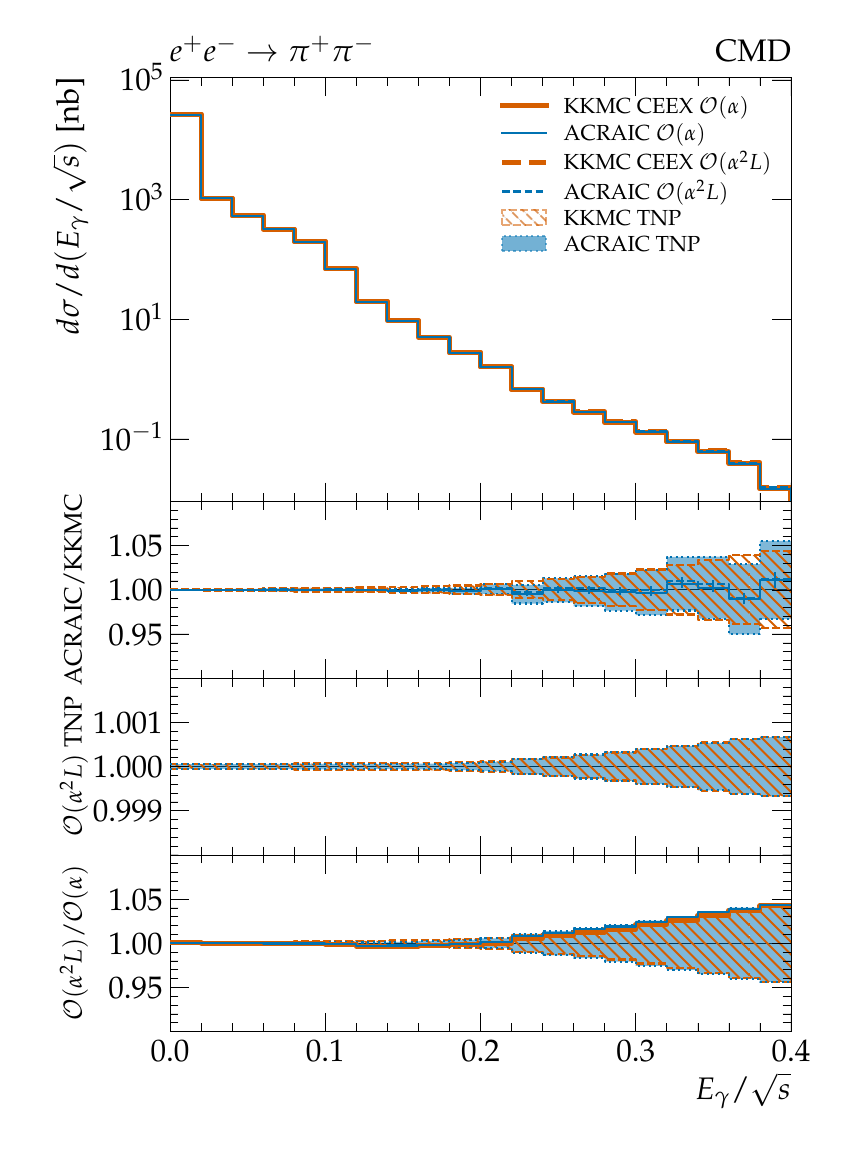}}\hfill
  \caption{The di-pion invariant mass $M_{\pi^+\pi^-}$ (a), the total
    photon energy fraction $E_\gamma/\sqrts$ (b) for the \CMD{} fiducial region. Curves, bands and panels as in~\Cref{fig:cmd_costheta}.}
  \label{fig:cmd_radiative}
\end{figure}
In the observables most sensitive to the radiation itself,
$M_{\pi^+\pi^-}$ and $E_\gamma/\sqrts$ (\cref{fig:cmd_radiative}), the two
$\mathcal{O}(\alpha)$ calculations agree within the statistical precision,
including the hard-photon tail.

The $\mathcal{O}(\alpha^2L)$ corrections are $\sim0.03\%$ in both
generators, small because the \CMD{} cuts select events close to the elastic
peak, and flat in the angular distributions, where the two generators again
agree to better than $0.1\%$ per bin. Where a hard photon is emitted, at
$M_{\pi^+\pi^-}<500$~MeV and at $E_\gamma/\sqrts\approx0.4$, they grow to a few
per cent. These tails carry less than $0.01\%$ of the fiducial cross section,
and there the two generators agree within the statistical precision of
\ACRAIC. With the parameters of the incomplete second order for
pions of \cref{sec:tnp}, the $\mathcal{O}(\alpha^2L)$ band is $\sim0.01\%$ in
the angular distributions and $\sim0.1\%$ where a hard photon is emitted (third
panels of \crefrange{fig:cmd_costheta}{fig:cmd_radiative}).
On the fiducial cross section the step reduces the band from $\sim0.05\%$, dominated by the $\tilde\beta_0$ parameters, to
$\sim0.005\%$, dominated by the incomplete second order.

\begin{figure}[htb]
  \centering
  \includegraphics[width=0.6\textwidth]{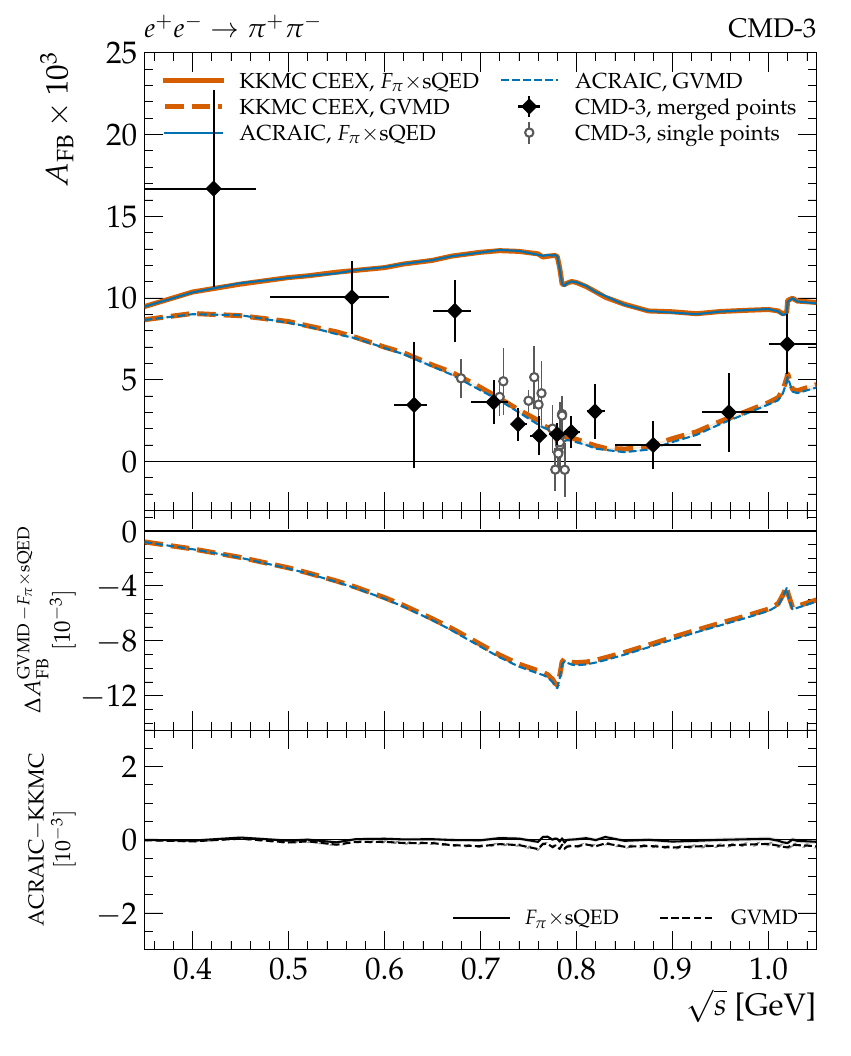}
  \caption{Charge asymmetry $A_{\rm FB}$ in the \CMDthree{} selection as
  a function of $\sqrts$, at $\mathcal{O}(\alpha^2L)$, for \kkmc CEEX (thick
  orange) and \Sherpa \ACRAIC (thin blue), with
  point-like loops in the factorised approach (solid) and
  with \GVMD{} in the loops (dashed).  The black diamonds and open circles are the
  \CMDthree{} measurement. The horizontal bars of the diamonds mark the energy
  points merged into each and are not uncertainties. The middle panel shows the shift
  $A_{\rm FB}^{\rm GVMD}-A_{\rm FB}^{\rm pt}$, the bottom panel \ACRAIC minus
  \kkmc for both treatments, all in units of $10^{-3}$. The $\mathcal{O}(\alpha)$
  predictions differ from those shown by $\lesssim2\times10^{-4}$ and are
  not drawn.}
  \label{fig:afb_sqrts}
\end{figure}

\Cref{fig:afb_sqrts} shows the charge asymmetry
\beq
A_{\rm FB}=(N_{\theta<\pi/2}-N_{\theta>\pi/2})/(N_{\theta<\pi/2}+N_{\theta>\pi/2}),
\eeq
with $\theta=\theta_{\rm avg}$, in the \CMDthree{} selection with
momentum cuts scaled to the beam energy, at 43 energies in
$0.35<\sqrts<1.05$~GeV with finer steps around the $\omega$
($0.765$--$0.795$~GeV) and in the $\phi$ region
($1.01$--$1.05$~GeV)~\cite{CMD-3:2023alj}.
With point-like loops in the factorised approach the asymmetry is $\sim10^{-2}$
over the whole range. \GVMD{} in the loops lowers it at every energy, by
$0.8\times10^{-3}$ at $\sqrts=0.35$~GeV and by up to $1.15\times10^{-2}$ at
$0.78$~GeV, where it removes most of the asymmetry. At the $\omega$ the
point-like asymmetry falls from $\sim13\times10^{-3}$ to $\sim11\times10^{-3}$
between $0.775$ and $0.785$~GeV in both generators, and the \GVMD{}
shift shrinks by $1.8\times10^{-3}$. In the $\phi$ region
the point-like asymmetry of both generators dips by $\sim3\times10^{-4}$ at $1.015$~GeV and rises
by $\sim10^{-3}$ above $1.02$~GeV. The \ACRAIC
\GVMD{} shift shrinks from $5.4\times10^{-3}$
at $1.01$~GeV to $4.2\times10^{-3}$ at $1.019$~GeV and grows again to
$5.7\times10^{-3}$ at $1.025$~GeV, and the \kkmc shift follows it within its errors
of $10^{-3}$. The two generators agree on the shift within the statistical
uncertainties, with $\chi^2/N=0.54$ for $N=43$ and a mean difference of
$0.04(14)\times10^{-3}$, and on the asymmetry itself. The
$\mathcal{O}(\alpha)\to\mathcal{O}(\alpha^2L)$ step changes it
by at most $3\times10^{-5}$ with point-like loops in both generators and, with
\GVMD{}, by up to $4\times10^{-5}$ in \kkmc and $1.8\times10^{-4}$ in \ACRAIC.

The asymmetry comes entirely from initial-final interference. None of
the parameters of \cref{sec:tnp_breakdown} is odd in the pion charge, so their
rescalings nearly cancel in the ratio, leaving only the small angular variation
of their templates, and alone they fall short of the step by one to two orders
of magnitude. We therefore add a parameter that rescales the initial-final
interference beyond the exponent, i.e.\ the initial-final box and the hard
one-photon interference. Its template, the part of the correction odd under a
sign flip of the final-state emissions minus the same part of $\tilde\beta_0$,
is obtained by flipping the final-state charges in \ACRAIC and from
interference on minus off in \kkmc, and the two agree to $\sim10^{-3}$ at
$0.7$~GeV. Its normalisation is $(g_e+g_\pi)^k/k!$ at $\mathcal{O}(\alpha^k)$,
with $g_f=(\alpha/\pi)(\ln(s/m_f^2)-1)$. For \kkmc at $\mathcal{O}(\alpha^2L)$
a further parameter of size $(g_e+g_\pi)/2$ accounts for the missing product of
the vertex correction and the pion box. The $\mathcal{O}(\alpha)$ band is
then $\sim10^{-4}$ in both generators with either treatment of the loops. It
covers the step at every energy with point-like loops ($|\theta_t|\le0.23$),
and with \GVMD{} the \kkmc step at every energy except $0.5$~GeV
($|\theta_t|=2.4$). Near $0.5$~GeV the \GVMD{} template passes through zero,
as the box and the hard-photon interference cancel, and the band misses the
\ACRAIC step between $0.45$ and $0.57$~GeV, with $|\theta_t|$ up to $7.4$ at
$0.5$~GeV. In the $\phi$ region the \ACRAIC step lies just outside the again
small \GVMD{} band at six of the seven energies ($1.0<|\theta_t|<2.0$). At
$\mathcal{O}(\alpha^2L)$ the band is $\sim10^{-5}$ for \ACRAIC and up to
$\sim10^{-4}$ for \kkmc, whose band covers the difference between the two
generators' steps with point-like loops ($|\theta_t|\le0.33$) but not with
\GVMD{} ($|\theta_t|>1$ at every energy above $0.4$~GeV), because the
hard-photon interference in its template has the opposite sign to the \GVMD{}
box and dilutes it. All bands are small
against the \GVMD{} shift, the $\mathcal{O}(\alpha)$ band at most $6\%$ of it and
the $\mathcal{O}(\alpha^2L)$ band less, so they are narrower than the
lines in \cref{fig:afb_sqrts} and not drawn, and the form-factor modelling
dominates the theoretical uncertainty of the asymmetry. In these predictions the tree-level amplitudes
use the STRONG2020 Gounaris--Sakurai form factor of \cref{eq:pion_ff}, and the
\CMDthree{} Breit--Wigner fit enters only inside the \GVMD{} loops.

\subsection{\KLOEI{}}\label{sec:kloeI_scen}
\begin{figure}[htb]
  \centering
  \subfloat[]{\includegraphics[width=0.5\textwidth]{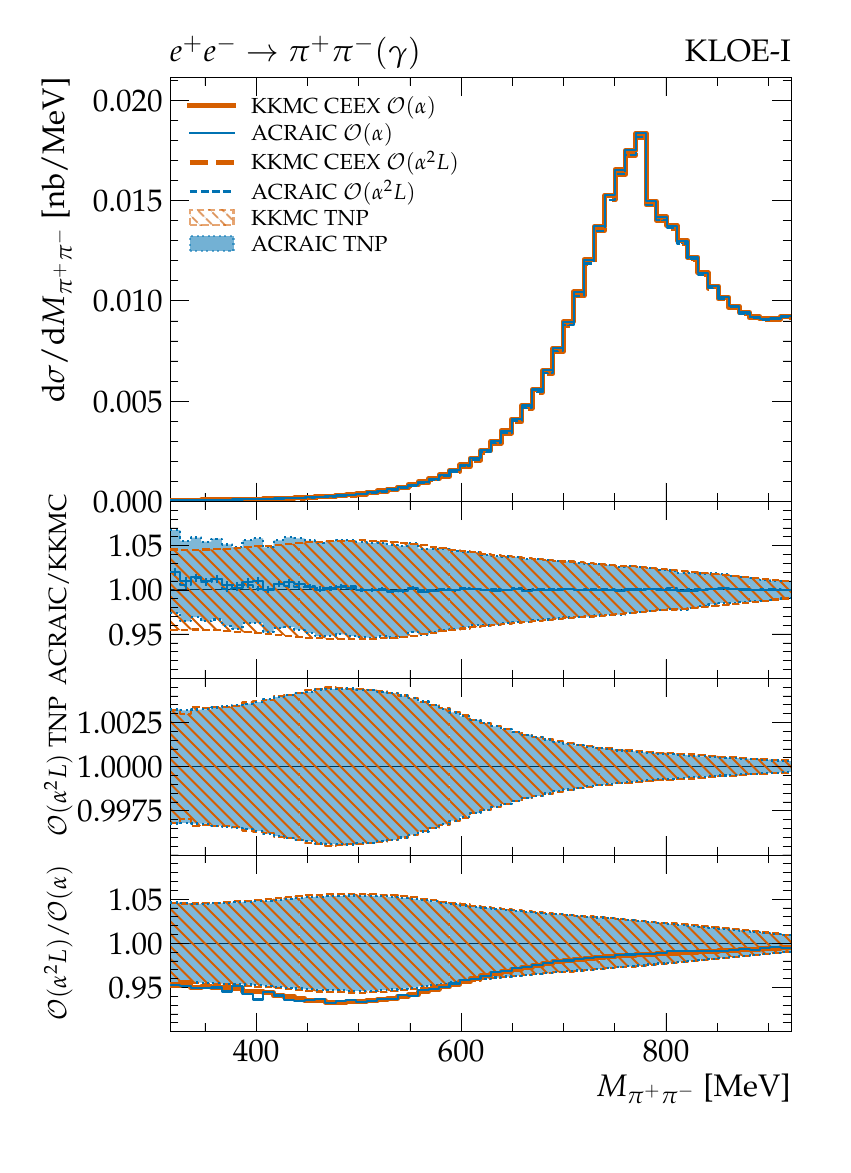}}\hfill
  \subfloat[]{\includegraphics[width=0.5\textwidth]{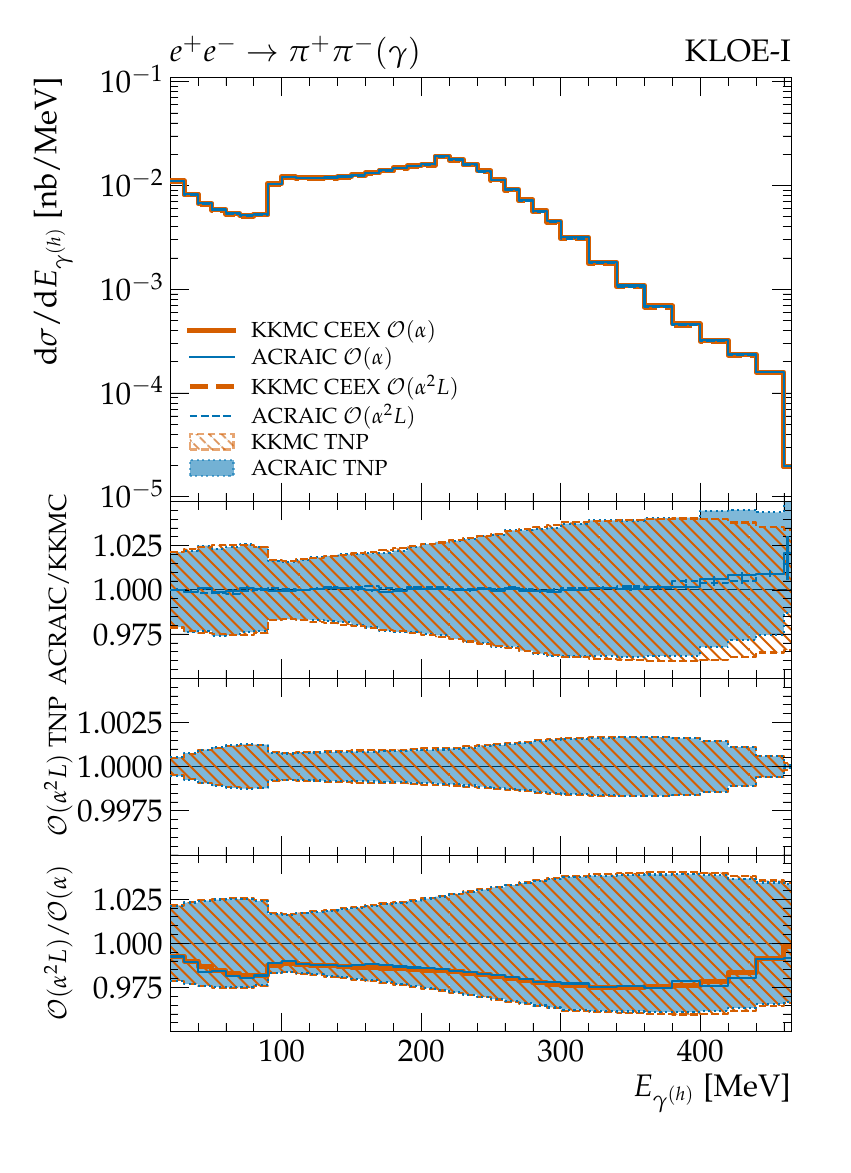}}
  \caption{The pion-pair mass $M_{\pi^+\pi^-}$ (left) and the energy of the
    hardest tagged photon (right) in the \KLOEI{} fiducial region. Curves and panels
    as in \cref{fig:cmd_costheta}.}
  \label{fig:kloeI_radiative}
\end{figure}

\begin{figure}[htb]
  \centering
  \subfloat[]{\includegraphics[width=0.5\textwidth]{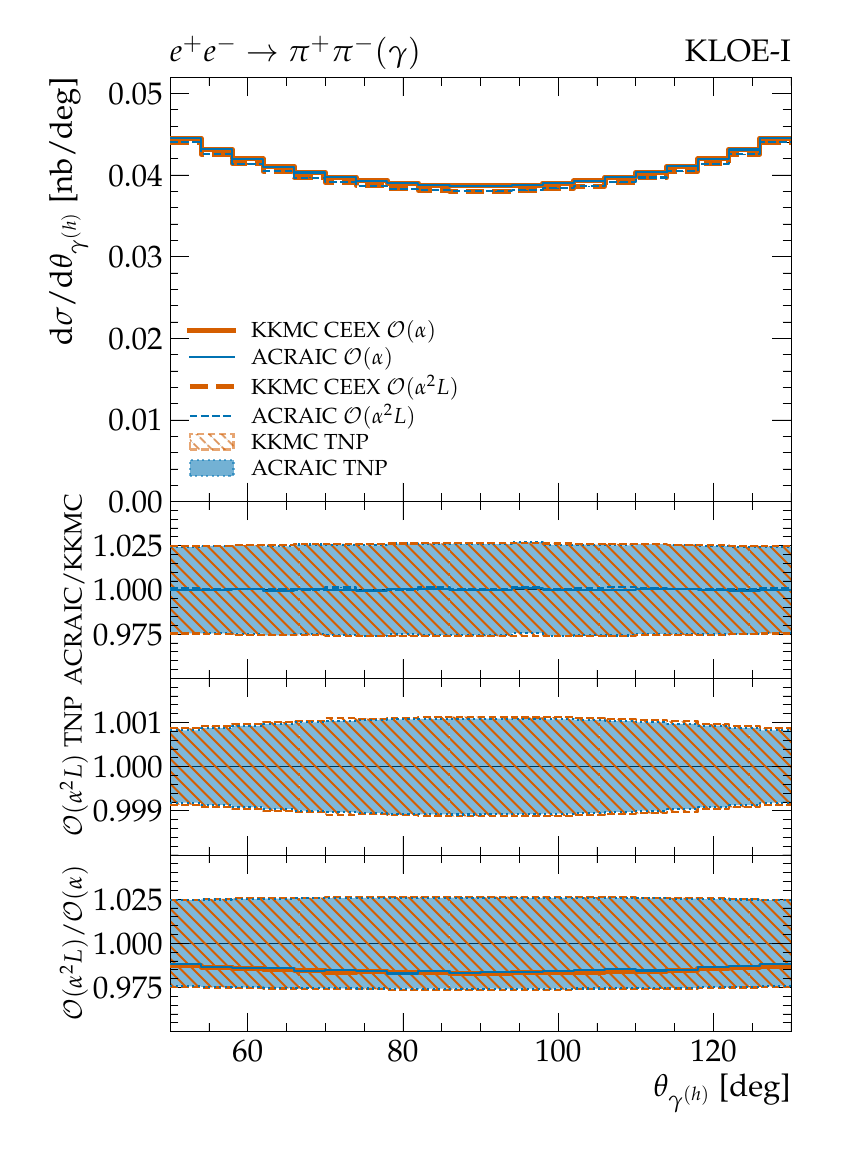}}\hfill
  \subfloat[]{\includegraphics[width=0.5\textwidth]{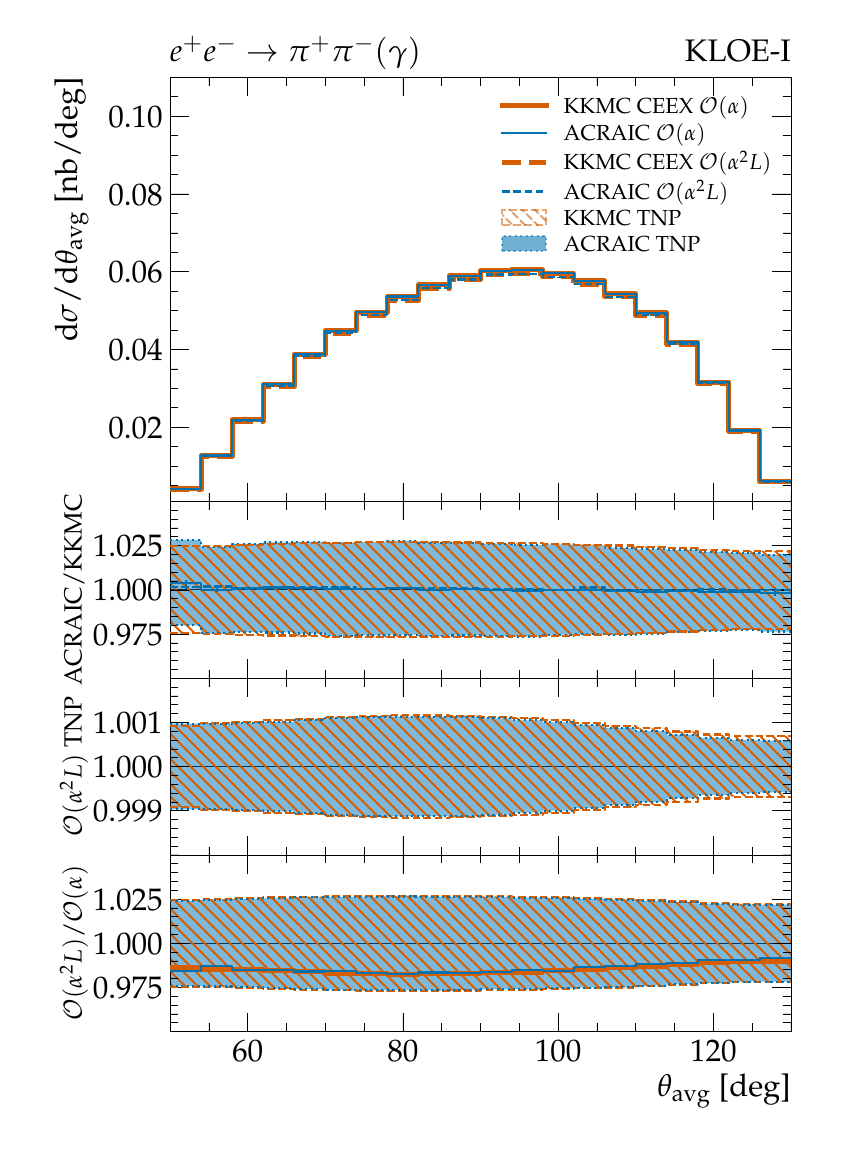}}\\
  \caption{The polar angle of the hardest tagged photon (left), the
    average pion angle $\theta_{\rm avg}$ (right) in the \KLOEI{}
    fiducial region. Curves and bands as in~\Cref{fig:kloeI_radiative}.}
  \label{fig:kloeI_photon}
\end{figure}

\begin{figure}[htb]
  \centering
  \subfloat[]{\includegraphics[width=0.5\textwidth]{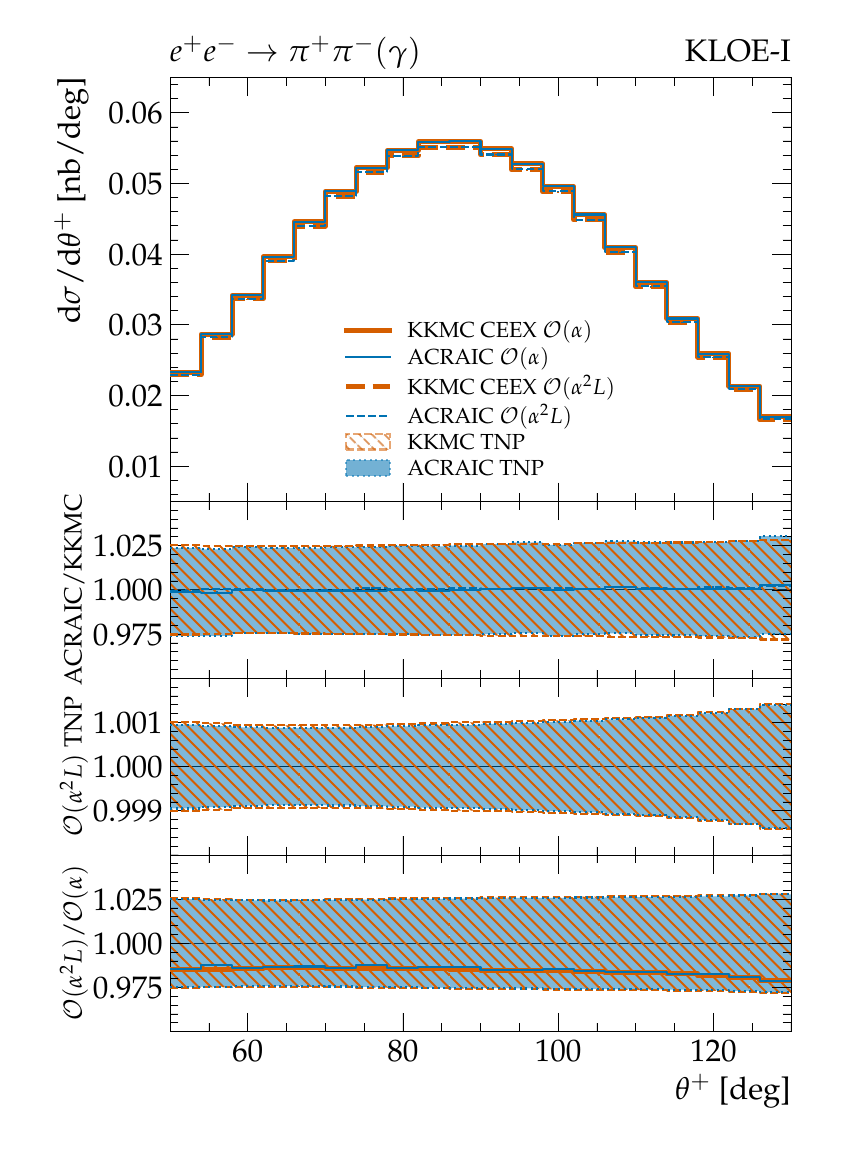}}\hfill
  \subfloat[]{\includegraphics[width=0.5\textwidth]{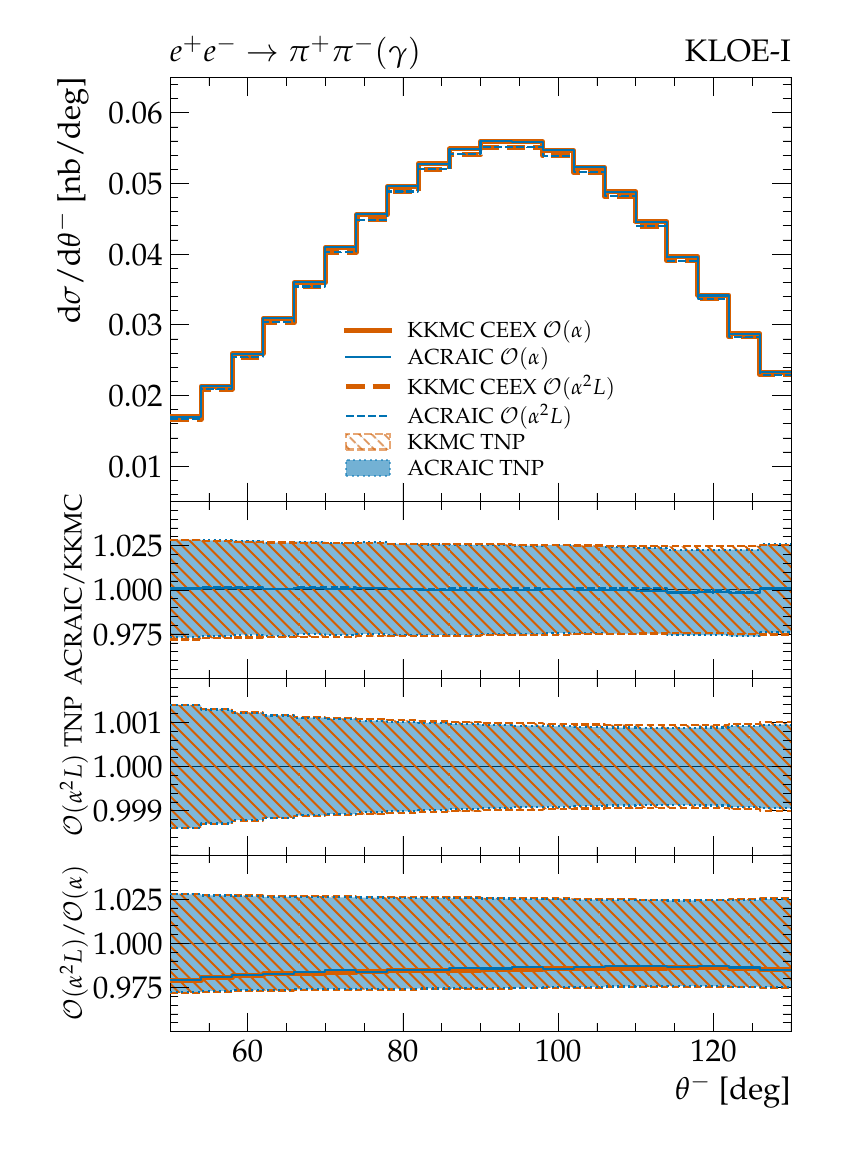}}\hfill
  \caption{The polar angle of the $\pi^\pm$ in the \KLOEI{} fiducial region.}
  \label{fig:kloeI_angles}
\end{figure}

\begin{figure}[htb]
  \centering
  \subfloat[]{\includegraphics[width=0.65\textwidth]{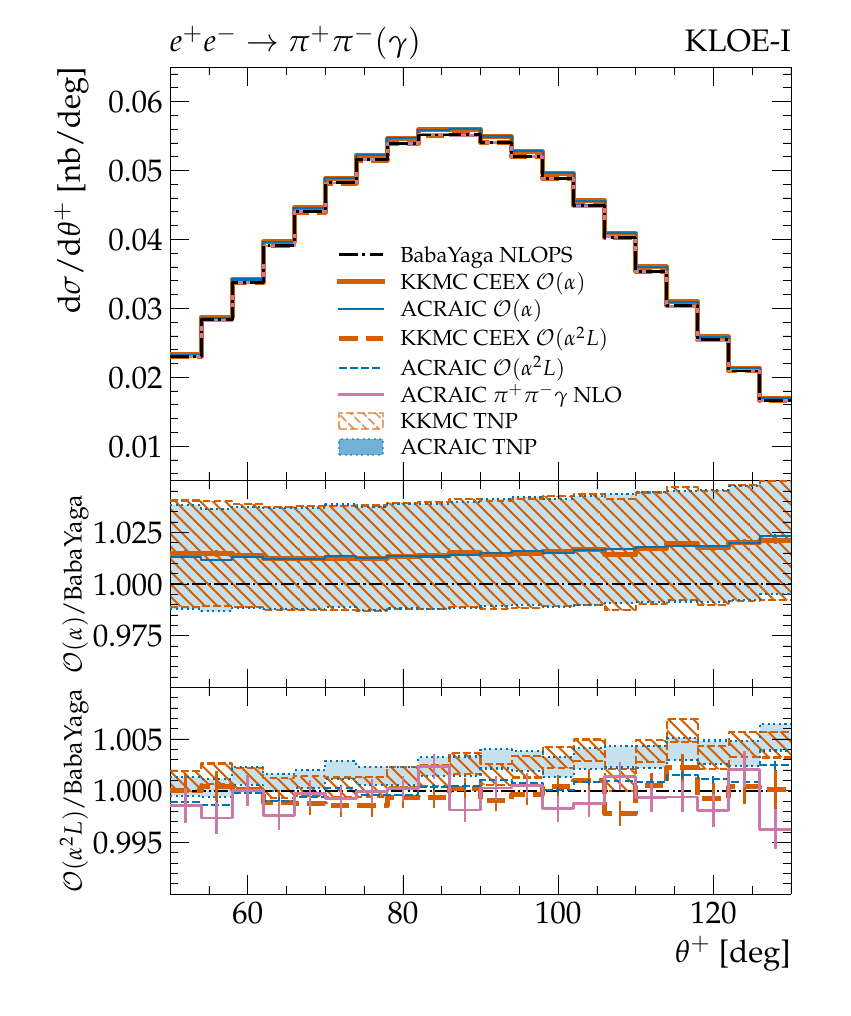}}
  \caption{Both generators against \BabaYaga as in \Cref{fig:bby_angles},
    at $\mathcal{O}(\alpha)$ (middle) and $\mathcal{O}(\alpha^2L)$ (lower panel) with their
    theory-nuisance-parameter bands, the lower panel also with the $\pi^+\pi^-\gamma$
    prediction of \ACRAIC at $\mathcal{O}(\alpha)$ (NLO) and its statistical uncertainty.}
  \label{fig:kloeI_bby_angles}
\end{figure}

At \KLOEI{} every event carries a large-angle photon with
$E_\gamma>20$~MeV, a direct test of the
one-photon amplitudes of $\tilde\beta_1$ with $F_\pi$ at the reduced virtuality
of each partition. At $\mathcal{O}(\alpha)$ the fiducial cross sections agree
within their statistical uncertainty of $\sim0.01\%$, and the photon and pion
angles (\cref{fig:kloeI_radiative,fig:kloeI_photon,fig:kloeI_angles}) within
the statistical precision without a slope. The $\mathcal{O}(\alpha)$ bands,
which include the radiative-return parameter of \cref{sec:tnp_breakdown}, are a
few per cent on the fiducial cross
section and in every $M_{\pi^+\pi^-}$ range in both generators. At
$\mathcal{O}(\alpha^2L)$ the fiducial band is $\sim0.1\%$, again mostly from
the radiative-return parameter, whose $\mathcal{O}(\alpha^2)$ correction is
dominated by the one-loop correction to the hard-photon emission, included
exactly in both generators. They agree on the $\mathcal{O}(\alpha)\to\mathcal{O}(\alpha^2L)$ step to
$\sim0.06\%$, within the $\mathcal{O}(\alpha^2L)$ band but several times the
statistical uncertainty, and bin by bin. Against \BabaYaga
(\cref{app:babayaga}) both generators lie $\sim1\%$ above at
$\mathcal{O}(\alpha)$ and below at $\mathcal{O}(\alpha^2L)$, by $\sim0.02\%$
(\ACRAIC) and $\sim0.09\%$ (\kkmc). In the polar angle of the $\pi^+$
(\cref{fig:kloeI_bby_angles}) both follow the shape of \BabaYaga and at
$\mathcal{O}(\alpha^2L)$ rise relative to it by a few tenths of a per cent
across $\theta^+$, more than their bands. With an uncertainty of the size of our
$\mathcal{O}(\alpha^2L)$ bands, not quoted for \BabaYaga, the predictions
agree. For $\pi^+\pi^-\gamma$ in the same scenario the exact
next-to-leading-order calculations of \BabaYaga and \Phokhara differ by a slope
of a few per cent across $\theta^+$~\cite{Budassi:2026lmr}, an order of
magnitude more than the difference found here, so our results support the
\BabaYaga prediction in this distribution.

\subsection{\KLOEII{}}

\begin{figure}[htb]
  \centering
  \subfloat[]{\includegraphics[width=0.48\textwidth]{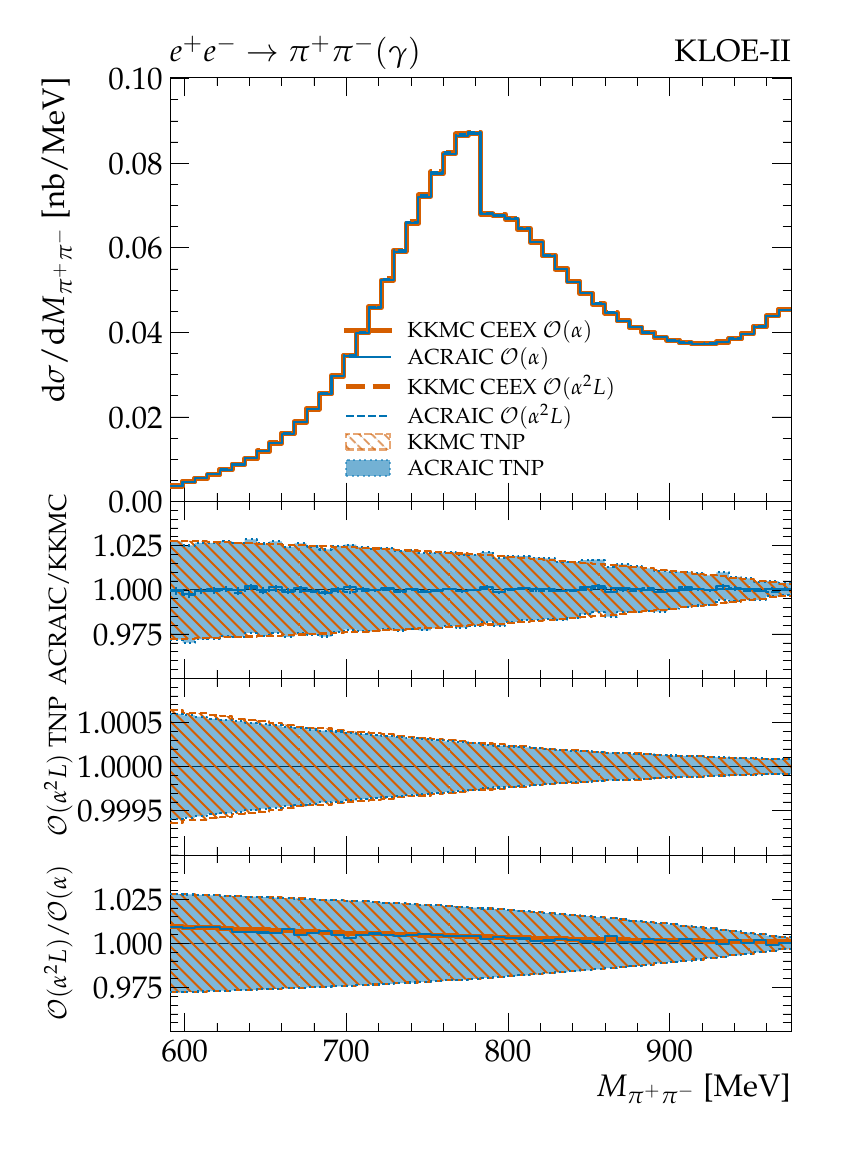}}\hfill
  \subfloat[]{\includegraphics[width=0.48\textwidth]{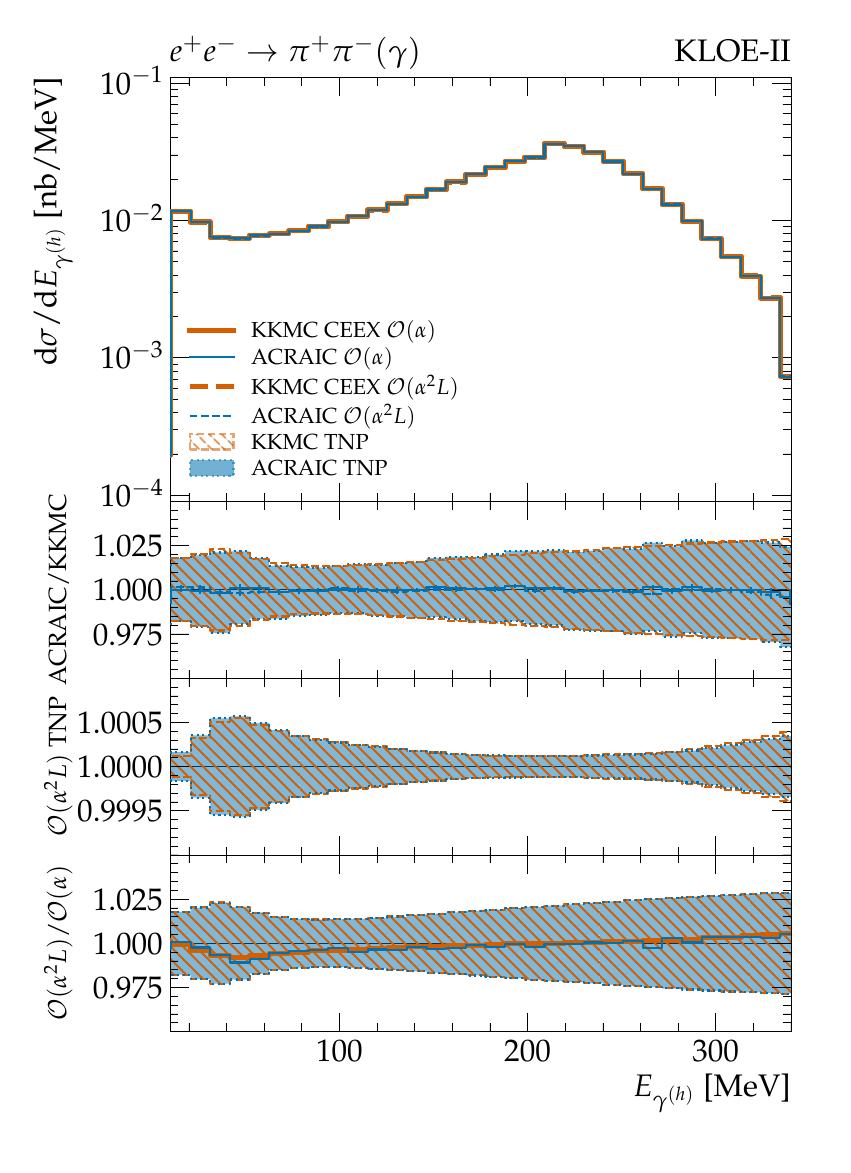}}
  \caption{The pion-pair mass $M_{\pi^+\pi^-}$ (left), the energy of the
    hardest photon (right) in the
    \KLOEII{} fiducial region. Curves and bands as in~\Cref{fig:kloeI_radiative}.}
  \label{fig:kloeII_radiative}
\end{figure}

\begin{figure}[htb]
  \centering
  \subfloat[]{\includegraphics[width=0.5\textwidth]{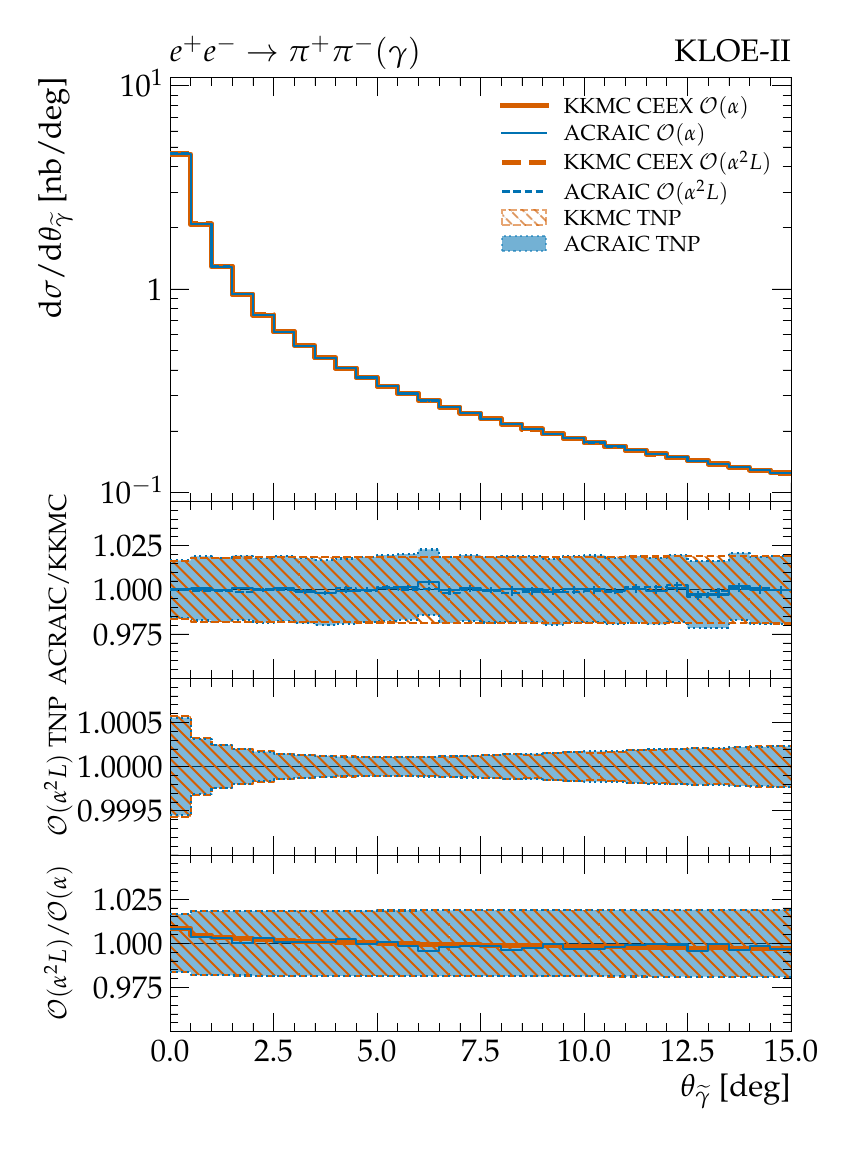}}\hfill
  \subfloat[]{\includegraphics[width=0.5\textwidth]{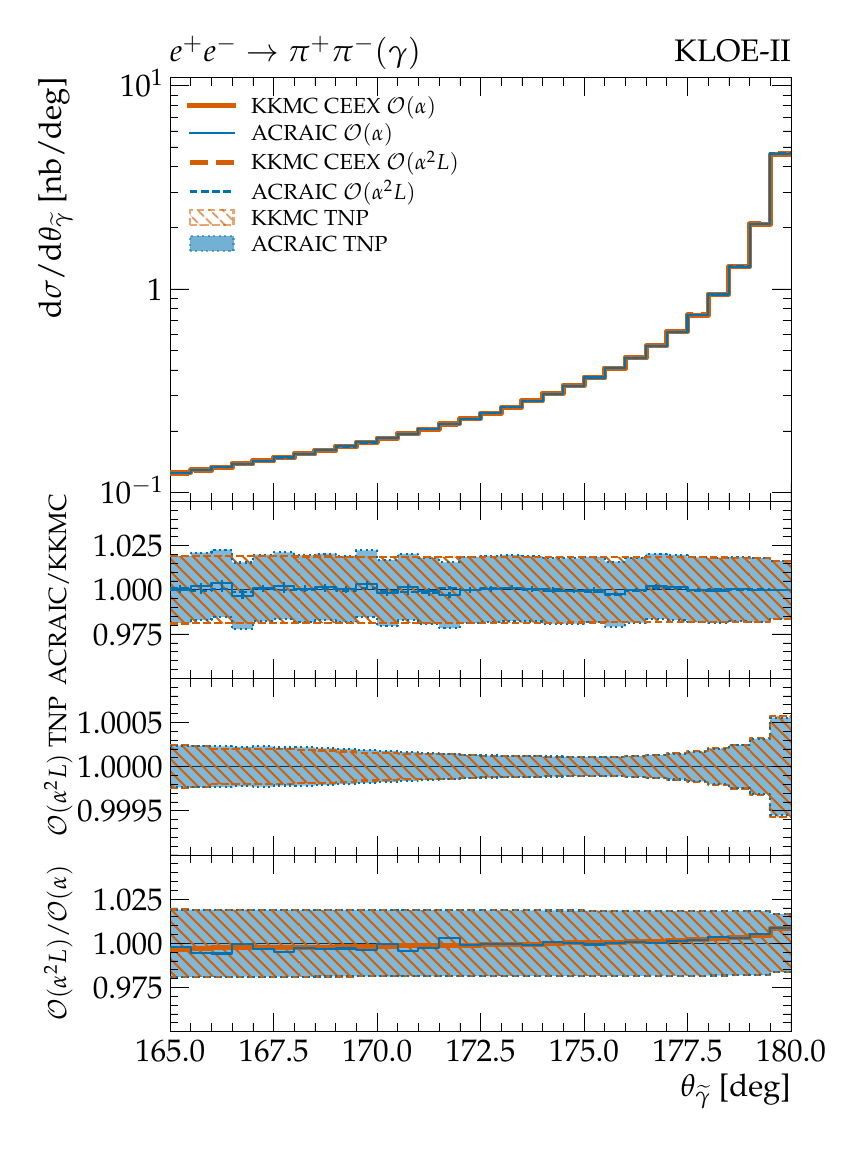}}\\
  \caption{The polar angle $\theta_{\widetilde\gamma}$ of the direction
    opposite the pion pair, near the two beams in the
    \KLOEII{} fiducial region. Curves and bands as in~\Cref{fig:kloeI_radiative}.}
  \label{fig:kloeII_photon}
\end{figure}

\begin{figure}[htb]
  \centering
  \subfloat[]{\includegraphics[width=0.5\textwidth]{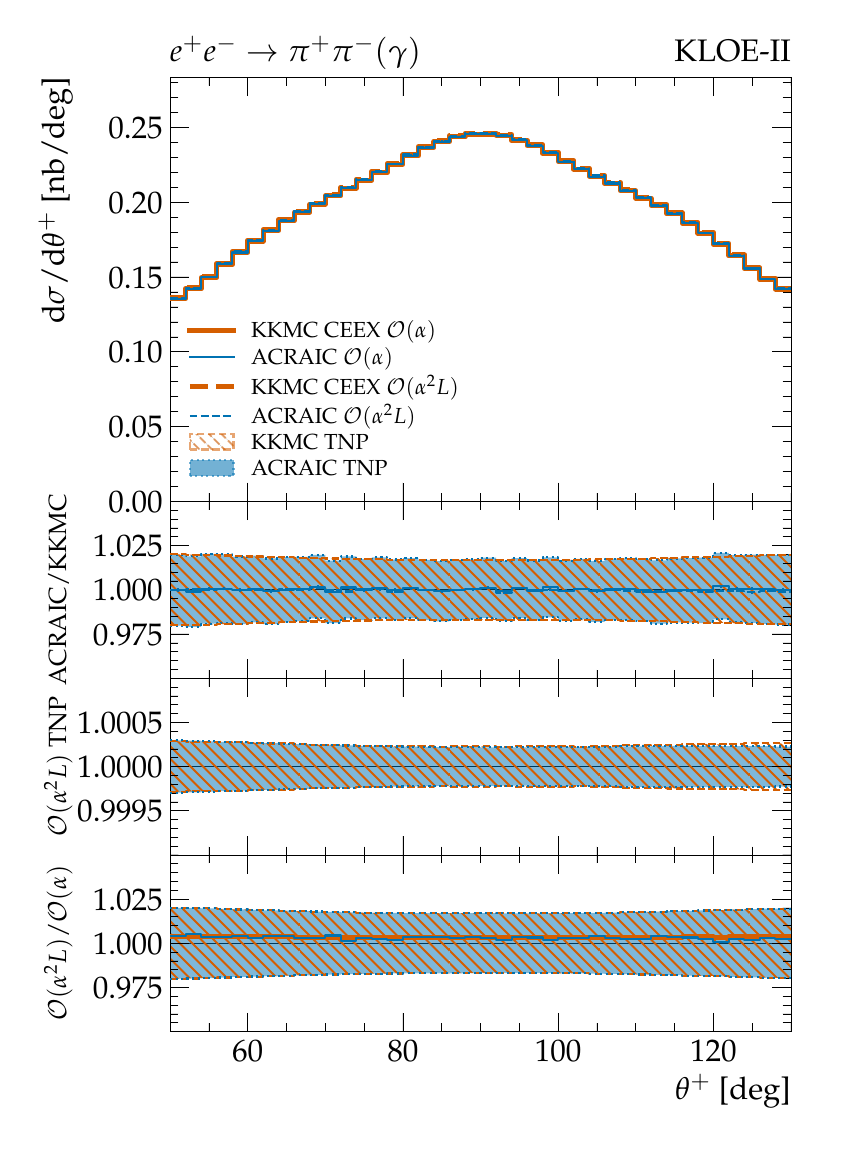}}\hfill
  \subfloat[]{\includegraphics[width=0.5\textwidth]{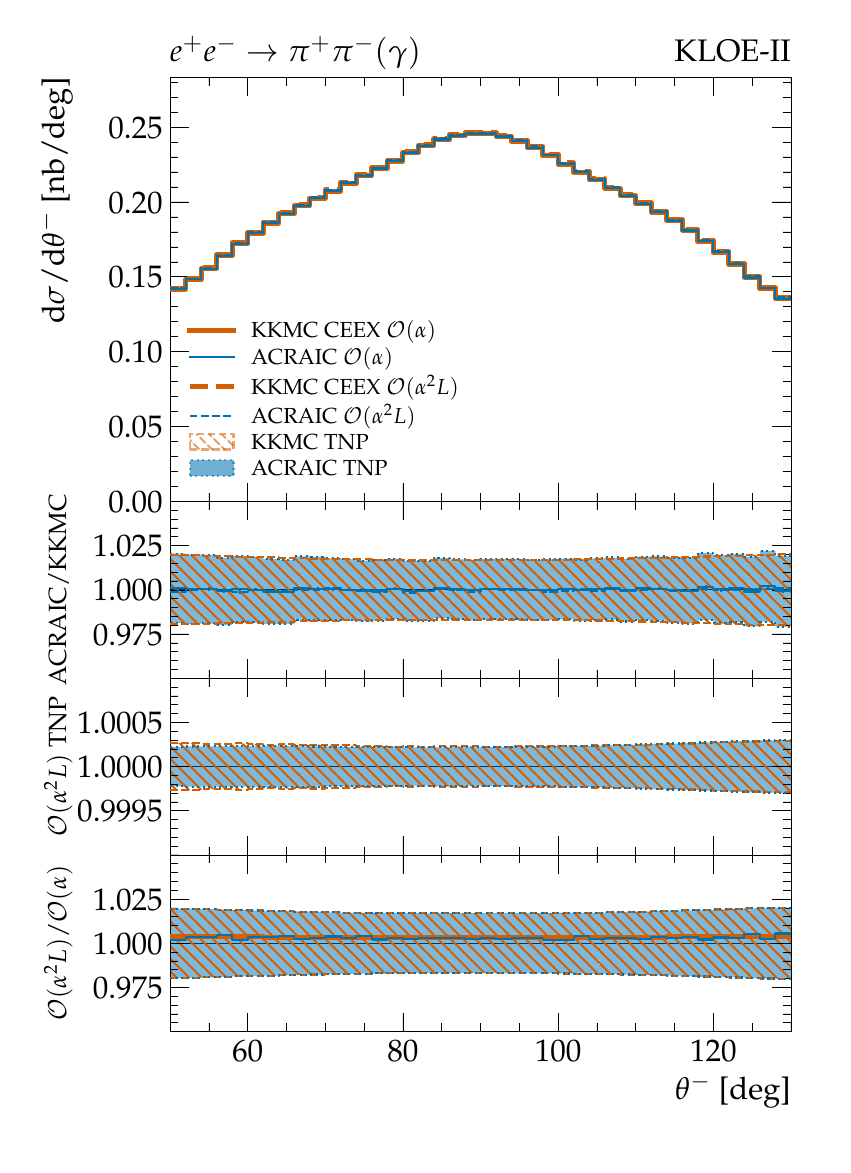}}
  \caption{The polar angles
    of the $\pi^+$ (left) and the $\pi^-$ (right) in the
    \KLOEII{} fiducial region. Curves and bands as in~\Cref{fig:kloeI_radiative}.}
  \label{fig:kloeII_angles}
\end{figure}

At \KLOEII{} the photon is not detected and the missing
momentum must point within $15^\circ$ of a beam, so collinear initial-state
radiation dominates. At $\mathcal{O}(\alpha)$ the fiducial cross sections agree
within their statistical uncertainty of $\sim0.01\%$. In the $M_{\pi^+\pi^-}$
ranges below $700$~MeV, $700$--$850$~MeV and above $850$~MeV
(\cref{fig:kloeII_radiative,fig:kloeII_photon,fig:kloeII_angles}) the ratio
agrees with one to $\sim0.1\%$, as does the missing-momentum angle near both
beams, and the pion angles agree to $\sim0.01\%$. The $\mathcal{O}(\alpha)$
bands, including the radiative-return parameter, are $\sim2\%$ on the fiducial cross
section in both generators and shrink to $\sim0.02\%$ at
$\mathcal{O}(\alpha^2L)$, because the radiative-return parameter scales with the
step between the orders. This step, $\sim0.3\%$, comes from the real-virtual
correction of collinear initial-state photons, and the generators agree on it to
$\sim0.02\%$ and on the fiducial cross section within the statistical
uncertainty.

\subsection{\BES{}}

\begin{figure}[htbp]
  \centering
  \subfloat[]{\includegraphics[width=0.45\textwidth]{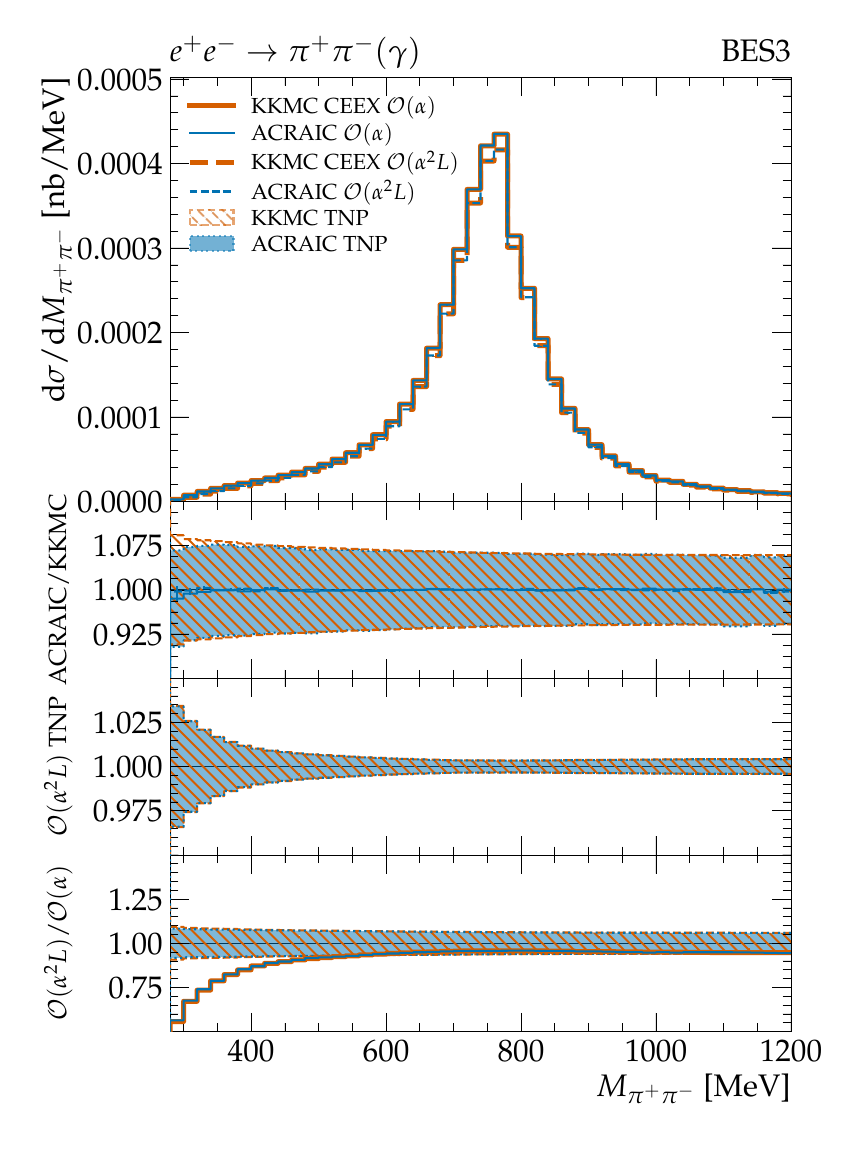}}
  \subfloat[]{\includegraphics[width=0.45\textwidth]{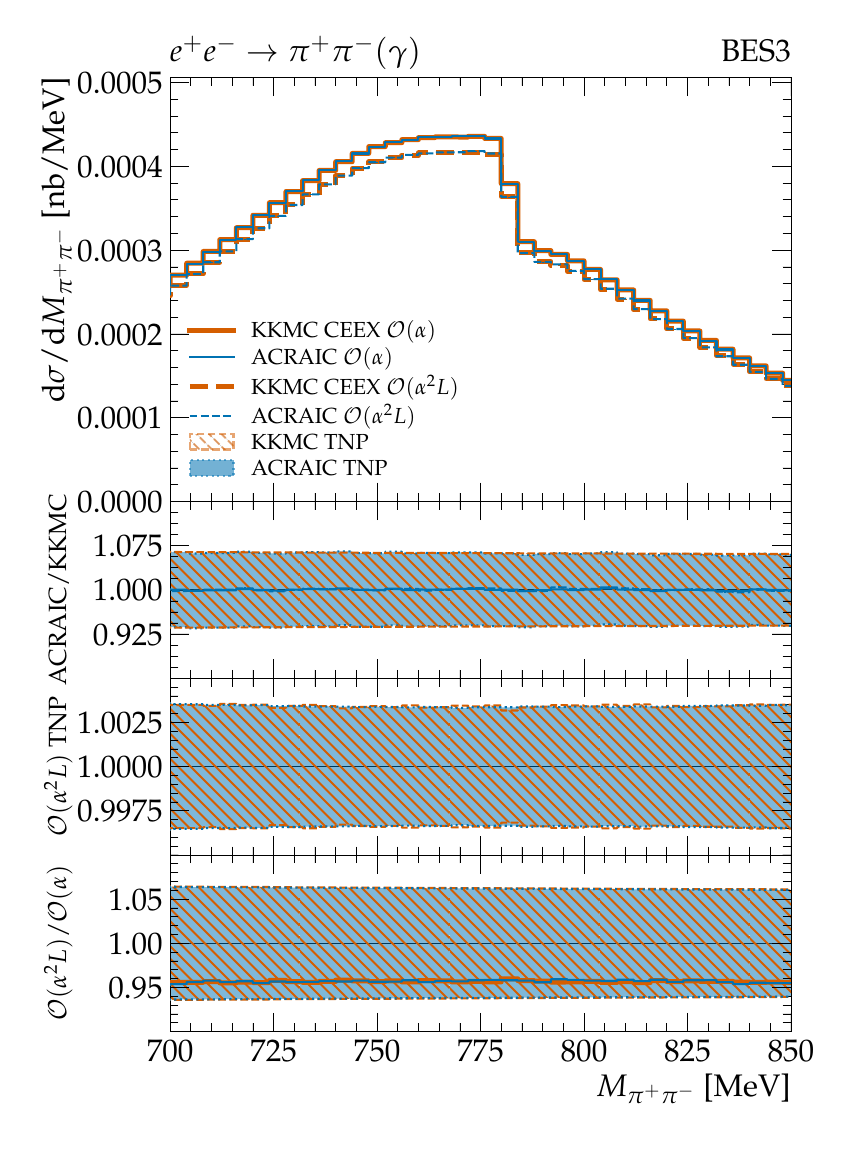}}\hfill
  \subfloat[]{\includegraphics[width=0.45\textwidth]{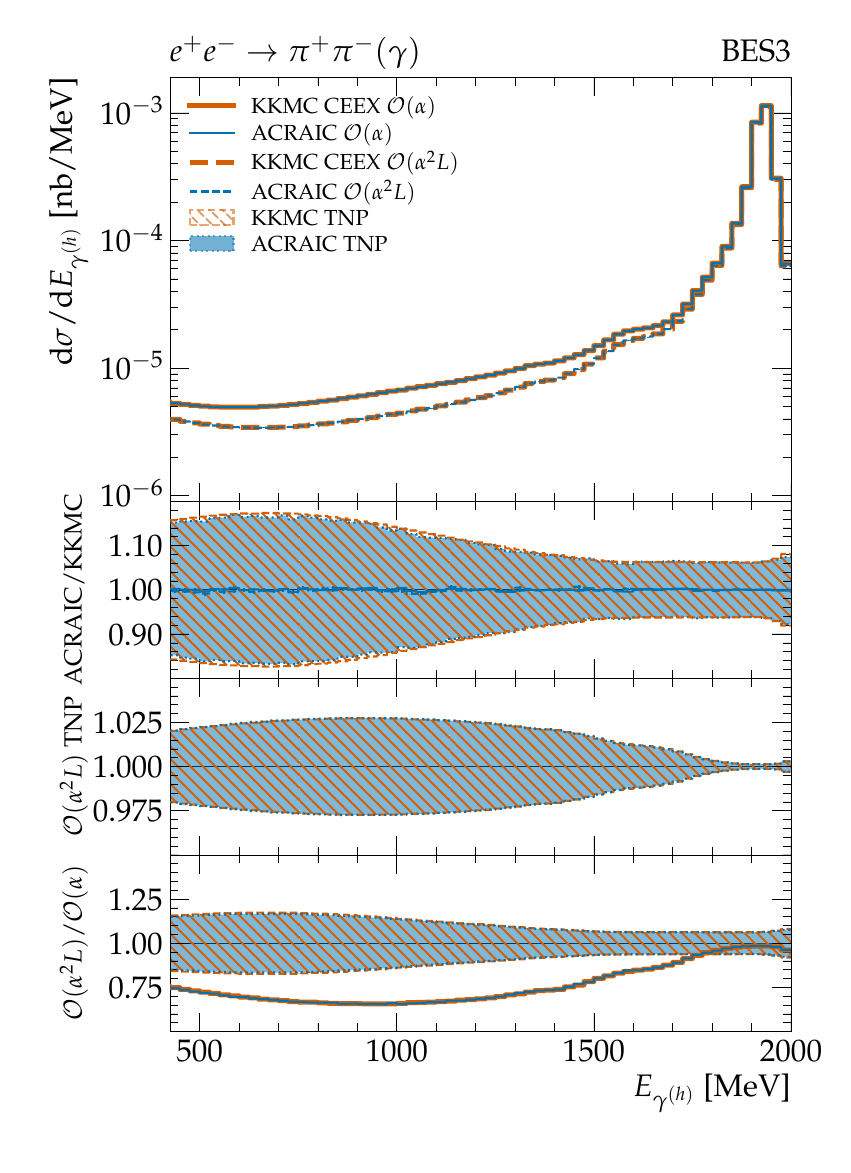}}
  \caption{The pion-pair mass $M_{\pi^+\pi^-}$ in $20$~MeV bins (a) and in the
    $\rho$--$\omega$ region in $4$~MeV bins on a linear scale (b), the
    energy (c) in the \BES{} fiducial region. Curves and bands as in~\Cref{fig:kloeI_radiative}.}
  \label{fig:bes3}
\end{figure}

At \BES{} the tagged photon carries most of the beam energy, with
$E_\gamma>0.4\sqrts$ for most of the fiducial cross section. At
$\mathcal{O}(\alpha)$ the fiducial cross sections agree to $\sim0.003\%$, within
the statistical uncertainty of the \kkmc sample. In \cref{fig:bes3} the ratio
agrees with one to $\sim0.1\%$ in the $M_{\pi^+\pi^-}$ ranges below $600$~MeV,
$600$--$900$~MeV and $900$--$1200$~MeV and in the photon energy, and the
$\rho$--$\omega$ interference near $782$~MeV is reproduced bin by bin. At
$\mathcal{O}(\alpha^2L)$, with the exact two-photon matrix element in both
generators, the fiducial cross sections agree to $\sim0.03\%$, within the
statistical uncertainty, and the step, $\sim-5\%$, agrees bin by bin. It reduces
the fiducial band, dominated at both orders by the radiative-return parameter,
from $\sim6\%$ to $\sim0.4\%$. The step is largest at low $M_{\pi^+\pi^-}$,
where the photon is hardest, and below $M_{\pi^+\pi^-}\sim500$~MeV and below the
peak of the photon energy it reaches tens of per cent and leaves the
$\mathcal{O}(\alpha)$ bands. These events need a second hard photon sharing the
energy~\cite{BaBar:2023xiy}, a leading-order effect of the $\pi^+\pi^-\gamma$
final state rather than a missing higher order (\cref{sec:tnp_breakdown}), which
also places the $\mathcal{O}(\alpha)$ predictions outside their bands against
\BabaYaga in the same regions (\cref{app:babayaga}). At
$\mathcal{O}(\alpha^2L)$ the generators agree within their bands in every
distribution.

\begin{figure}[htb]
  \centering
  \subfloat[]{\includegraphics[width=0.45\textwidth]{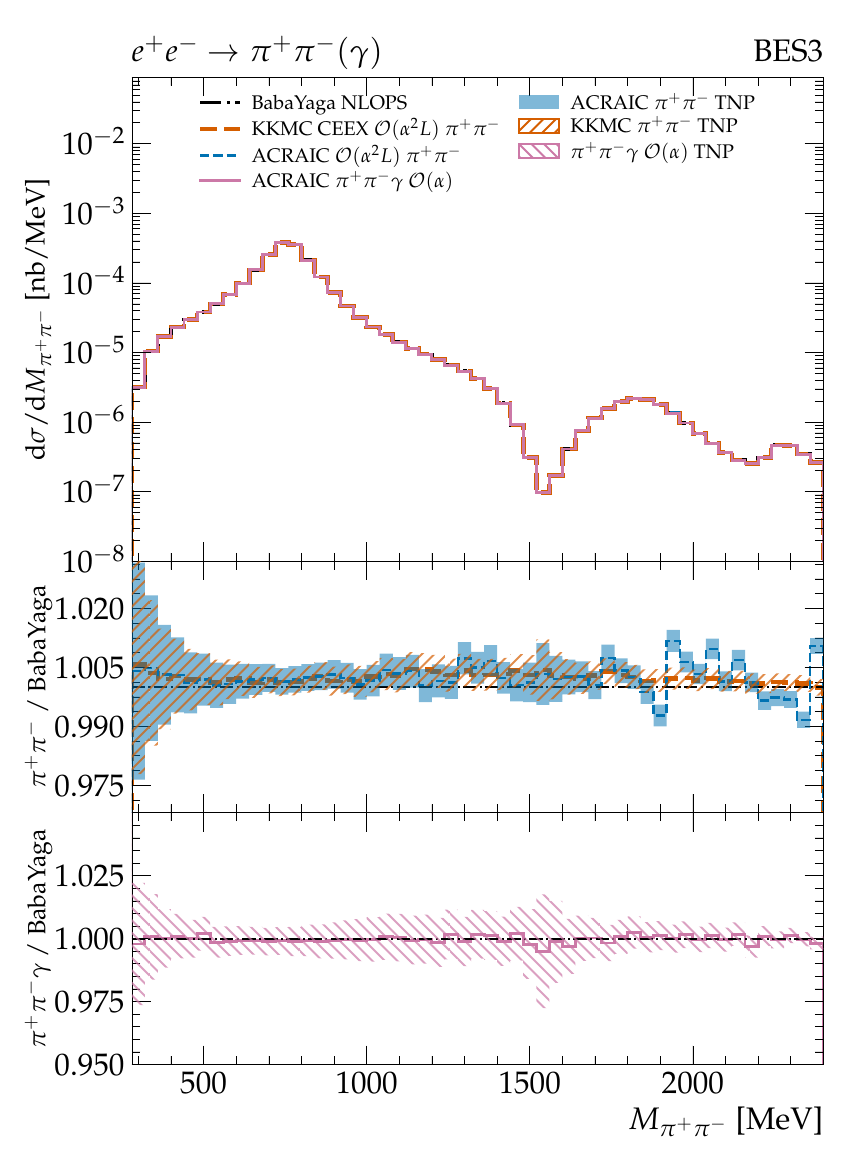}}\hfill
  \subfloat[]{\includegraphics[width=0.45\textwidth]{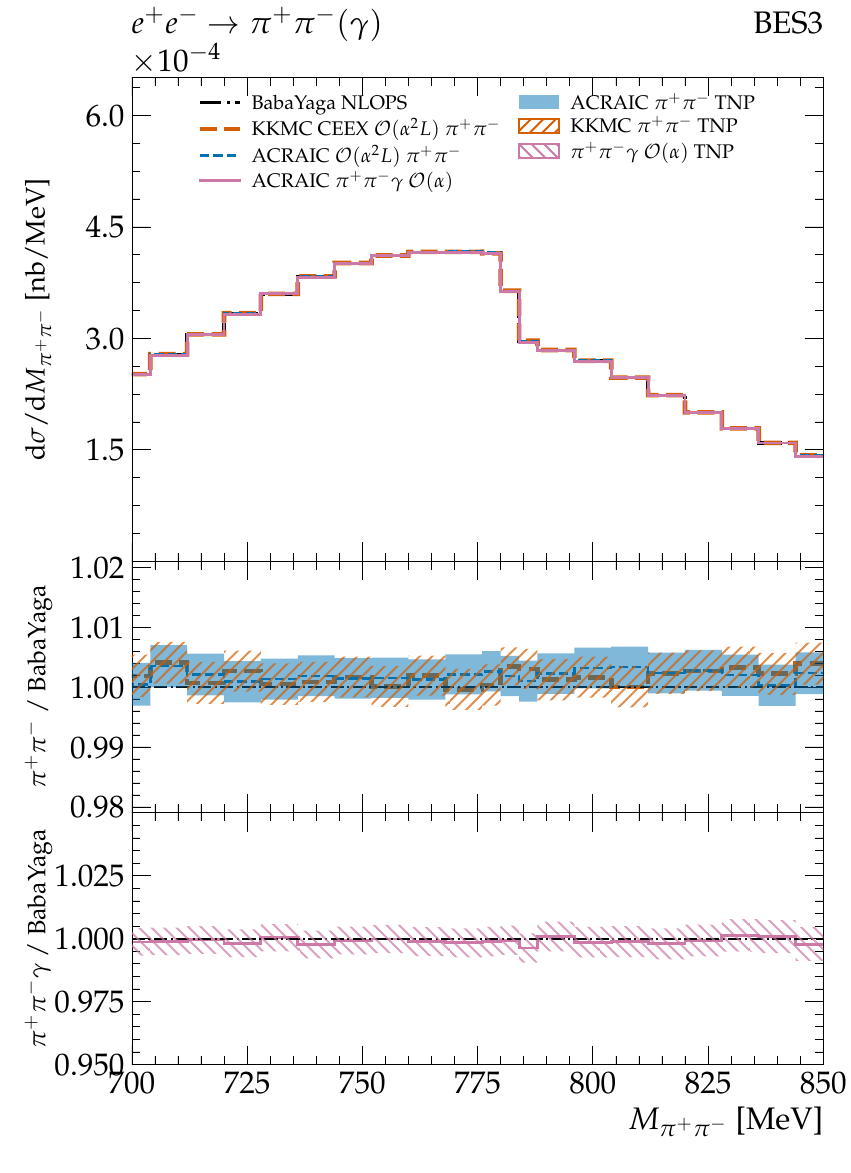}}
  \caption{The pion-pair mass (a) and in the
    $\rho$--$\omega$ region on a linear scale (b) in the \BES{} fiducial region. The upper panels show
    \BabaYaga (NLOPS), the $\pi^+\pi^-$ predictions of \kkmc CEEX
    and \Sherpa \ACRAIC at $\mathcal{O}(\alpha^2L)$ and the
    $\pi^+\pi^-\gamma$ prediction of \ACRAIC at $\mathcal{O}(\alpha)$. The middle and
    lower panels show the $\pi^+\pi^-$ and the $\pi^+\pi^-\gamma$ predictions with their
    theory-nuisance-parameter bands as ratios to \BabaYaga.}
  \label{fig:bes3_pipig}
\end{figure}

\Cref{fig:bes3_pipig} compares the $\pi^+\pi^-$ predictions with
\BabaYaga and with the explicit $\pi^+\pi^-\gamma$ calculation of \ACRAIC at
next-to-leading order, which includes the tagged photon and the one-loop
correction to its emission in the Born. The further
photons, matched to the \Comix amplitudes, change the $\pi^+\pi^-\gamma$ cross
section by at most a few tenths of a per cent and are included. This prediction agrees with \BabaYaga to $\sim1\%$ in all distributions and to
$\sim0.1\%$ in most bins around the $\rho$ peak, within its band in every bin.
Its next-to-leading-order correction reduces the fiducial band from several per cent at leading order, dominated by the
radiative-return parameter, to a few tenths of a per cent, comparable with the
$\mathcal{O}(\alpha^2L)$ band of the $\pi^+\pi^-$ predictions. The
$\pi^+\pi^-$ and $\pi^+\pi^-\gamma$ predictions agree on the fiducial cross
section to $\sim0.1\%$ at strict first order, where they are the same
calculation (\cref{app:babayaga}), and for \ACRAIC at strict second order.

The $\pi^+\pi^-$ predictions draw the tagged photon from the
YFS crude distribution, which follows the eikonal factors and rarely gives a
hard photon at a large angle, and $\sim10\%$ of the generated events of
\ACRAIC are effective, against a few tenths for the $\pi^+\pi^-\gamma$
calculation, which samples the tagged photon directly. For tagged-photon
measurements the latter is therefore both the more natural and the more
efficient description.
\subsection{\Bscen{}}
\begin{figure}[htb]
  \centering
  \subfloat[]{\includegraphics[width=0.5\textwidth]{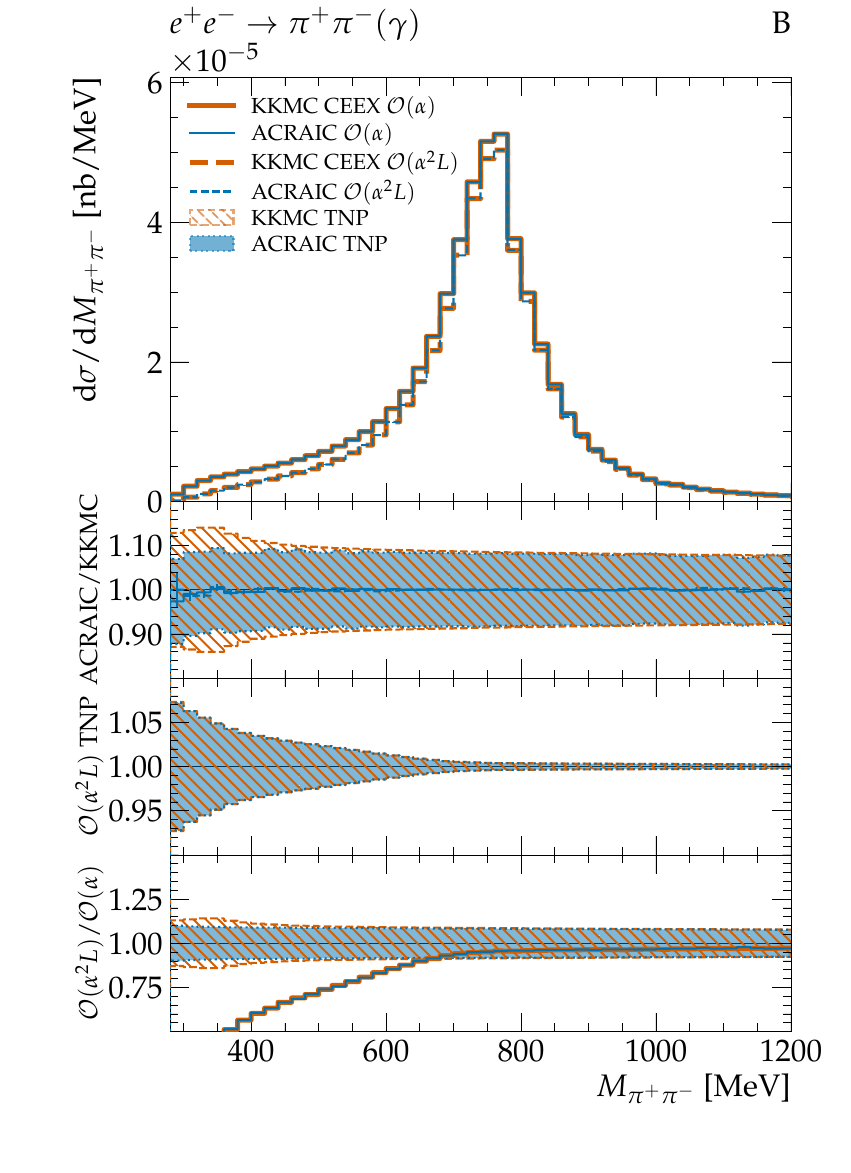}}\hfill
  \subfloat[]{\includegraphics[width=0.5\textwidth]{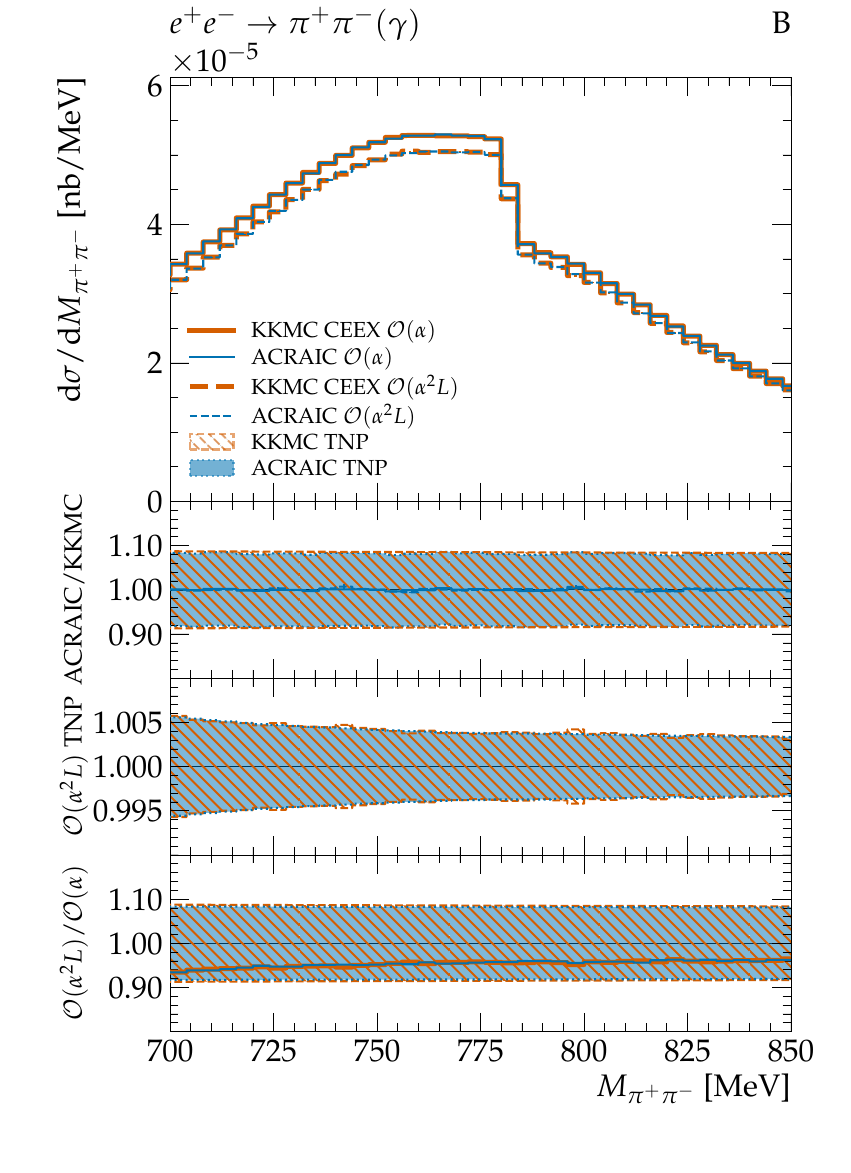}}\hfill
  \caption{The pion-pair mass $M_{\pi^+\pi^-}$ (a) and in the
    $\rho$--$\omega$ region on a linear scale (b) in the \Bscen{} fiducial region. Curves and bands as in~\Cref{fig:kloeI_radiative}.}
  \label{fig:b}
\end{figure}

\begin{figure}[htbp]
  \centering
  \subfloat[]{\includegraphics[width=0.45\textwidth]{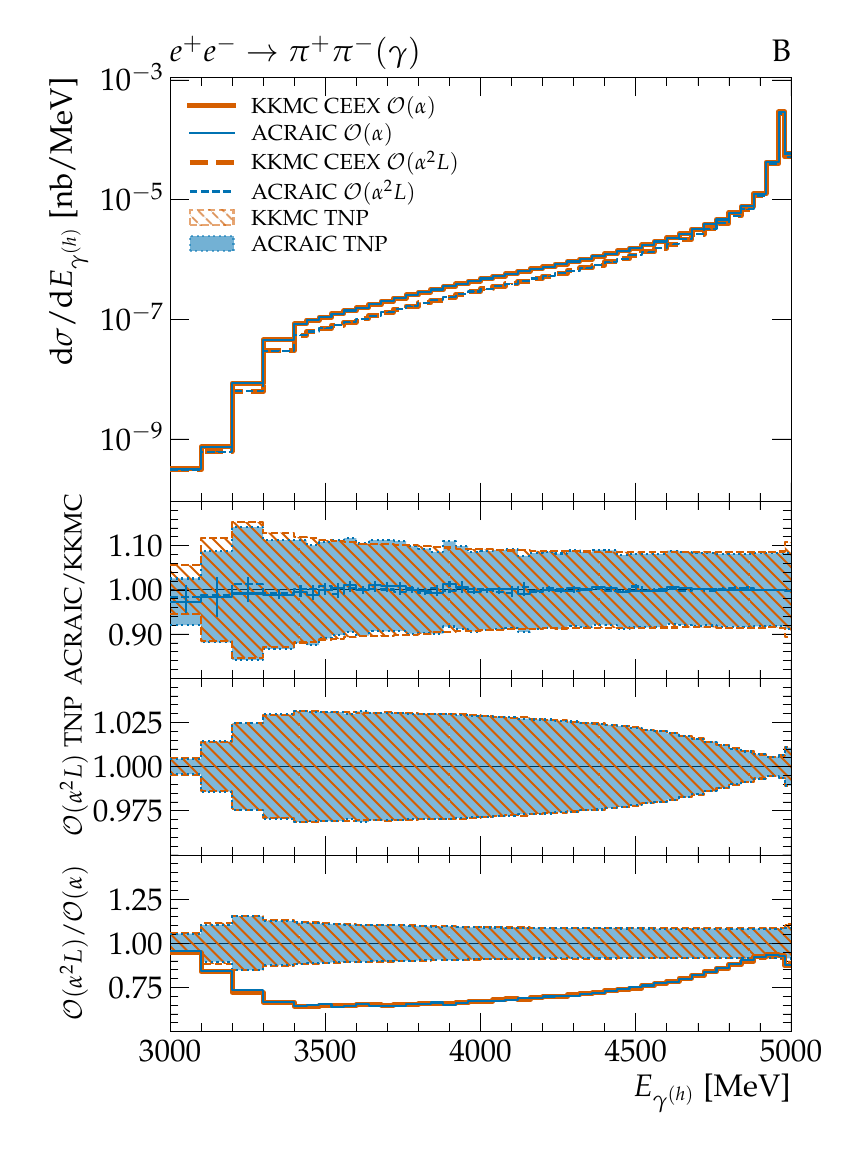}}\hfill
  \subfloat[]{\includegraphics[width=0.45\textwidth]{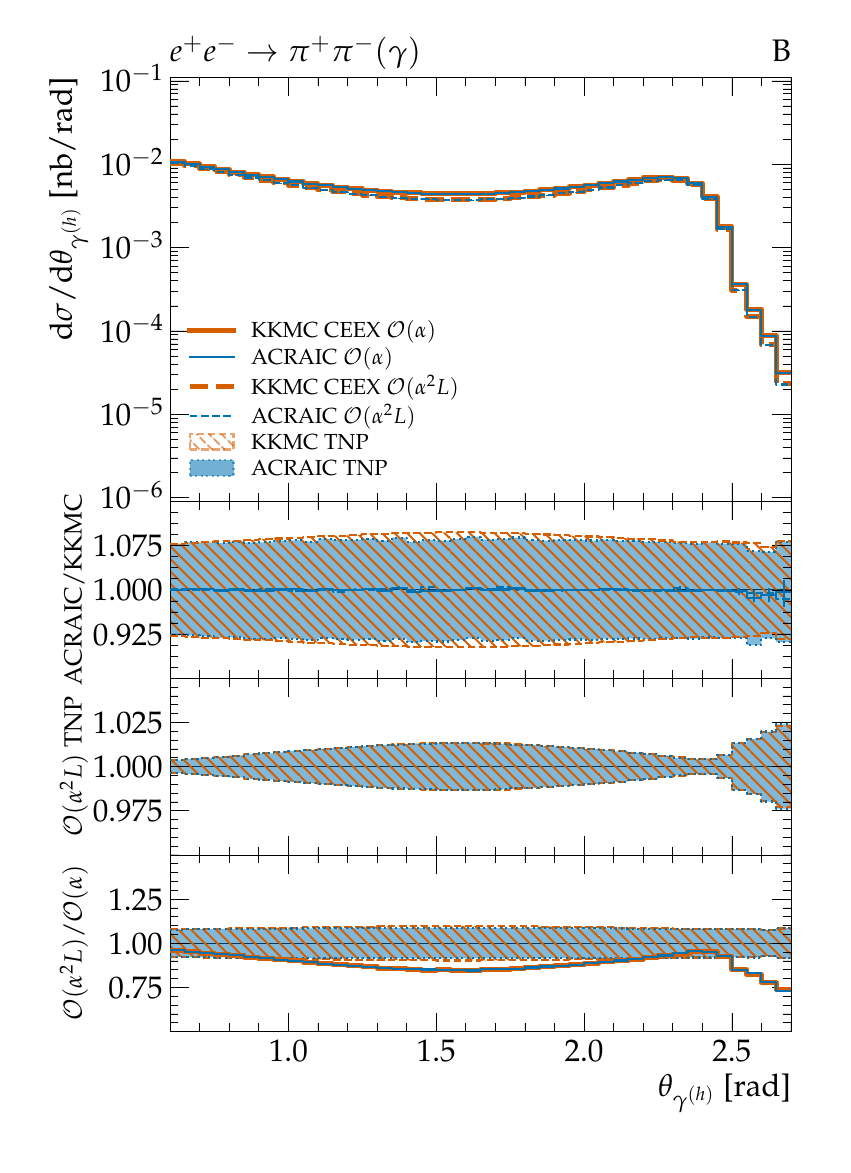}}\hfill
  \subfloat[]{\includegraphics[width=0.45\textwidth]{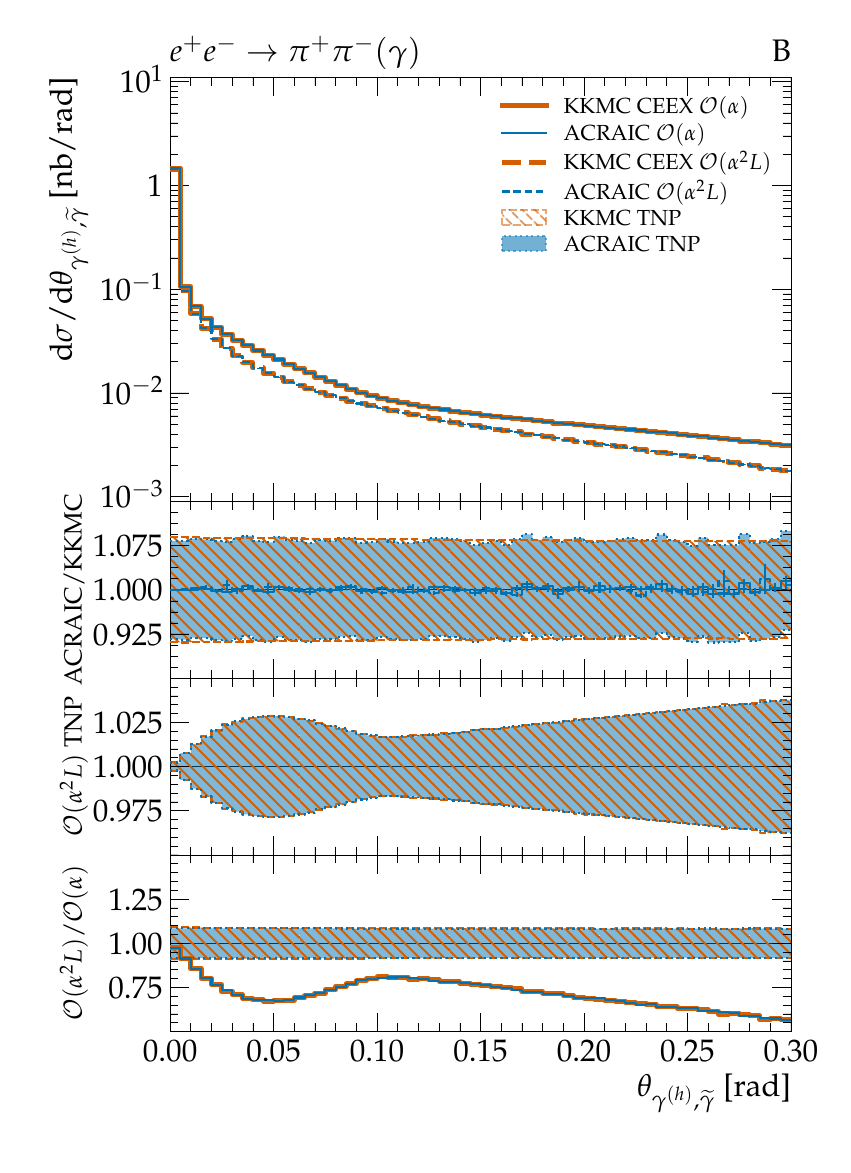}}
  \caption{The energy (a) and the polar angle (b) of the hardest photon, and its angle
    to the direction opposite the pion pair (c) in the \Bscen{} fiducial region. Curves and bands as in
    ~\Cref{fig:kloeI_radiative}.}
  \label{fig:b_photon}
\end{figure}

At \Bscen{} the photon carries more than $60\%$ of the beam energy in
every event. At $\mathcal{O}(\alpha)$ the fiducial cross sections agree to
$\sim0.03\%$, within the statistical uncertainty of the \kkmc sample, and the
distributions (\cref{fig:b,fig:b_photon}) within the statistical precision,
$\sim0.1\%$ where most events lie. At $\mathcal{O}(\alpha^2L)$ they agree to
$\sim0.02\%$, within the statistical uncertainty, and the steps, $\sim-10\%$,
agree bin by bin, reducing the fiducial band, dominated at both orders by the
radiative-return parameter, from $\sim9\%$ to $\sim0.9\%$. As at \BES{} the step leaves the $\mathcal{O}(\alpha)$
bands where a second hard photon matters, below $M_{\pi^+\pi^-}\sim650$~MeV,
below the peak of the photon energy and in both photon angles, by up to tens of
per cent in the lowest-mass bins. The angle to the direction opposite
the pion pair needs a second photon.
\begin{figure}[htb]
  \centering
  \subfloat[]{\includegraphics[width=0.45\textwidth]{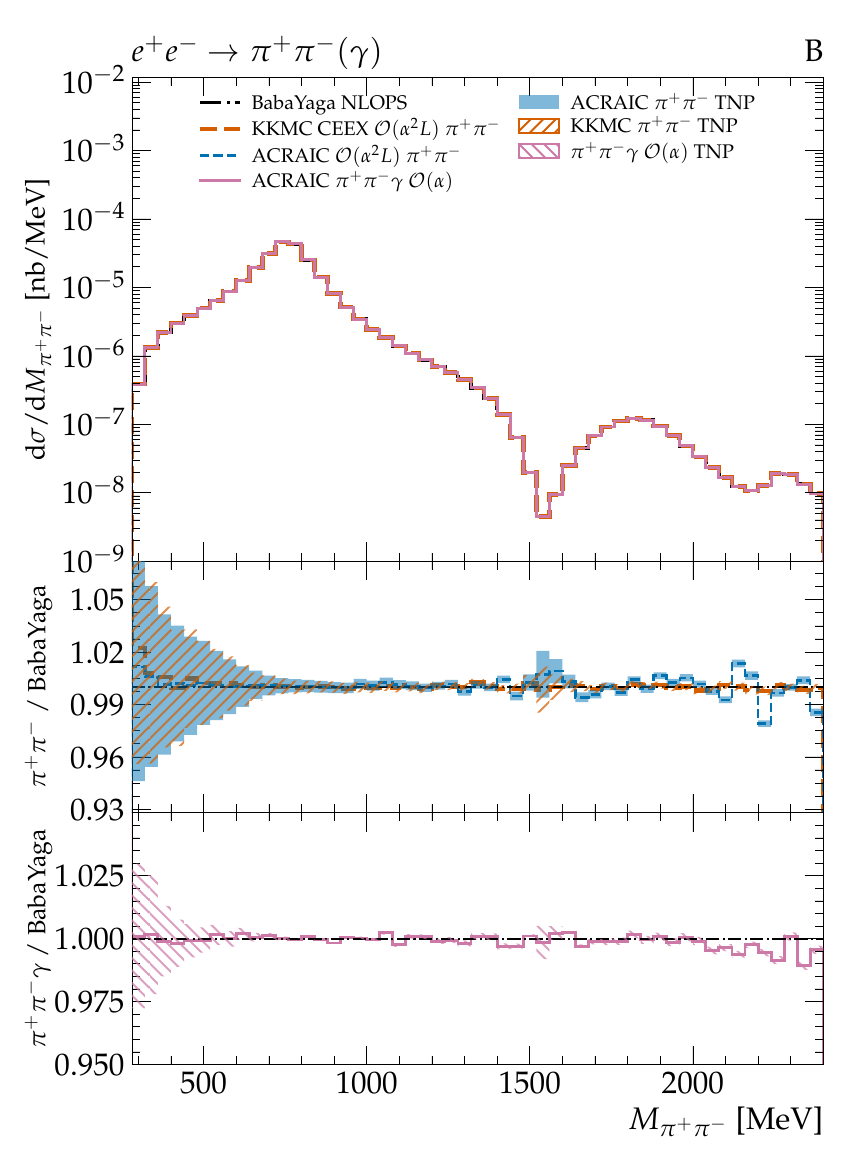}}\hfill
  \subfloat[]{\includegraphics[width=0.45\textwidth]{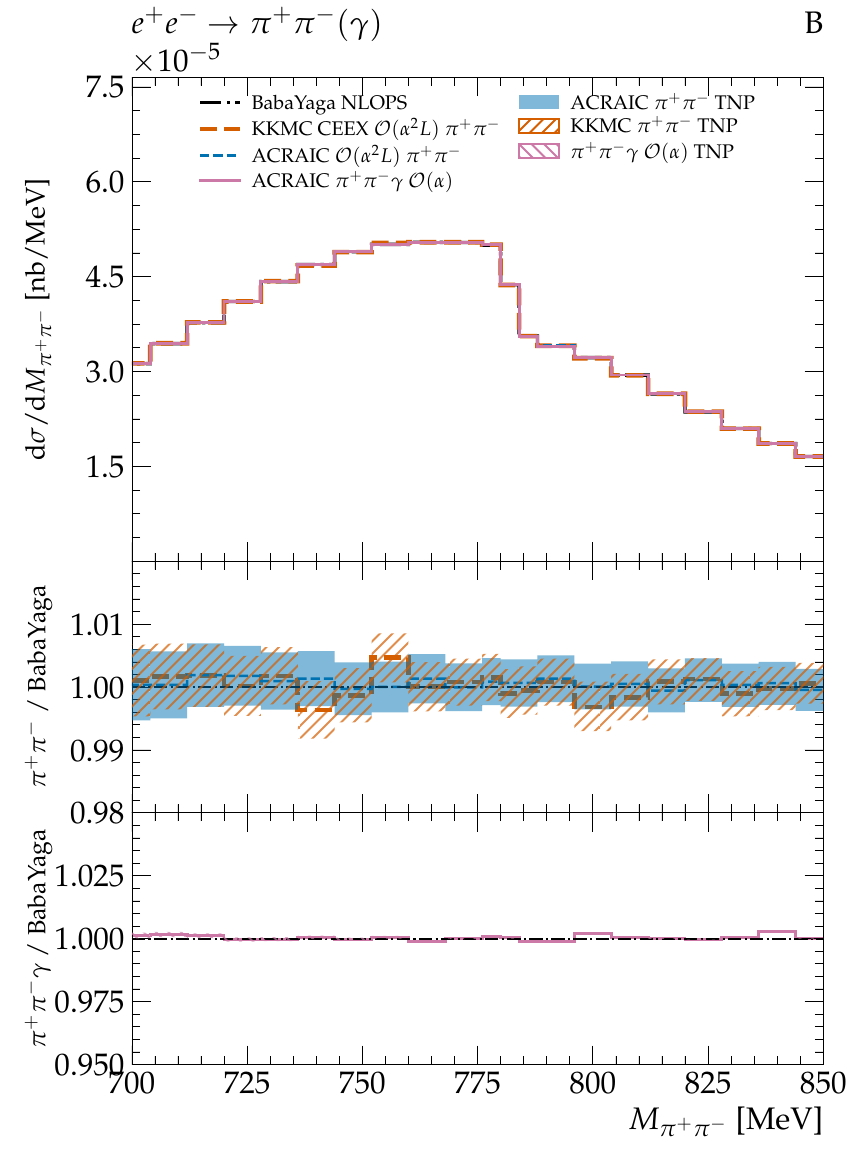}}\\[1em]
  \caption{The pion-pair mass $M_{\pi^+\pi^-}$ (a) and in the
    $\rho$--$\omega$ region on a linear scale (b) in the \Bscen{} fiducial region. Curves and panels as in \Cref{fig:bes3_pipig}.}
  \label{fig:b_pipig_a}
\end{figure}
\begin{figure}[htbp]
  \centering
  \subfloat[]{\includegraphics[width=0.45\textwidth]{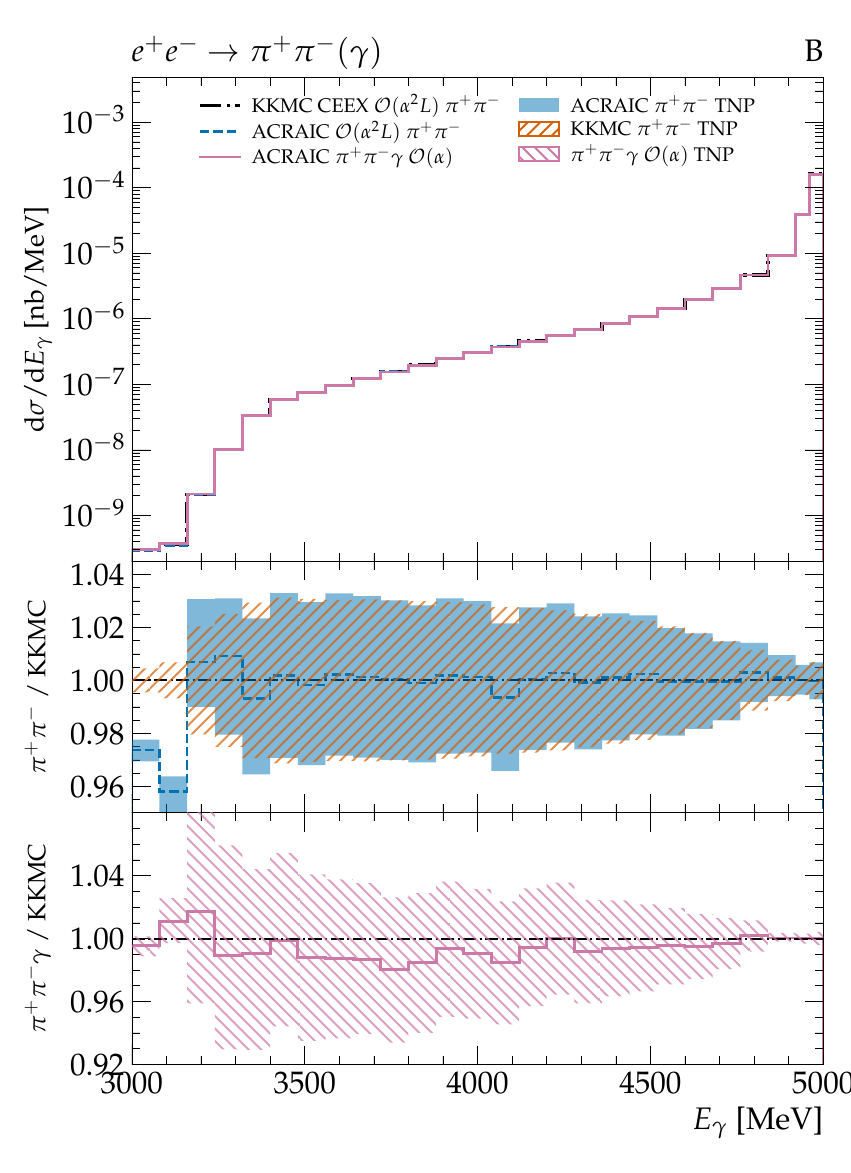}}\hfill
  \subfloat[]{\includegraphics[width=0.45\textwidth]{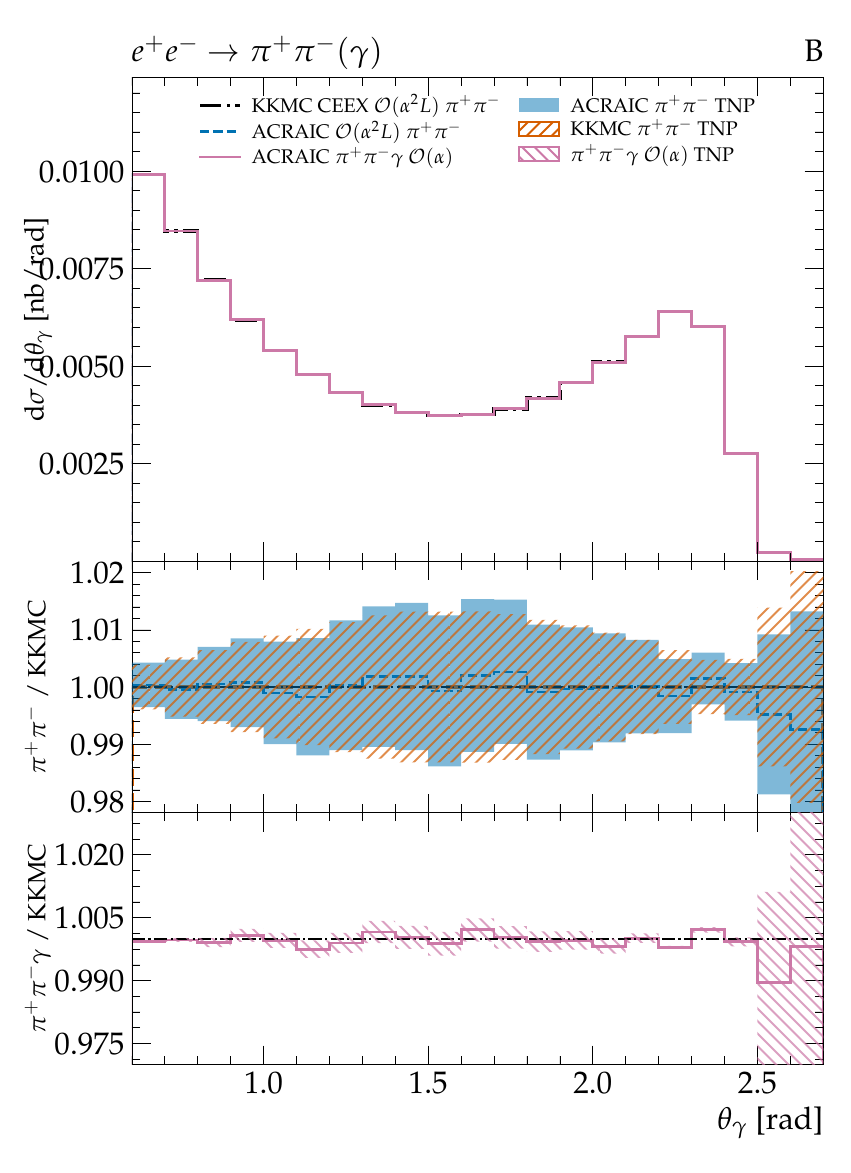}}\\
  \subfloat[]{\includegraphics[width=0.45\textwidth]{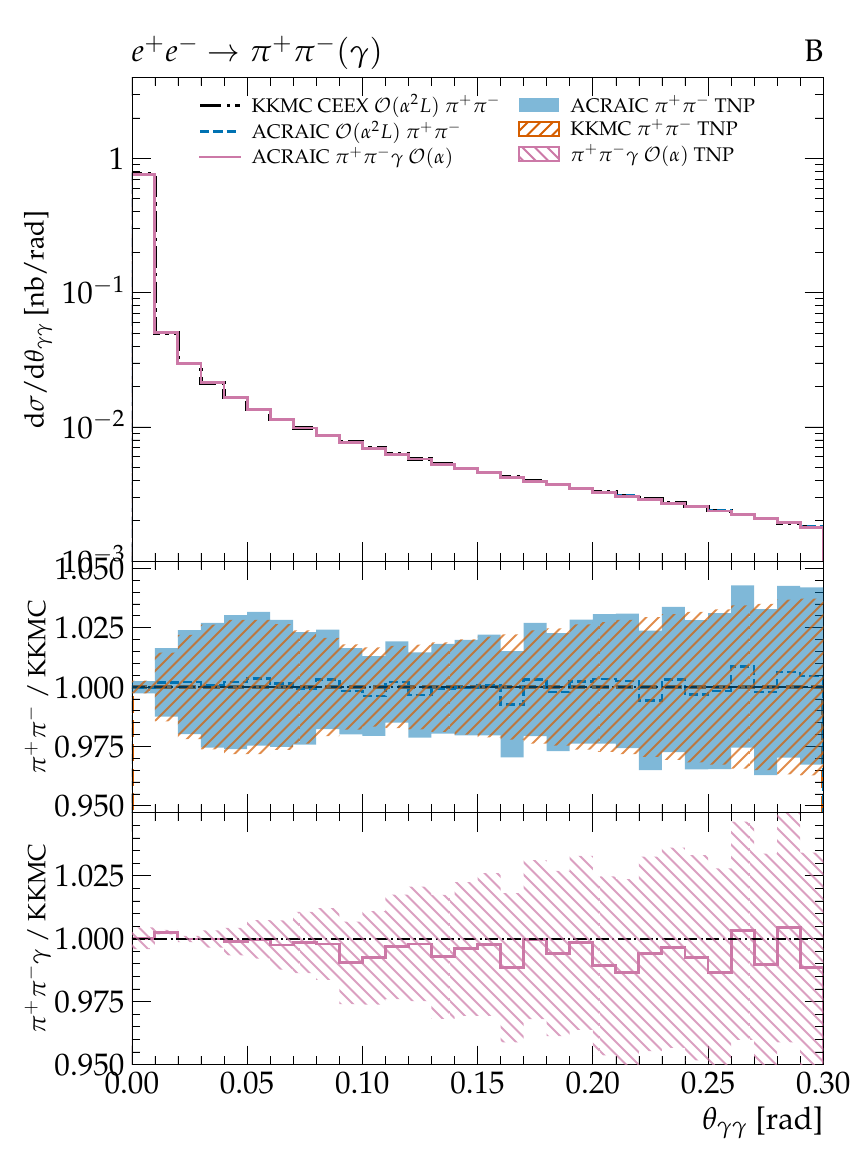}}
  \caption{The energy (a) and the polar angle (b) of the hardest photon, and its
    angle to the direction opposite the pion pair (c), in the \Bscen{} fiducial region. Curves
    and panels as in \Cref{fig:bes3_pipig}.}
  \label{fig:b_pipig_b}
\end{figure}
With the photon in the Born process (\cref{fig:b_pipig_a}), the
$\pi^+\pi^-\gamma$ prediction agrees with \BabaYaga to $\sim1\%$ in the
distributions and to $\sim0.1\%$ around the $\rho$ peak in $M_{\pi^+\pi^-}$, and
its next-to-leading-order correction reduces the fiducial band from several per
cent to $\sim0.1\%$, several times smaller than the $\mathcal{O}(\alpha^2L)$
band of the $\pi^+\pi^-$ predictions. Those agree with \BabaYaga and with the
$\pi^+\pi^-\gamma$ prediction to a few tenths of a per cent down to $400$~MeV
and to $\lesssim1\%$ in the lowest bins. In the photon energy and angles
(\cref{fig:b_pipig_b}) the two descriptions agree within the statistical
precision away from the fiducial edges.
\subsection{Fixed order Results}

\Cref{tab:fixed_order} compares the strict fixed-order
$\mathcal{O}(\alpha)$ prediction of \ACRAIC, defined in \cref{sec:fo_theory},
with the resummed $\mathcal{O}(\alpha)$ and $\mathcal{O}(\alpha^2L)$ ones. The
resummed predictions of \kkmc agree with these to $\lesssim0.07\%$
(\cref{tab:fid_xsec}), so resummation acts the same in both generators, and the
strict fixed order of \kkmc agrees with that of \ACRAIC at \BES{} and \Bscen{}
to $\sim0.1\%$. Resummation changes the cross section by $\sim0.5\%$ at \CMD{},
where the cuts select elastic events, and by $\sim10\%$ at \BES{} and
\Bscen{}, where a hard photon is required. In the tagged-photon scenarios,
including \KLOEI{} and \KLOEII{}, the fixed order is the leading order of
$\pi^+\pi^-\gamma$, and most of this change is its next-to-leading-order
correction (\cref{app:babayaga}). At \BES{} the second order partly
undoes the first, bringing the $\mathcal{O}(\alpha^2L)$ result closer to the
fixed order. The fixed-order $\pi^+\pi^-$ and $\pi^+\pi^-\gamma$ calculations
agree with the exact fixed order of \BabaYaga{} to $\sim0.1\%$.

\begin{table}[htb]
  \centering
  \renewcommand{\arraystretch}{1.3}
  \resizebox{\textwidth}{!}{
    \begin{tabular}{lccccc}
      \hline\hline
      Scenario  & Fixed order $\mathcal{O}(\alpha)$ [nb] & Resummed $\mathcal{O}(\alpha)$ [nb] & Resummed $\mathcal{O}(\alpha^2L)$ [nb] & $\mathcal{O}(\alpha)$/FO & $\mathcal{O}(\alpha^2L)$/FO \\
      \hline
      \CMD{}    & $567.42(12)$                           & $570.041(87)$                       & $570.263(87)$                          & $1.0046$                 & $1.0050$                    \\
      \KLOEI{}  & $2.6604(3)$                & $3.2513(2)$             & $3.2033(2)$                & $1.2221$     & $1.2041$        \\
      \KLOEII{} & $16.456(2)$                & $15.965(1)$             & $16.017(1)$                & $0.9702$     & $0.9733$        \\
      \BES{}    & $0.080115(15)$             & $0.088507(10)$          & $0.083911(9)$              & $1.1048$     & $1.0474$        \\
      \Bscen{}  & $0.010416(2)$              & $0.011157(2)$           & $0.010058(1)$              & $1.0712$     & $0.9656$        \\
      \hline\hline
    \end{tabular}}
  \caption{Fixed-order and resummed $\pi^+\pi^-$ fiducial cross sections of \ACRAIC.}
  \label{tab:fixed_order}
\end{table}

\begin{figure}[htb]
  \centering
  \subfloat[]{\includegraphics[width=0.45\textwidth]{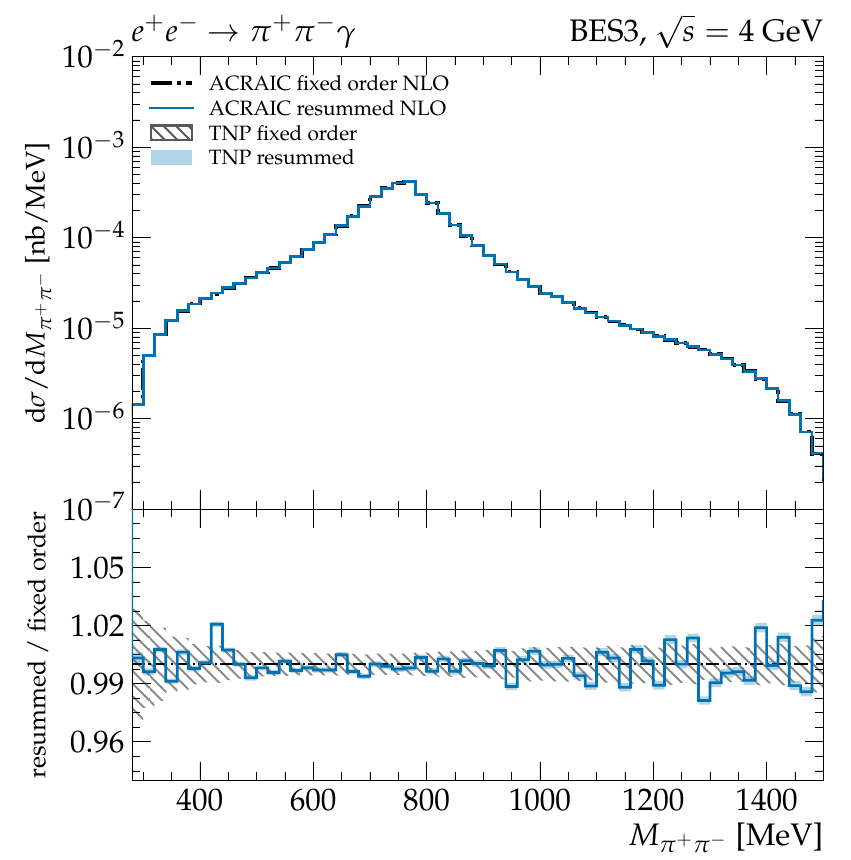}}\hfill
  \subfloat[]{\includegraphics[width=0.45\textwidth]{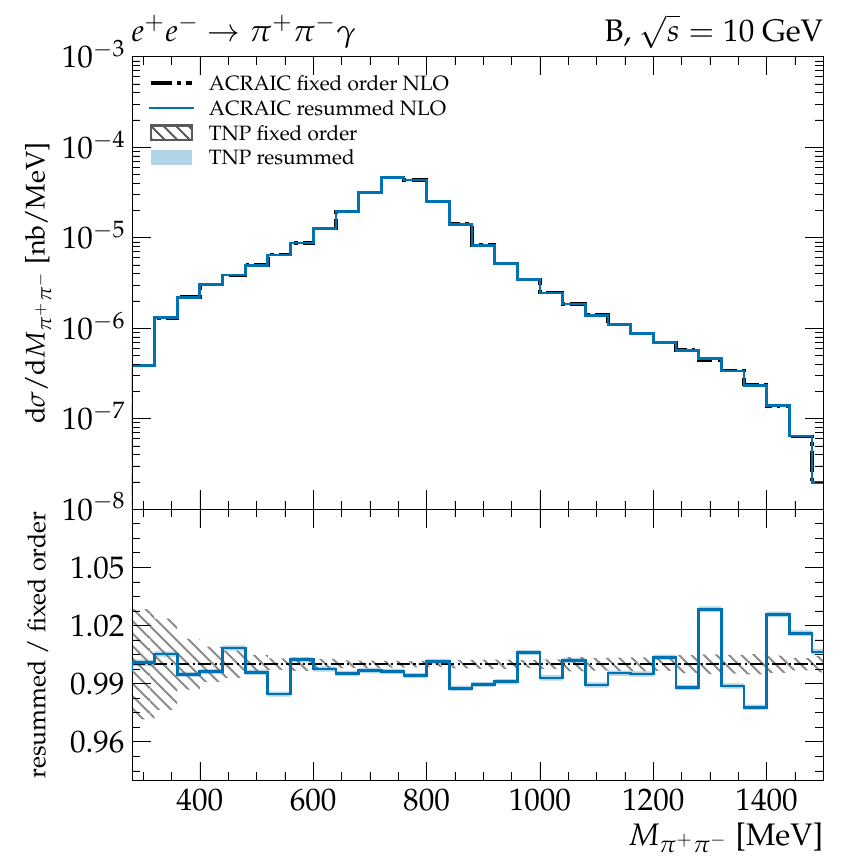}}
  \caption{The pion-pair mass $M_{\pi^+\pi^-}$ in the \BES{} (a) and \Bscen{} (b) fiducial
    regions. The $\pi^+\pi^-\gamma$ prediction of \ACRAIC at next-to-leading
    order at strict fixed order (dash-dotted) and resummed (solid), on the same events, and
    their ratio with the theory-nuisance-parameter bands of the fixed order (hatched) and of
    the resummed prediction (shaded).}
  \label{fig:fo_mxx}
\end{figure}

\begin{figure}[htb]
  \centering
  \subfloat[]{\includegraphics[width=0.45\textwidth]{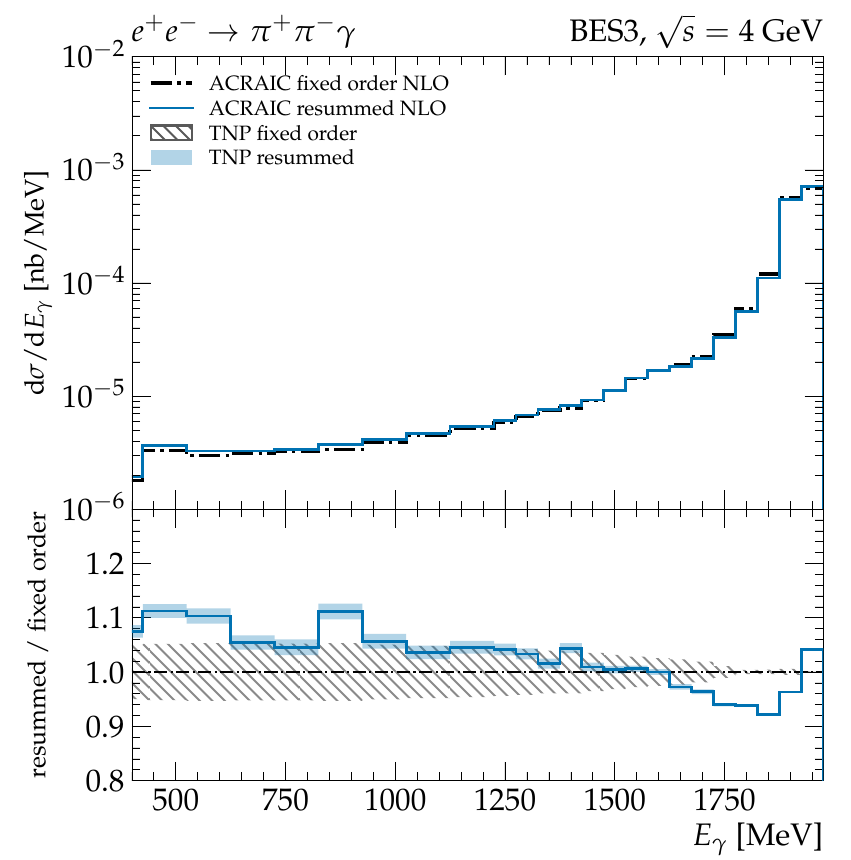}}\hfill
  \subfloat[]{\includegraphics[width=0.45\textwidth]{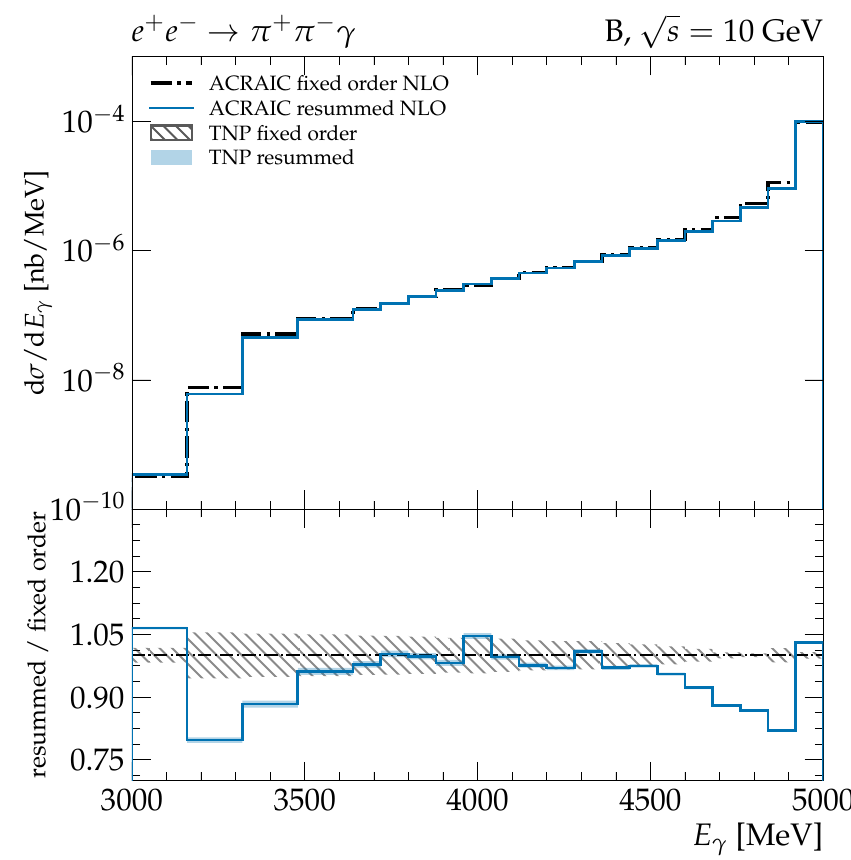}}
  \caption{The energy of the hardest photon $E_\gamma$ in the \BES{} (a) and \Bscen{} (b)
    fiducial regions. Curves and bands as in \cref{fig:fo_mxx}.}
  \label{fig:fo_eg}
\end{figure}

For the $\pi^+\pi^-\gamma$ prediction at next-to-leading order
(\cref{fig:fo_mxx,fig:fo_eg}), resummation changes the pion-pair mass by
$\lesssim1\%$, within the band of the fixed order, and the photon energy by up
to $\sim10\%$. Just below
its upper end the resummed prediction lies below the fixed order, beyond its
band, as the phase space for further radiation closes and its logarithms, which
the fixed order does not contain, grow.

\begin{figure}[htb]
  \centering
  \includegraphics[width=0.45\textwidth]{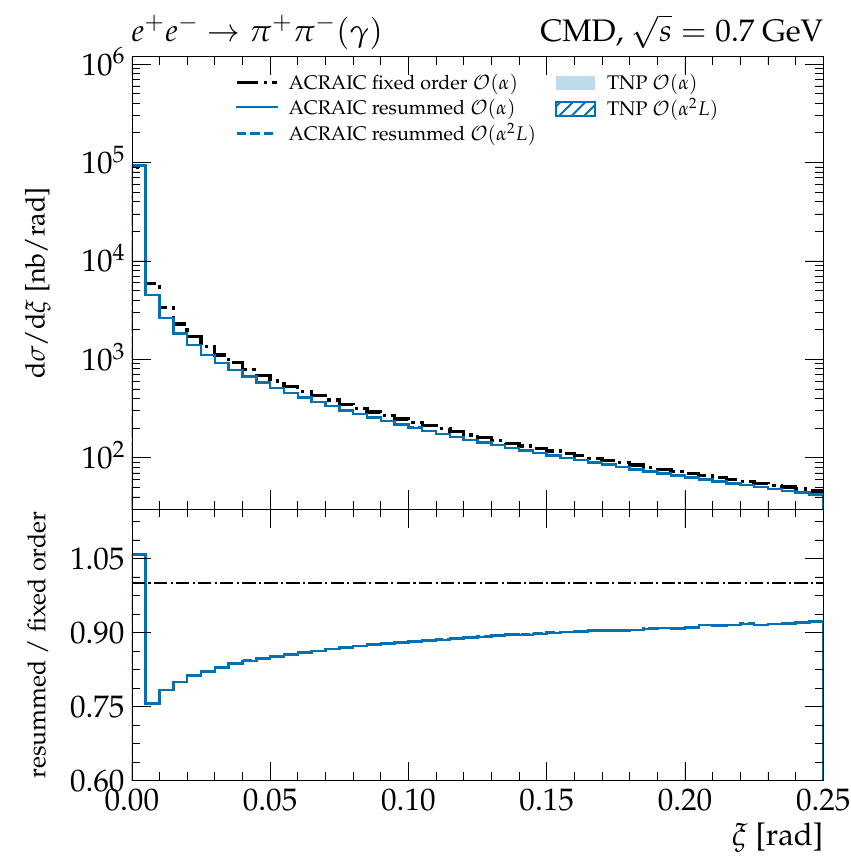}
  \caption{The acollinearity $\xi$ of the pion pair in the \CMD{} fiducial region.
    Curves and bands as in \cref{fig:fo_mxx}. All predictions use the same events
    of an independent \ACRAIC sample.}
  \label{fig:fo_xi}
\end{figure}

At \CMD{} the acollinearity $\xi$ probes the soft limit instead
(\cref{fig:fo_xi}). At fixed order the events without a real photon sit at
$\xi=0$ and one real photon gives the tail, and the exponentiated soft-photon
factor raises the first bin by several per cent and lowers the tail next to it
by $\sim10\%$.

Truncating the resummed result at second order changes it by up to
$\sim0.7\%$ (\cref{tab:fixed_order_tower}), almost all of it the third-order
term, and the remainder beyond third order lies inside the
$\mathcal{O}(\alpha^2L)$ band in every scenario. This tests the higher orders
the resummation contains, while the third-order corrections it does not
contain, such as the hard three-photon matrix element and the three-loop
virtual correction, are estimated by the theory nuisance parameters alone.

\begin{table}[htb]
  \centering
  \renewcommand{\arraystretch}{1.3}
  \resizebox{\textwidth}{!}{
    \begin{tabular}{lccccc}
      \hline\hline
      Scenario  & $\mathcal{O}(\alpha^2)$/resummed & $\mathcal{O}(\alpha^3)$/resummed & Third-order term                & Beyond third order              & TNP band                       \\
      \hline
      \CMD{}    & $1.0001(1)$                      & $1.0000(1)$                      & $-3\times10^{-5}$               & $-2\times10^{-5}$               & $5.4\times10^{-5}$             \\
      \KLOEI{}  & $1.0073(1)$          & $1.0004(1)$          & $-6.9\times10^{-3}$ & $-3.9\times10^{-4}$ & $9.8\times10^{-4}$ \\
      \KLOEII{} & $1.0024(1)$          & $0.9998(1)$          & $-2.6\times10^{-3}$ & $+1.6\times10^{-4}$ & $2.3\times10^{-4}$ \\
      \BES{}    & $0.9981(2)$          & $1.0001(2)$          & $+2.0\times10^{-3}$ & $-1.2\times10^{-4}$ & $4.1\times10^{-3}$ \\
      \Bscen{}  & $1.0022(3)$          & $0.9999(2)$          & $-2.3\times10^{-3}$ & $+0.9\times10^{-4}$ & $8.6\times10^{-3}$ \\
      \hline\hline
    \end{tabular}}
  \caption{The resummed $\mathcal{O}(\alpha^2L)$ \ACRAIC prediction truncated at
    second and third order in $\alpha$, the third-order term and the remainder beyond it, all
    relative to the resummed fiducial cross section on the same events, and the
    $\mathcal{O}(\alpha^2L)$ band of \ACRAIC from \Cref{tab:tnp_breakdown}.}
  \label{tab:fixed_order_tower}
\end{table}
\subsection{Theory-nuisance-parameter breakdown}
\label{sec:tnp_breakdown}
\begin{table}[htb]
  \centering
  \small
  \renewcommand{\arraystretch}{1.25}
  \setlength{\tabcolsep}{4pt}
  \begin{tabular}{ll|cccccc|cccc}
    \hline\hline
              &                           & \multicolumn{6}{c|}{$\mathcal{O}(\alpha)$} & \multicolumn{4}{c}{$\mathcal{O}(\alpha^2L)$}                                                                                                                                            \\
    Scenario  &                           & $\tilde\beta_0$                            & $\tilde\beta_1^{\rm ISR}$                    & $\tilde\beta_1^{\rm FSR}$ & $\tilde\beta_2$  & RR   & total
              & $\tilde\beta_1^{\rm ISR}$ & $\tilde\beta_0^{\rm 2L}$                   & RR                                           & total                                                                                                                                    \\
    \hline
    \CMD{}    & \kkmc                     & 0.52                                       & 0.13                                         & 0.004                     & 0.001            & --   & 0.53             & 0.002 & 0.054             & --               & 0.054            \\
              & \ACRAIC                   & 0.52                                       & 0.09                                         & 0.000                     & 0.002            & --   & 0.52             & 0.001 & 0.053             & --               & 0.054            \\
    \KLOEI{}  & \kkmc                     & 0.74                                       & 7.87                                         & 0.14                      & 4.23             & 24.1 & 25.7             & 0.13  & 0.068 & 1.01             & 1.03  \\
              & \ACRAIC                   & 0.71                                       & 6.40                                         & 0.09                      & 4.09             & 24.1 & 25.3             & 0.11  & 0.065             & 0.97 & 0.98 \\
    \KLOEII{} & \kkmc                     & 0.68                                       & 4.83                                         & 0.005                     & 0.42 & 17.2 & 17.8 & 0.08  & 0.061             & 0.22             & 0.24  \\
              & \ACRAIC                   & 0.67                                       & 4.82                                         & 0.001                     & 0.42             & 17.2 & 17.9             & 0.08  & 0.060 & 0.21 & 0.23 \\
    \BES{}    & \kkmc                     & 1.18                                       & 10.5                             & 0.03                      & 9.90 & 63.7 & 65.3             & 0.17  & 0.065             & 4.11             & 4.12  \\
              & \ACRAIC                   & 1.08                                       & 6.60                                         & 0.05                      & 9.45             & 63.7 & 64.7             & 0.16  & 0.064 & 4.08 & 4.09 \\
    \Bscen{}  & \kkmc                     & 2.01                                       & 29.4                                         & 0.67                      & 4.87 & 82.9 & 88.2             & 0.20  & 0.065             & 8.63             & 8.63  \\
              & \ACRAIC                   & 1.34                                       & 6.88                                         & 0.79                      & 4.71             & 82.9 & 83.4             & 0.18  & 0.063             & 8.59 & 8.59 \\
    \hline\hline
  \end{tabular}
  \caption{Theory-nuisance-parameter contributions
    $N_t|T_t|/\sigma_{\rm fid}$ to the fiducial cross section, in units of
    $10^{-3}$. $\tilde\beta_0$ combines its initial-state, final-state and
    cross parameters in quadrature. At $\mathcal{O}(\alpha)$ the
    $\tilde\beta_2$ column is the corrections missing entirely at this order, at
    $N=1$. RR is
    the radiative-return parameter. At $\mathcal{O}(\alpha^2L)$, $\tilde\beta_0^{\rm 2L}$
    is the parameter of the incomplete second order. The final-state, $\tilde\beta_0$ and
    $\tilde\beta_2$ parameters at $\mathcal{O}(\alpha^2L)$ are below $10^{-4}$
    and omitted from the columns, but included in the totals.}
  \label{tab:tnp_breakdown}
\end{table}

\Cref{tab:tnp_breakdown} breaks the fiducial bands into their
parameters. The templates $T_c$ of \cref{sec:tnp} of a generator sum to twice
its cross section, and the two generators share this total between the
corrections differently in two places. First, the one-photon residual
$\tilde\beta_1$, the one-photon amplitude minus the eikonal factor times the
Born amplitude, depends on where this subtraction is made, at its own reduced
point in \kkmc and at the Born point of the stage in \ACRAIC. The initial-state
template of \kkmc is therefore larger at
\CMD{}, \KLOEI{} and \BES{} and about four times larger at \Bscen{}, with a
correspondingly larger $\tilde\beta_0$ template except at \CMD{}. Second, \kkmc
assigns the photons of a two-photon amplitude to the initial or the final state
partition by partition and \ACRAIC by the emitting line. The sum of the
two-photon templates is the same in both, negative and a few tenths of a per
cent at \KLOEI{}, but its split between the initial-state and the mixed
corrections is not. As independent parameters the three two-photon corrections
would add pieces of opposite sign in quadrature and enlarge the \kkmc band, by
tens of per cent at $M_{\pi^+\pi^-}=420$--$500$~MeV at \CMD{}, so they share
one parameter. The bands of the two generators then agree where initial-state
radiation dominates, at \KLOEII{} and in the \CMD{} elastic peak, and elsewhere
differ by the one-photon split, by at most $\sim10\%$ of the band on the
fiducial cross sections.

The radiative-return parameter RR of \cref{sec:tnp} is needed in the
four radiative-return scenarios, where the eikonal approximation of the hard
photon, all of $\mathcal{O}(\alpha^0)$, overestimates the cross section and the
first order corrects it by $r_1$, a few tens of per cent at \KLOEII{} and
\KLOEI{} and of order $-1$ at \BES{} and \Bscen{}. The $\mathcal{O}(\alpha)$
prediction is then effectively the leading order of
$e^+e^-\to\pi^+\pi^-\gamma$, and its missing higher orders are the full
next-order correction of that process, which the other templates, $\sim1$--$3\%$ at \BES{} and \Bscen{}, would not
cover because their normalisations assume a well-behaved series in $\alpha L$.
As $r_1$ agrees between the generators to four digits, RR does not depend on
how they split the corrections into classes. At $\mathcal{O}(\alpha)$ it
dominates the bands and covers the $\mathcal{O}(\alpha^2L)$ correction in all
four scenarios, with $|\theta_t|\lesssim1$, but not at the low pion-pair masses
of \BES{} and \Bscen{} (\cref{fig:bes3,fig:b}), whose second hard photon the
$\mathcal{O}(\alpha)$ $\pi^+\pi^-$ prediction contains only in the eikonal
approximation. This gap is a leading-order effect of the $\pi^+\pi^-\gamma$
final state rather than a missing higher order that a parameter could
estimate. The $\pi^+\pi^-\gamma$
calculation of \ACRAIC, which has this final state at Born level, shows no such gap,
and in the pion-pair mass the matched photons change it by $\sim0.1\%$. At \CMD{} $r_1\sim+1\%$ is the soft and virtual
correction of the elastic events, which the $\tilde\beta_0$ parameters already
cover, and RR is not used.

The band of the explicit-photon $\pi^+\pi^-\gamma$ prediction (lower
panels of \cref{fig:bes3_pipig,fig:b_pipig_a}) takes as RR parameter $\gamma$
times the first-order correction $r_1^{\pi\pi\gamma}$ of its leading order, the
tree-level $\pi^+\pi^-\gamma$ process. On the fiducial cross section
$r_1^{\pi\pi\gamma}$ is $\sim+5\%$ at \BES{} and $\sim-3\%$ at \Bscen{}
(\cref{app:babayaga}), so the RR band is a few tenths of a per cent in both,
with the other parameters far smaller. Per bin $r_1^{\pi\pi\gamma}$ varies more
and the bands follow it. Being proportional to the net correction, RR
underestimates the missing higher orders wherever the real and virtual
corrections cancel.
\FloatBarrier

\FloatBarrier
\section{Conclusion}\label{sec:conclusion}

We have presented QED radiative corrections to
$e^{+}e^{-}\rightarrow\pi^{+}\pi^{-}(\gamma)$ with amplitude-level YFS
resummation for the first time, implemented independently in \Sherpa (via
\ACRAIC) and \kkmc (via CEEX). Pions are point-like \sQED{} scalars dressed with an
electromagnetic form factor, either factorised (\FxsQED{}) or via \GVMD{} in the loops. Both
generators match the YFS resummation to the complete first order, including
exact one- and two-photon tree-level matrix elements and the initial-final
boxes and interference, and to the logarithmically enhanced second order
through the two-loop photonic vertices and the exact one-loop correction to
one-photon emission. Theory-nuisance-parameter bands estimate the
missing higher orders, extending Ref.~\cite{Price:2026qop} with a
radiative-return parameter for scenarios in which the first-order correction
is of order one and with parameters for the omitted parts of the second order.
Across the five STRONG2020 scenarios (\CMD{}, \KLOEI{}, \KLOEII{}, \BES{} and
\Bscen{}) the first-order fiducial cross sections of \Sherpa and \kkmc agree
to $\lesssim0.03\%$ and the differential distributions within the statistical
uncertainties. At $\mathcal{O}(\alpha^2L)$ the bands on the fiducial cross
sections are $\sim0.005\%$ for the elastic \CMD{} scenario, $\sim0.1\%$ at
\KLOEI{} and $\sim0.02\%$ at \KLOEII{}, and in the tagged-photon scenarios
$\sim0.4\%$ at \BES{} and $\sim0.9\%$ at \Bscen{}, dominated by the
radiative-return parameter. The \GVMD{} form factor inside the loops, instead of
point-like loops, reduces the forward-backward asymmetry
for \CMDthree{} kinematics by up to $\sim1\%$ near the $\rho$ resonance.

Against \BabaYaga the $\mathcal{O}(\alpha^2L)$ fiducial cross
sections of \CMD{}, \KLOEI{} and \KLOEII{} agree within $0.01\%$ to $0.1\%$. In
the polar angles of the pions at \KLOEI{}, where the next-to-leading-order
calculations of \BabaYaga and \Phokhara differ by a slope of a few per
cent~\cite{Budassi:2026lmr}, both generators follow the shape of \BabaYaga to a
few tenths of a per cent and so support it.

The next steps are the two-loop boxes, the genuine two-loop
scalar-QED pion vertex and the $\mathcal{O}(\alpha^3)$ photonic corrections,
sized by the theory nuisance parameters. The boxes and the two-loop
pion vertex are non-logarithmic at second order, and the
$\tilde\beta_0^{\rm 2L}$ parameter sizes them at $\sim0.005\%$, relevant only
for the elastic \CMD{} scenario. The $\mathcal{O}(\alpha^3)$ photonic
corrections are sized at $\sim0.02\%$ by the one-photon and $\tilde\beta_0$
parameters, and by the radiative-return parameter in the radiative-return
scenarios. The limiting factor is the calculation of the amplitudes, and once they
are available the implementation in both generators is straightforward as
the YFS subtraction scheme is defined to all orders.

\section*{Acknowledgements}
We would like to thank Janusz Gluza for discussions and remarks.
We are grateful to Fedor Ignatov for providing the \CMDthree{} data and for useful discussions.
We are also grateful to the STRONG2020 working group for providing the scenarios and for useful discussions.
The work is supported in part by grant No. 2023/50/A/ST2/00224 of the National Science Centre (NCN), Poland.
A.P.\ gratefully acknowledges the Polish high-performance computing infrastructure PLGrid (HPC Centers: CI TASK, ACK Cyfronet AGH) for providing computer facilities and support within computational grant no. PLG/2024/017287.
F.K.'s work has been supported by STFC under grant agreement ST/P006744/1.
\FloatBarrier
\appendix
\section{Comparison with \BabaYaga}
\label{app:babayaga}

We compare the results of \cref{sec:results} with
\BabaYaga~\cite{Balossini:2006wc,Balossini:2008xr,Budassi:2024whw}, which
matches the exact $\mathcal{O}(\alpha)$ corrections to a leading-logarithmic
QED parton shower (NLOPS), using the RadioMonteCarLow~2
predictions~\cite{radiomontecarlow2-results} with the
same STRONG2020 pion form factor and without vacuum polarisation (NLOPSnovp).
A full comparison is left to the next STRONG2020 study.

\begin{figure}[htb]
  \centering
  \subfloat[]{\includegraphics[width=0.4\textwidth]{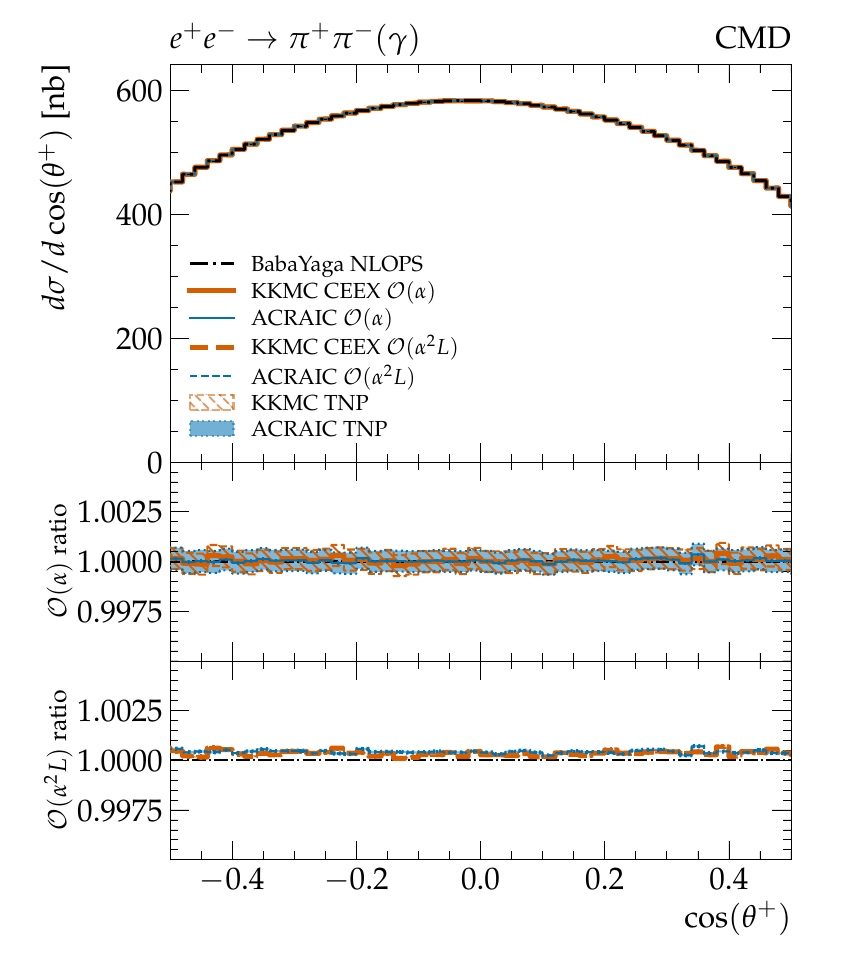}}
  \subfloat[]{\includegraphics[width=0.4\textwidth]{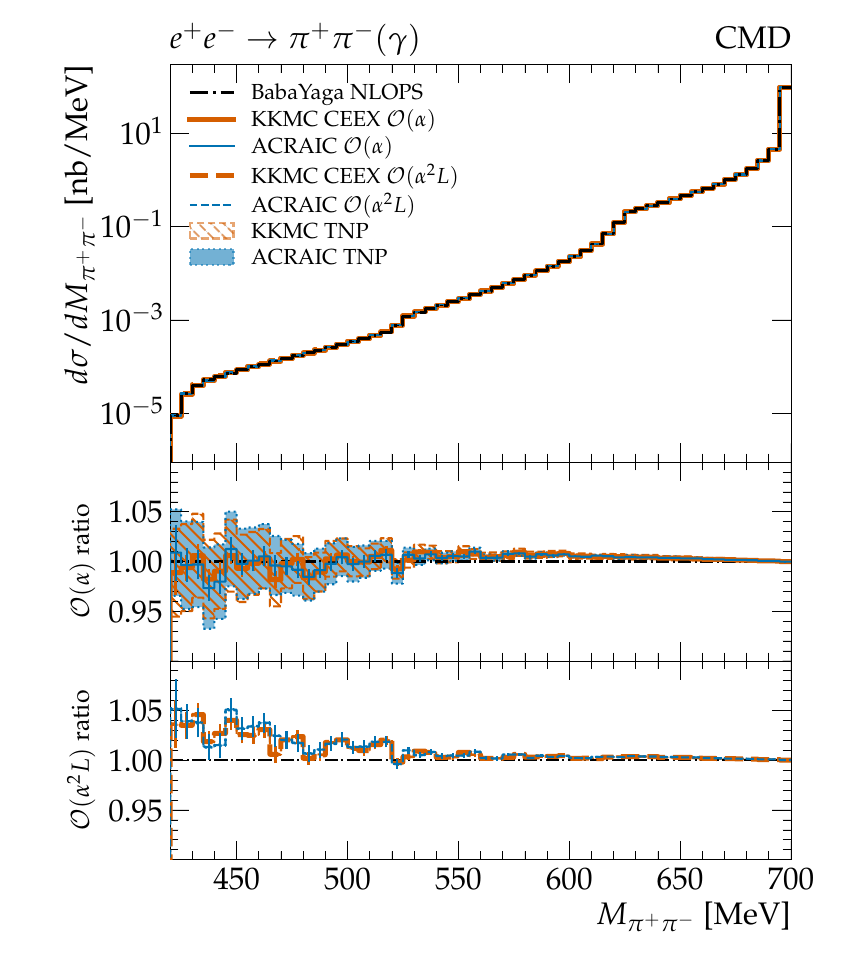}}
  \caption{\CMD{} scenario: $\cos\theta^+$ (a) and the pion-pair mass $M_{\pi^+\pi^-}$ (b). \BabaYaga
    (NLOPS, black, dash-dotted), \kkmc CEEX (thick orange) and \ACRAIC
    (thin blue) at $\mathcal{O}(\alpha)$ (solid) and $\mathcal{O}(\alpha^2L)$ (dashed). The middle and lower panels show the
      $\mathcal{O}(\alpha)$ and $\mathcal{O}(\alpha^2L)$ ratios to
      \BabaYaga, each with the same-order theory-nuisance-parameter bands of \kkmc (diagonal
      hatching, dashed edges) and \ACRAIC (shaded, dotted edges).}
  \label{fig:bby_angles}
\end{figure}

\paragraph{\CMD{}.}
The fiducial cross sections of \kkmc CEEX and \ACRAIC agree with \BabaYaga to
$\lesssim0.01\%$ at $\mathcal{O}(\alpha)$ and lie $\sim0.04\%$ above it at $\mathcal{O}(\alpha^2L)$,
the size of the photonic two-loop vertex corrections absent from the NLOPS prediction.
The angular distributions $\cos\theta^\pm$ and $\theta_{\rm avg}$ agree with \BabaYaga to
$\sim0.01\%$ in both generators (\cref{fig:bby_angles}a).

For the pion-pair mass (\cref{fig:bby_angles}b) we use version~2.1 of
\BabaYaga~\cite{Budassi:2026lmr} with the \CMD{} selection, which includes the
exact two-photon tree amplitude also in the $\pi^+\pi^-$ predictions.
Between $600$ and $690$~MeV, where the events carry one hard photon,
both generators lie a few tenths of a per cent above \BabaYaga at both orders,
beyond their $\mathcal{O}(\alpha)$ bands. As they agree with each other, the
difference lies between the YFS treatment and the parton shower. Below $520$~MeV, where every event carries two hard photons, they
agree with \BabaYaga at $\mathcal{O}(\alpha)$ and lie $\sim2\%$ above it at
$\mathcal{O}(\alpha^2L)$.

\begin{figure}[htb]
  \centering
  \subfloat[]{\includegraphics[width=0.4\textwidth]{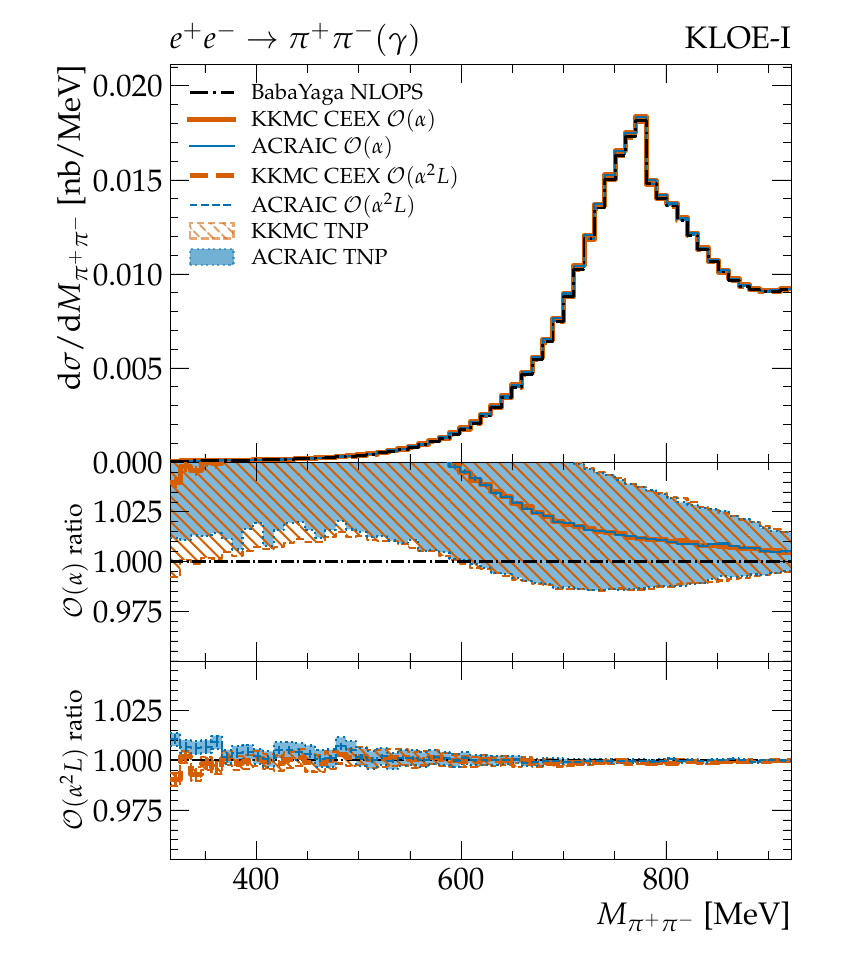}}
  \subfloat[]{\includegraphics[width=0.4\textwidth]{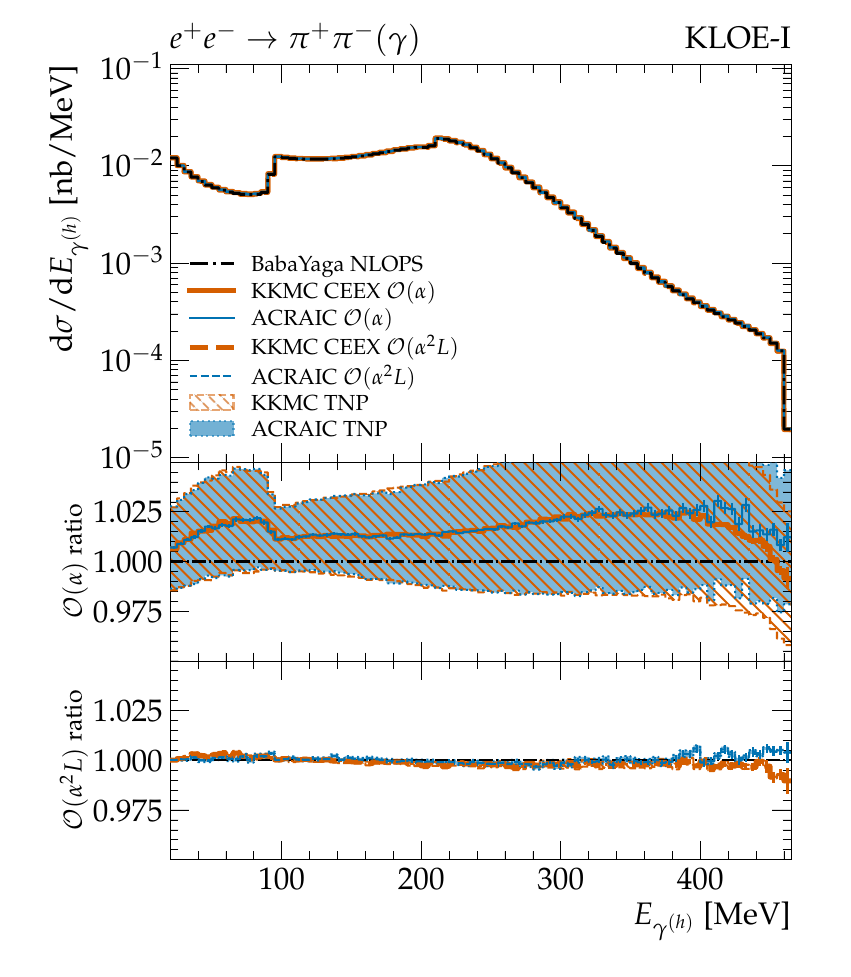}}
  \caption{\KLOEI{} scenario: the pion-pair mass $M_{\pi^+\pi^-}$ (a) and the
    energy of the hardest tagged photon (b). Curves and bands as in \Cref{fig:bby_angles}.}
  \label{fig:bby_kloeI}
\end{figure}

\paragraph{\KLOEI{}.}
With a tagged hard photon in every event the higher orders matter more.
Both $\mathcal{O}(\alpha)$ predictions lie $\sim1\%$ above
\BabaYaga, and \ACRAIC and \kkmc at $\mathcal{O}(\alpha^2L)$
$\sim0.02\%$ and $\sim0.09\%$ below it (\cref{fig:bby_kloeI}). The
$\mathcal{O}(\alpha)$ excess, flat in the photon and pion angles, grows with
the photon energy from $\sim1\%$ at $M_{\pi^+\pi^-}>700$~MeV to several per
cent below $500$~MeV. It lies inside the $\mathcal{O}(\alpha)$ band of a few per cent set by the
radiative-return parameter of~\cref{sec:tnp_breakdown}, and only slightly
outside it below $500$~MeV.
It is therefore dominated by the $\mathcal{O}(\alpha^2)$ corrections to
hard-photon emission, which \BabaYaga includes at leading-logarithmic accuracy
through its parton shower and \kkmc through $\tilde\beta_2$ and its
real-virtual correction. At $\mathcal{O}(\alpha^2L)$ \kkmc and \ACRAIC agree
with \BabaYaga to $\sim0.1\%$ across $M_{\pi^+\pi^-}$ and the photon energy,
and with each other to $\sim0.07\%$ on the fiducial cross section.

\begin{figure}[htb]
  \centering
  \subfloat[]{\includegraphics[width=0.4\textwidth]{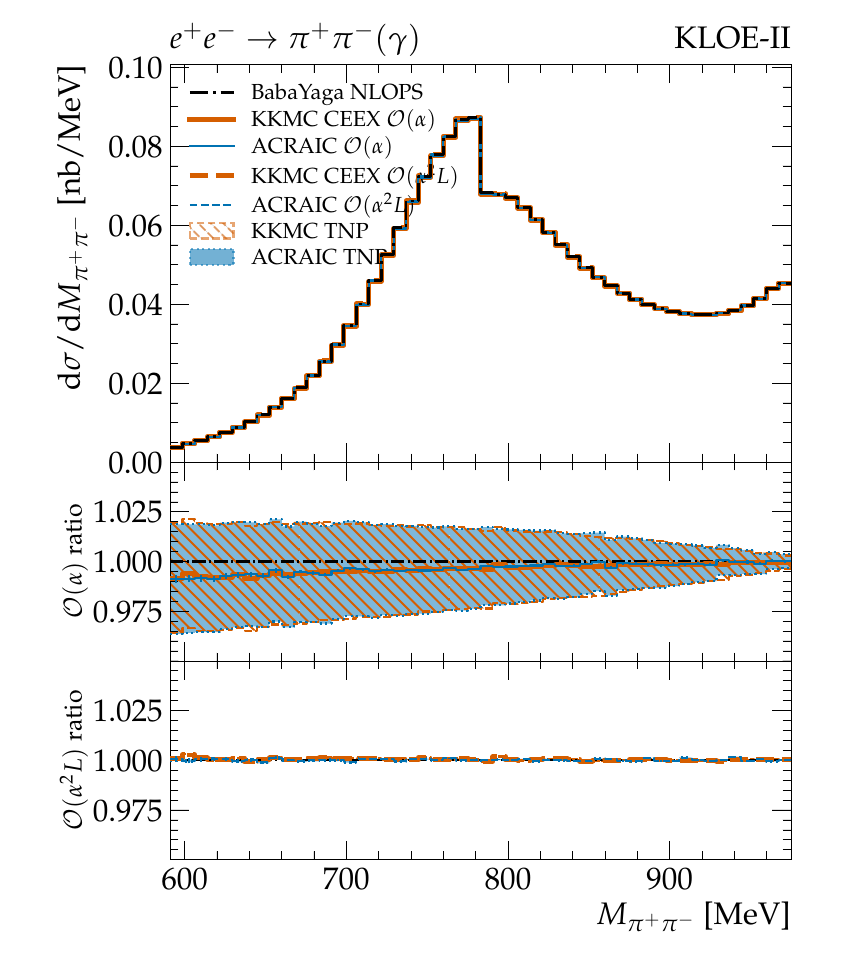}}
  \subfloat[]{\includegraphics[width=0.4\textwidth]{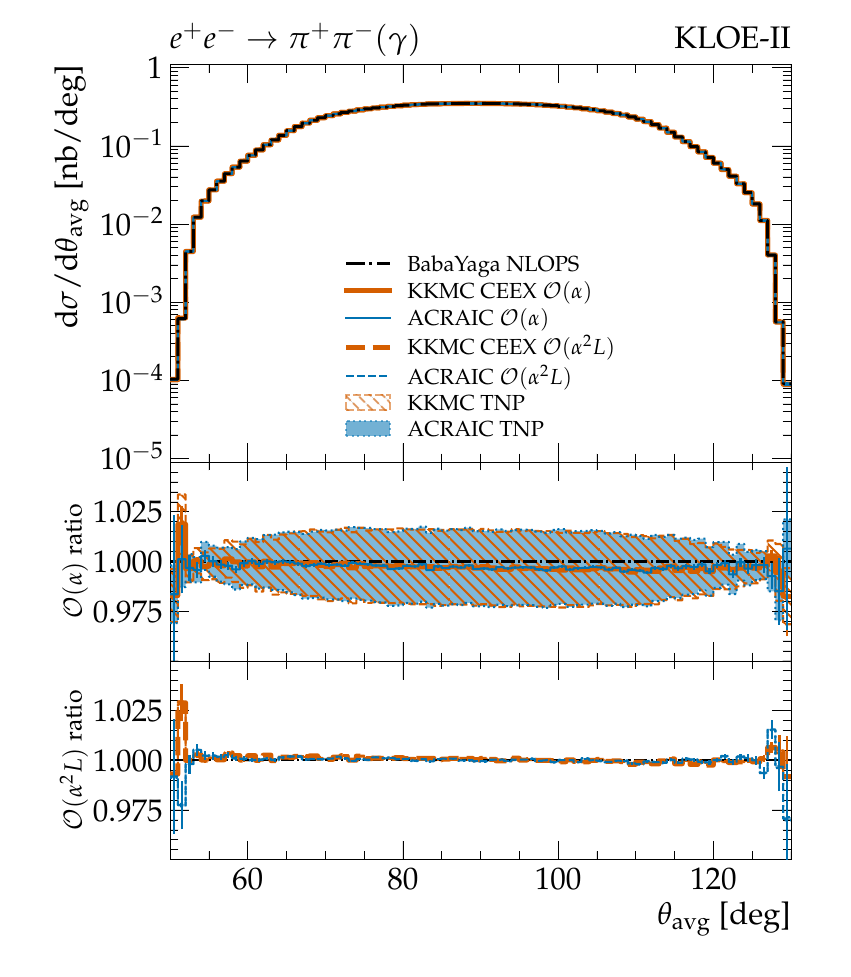}}
  \caption{\KLOEII{} scenario: the pion-pair mass $M_{\pi^+\pi^-}$ (a) and the
    average pion angle (b). Curves and bands as in \Cref{fig:bby_angles}.}
  \label{fig:bby_kloeII}
\end{figure}

\paragraph{\KLOEII{}.}
With the photon along a beam the corrections are smaller. The
$\mathcal{O}(\alpha)$ predictions lie $\sim0.3\%$ below \BabaYaga, inside the
$\mathcal{O}(\alpha)$ band, and \kkmc and \ACRAIC at
$\mathcal{O}(\alpha^2L)$ are $\sim0.03\%$ above (\cref{fig:bby_kloeII}).
The $\mathcal{O}(\alpha)$ deficit is largest, $\sim1\%$, below
$700$~MeV, where the $\mathcal{O}(\alpha^2L)$ predictions are $\sim0.1\%$
above \BabaYaga. The missing-momentum angle behaves alike near both beams.

\begin{figure}[htb]
  \centering
  \subfloat[]{\includegraphics[width=0.4\textwidth]{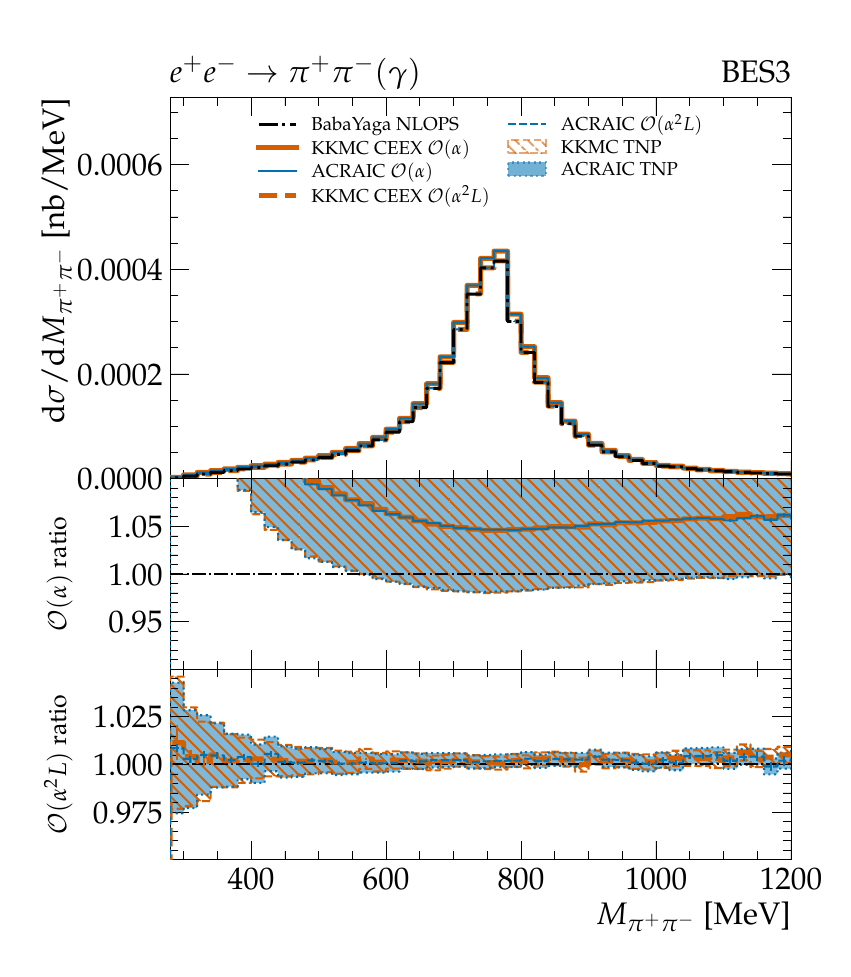}}
  \subfloat[]{\includegraphics[width=0.4\textwidth]{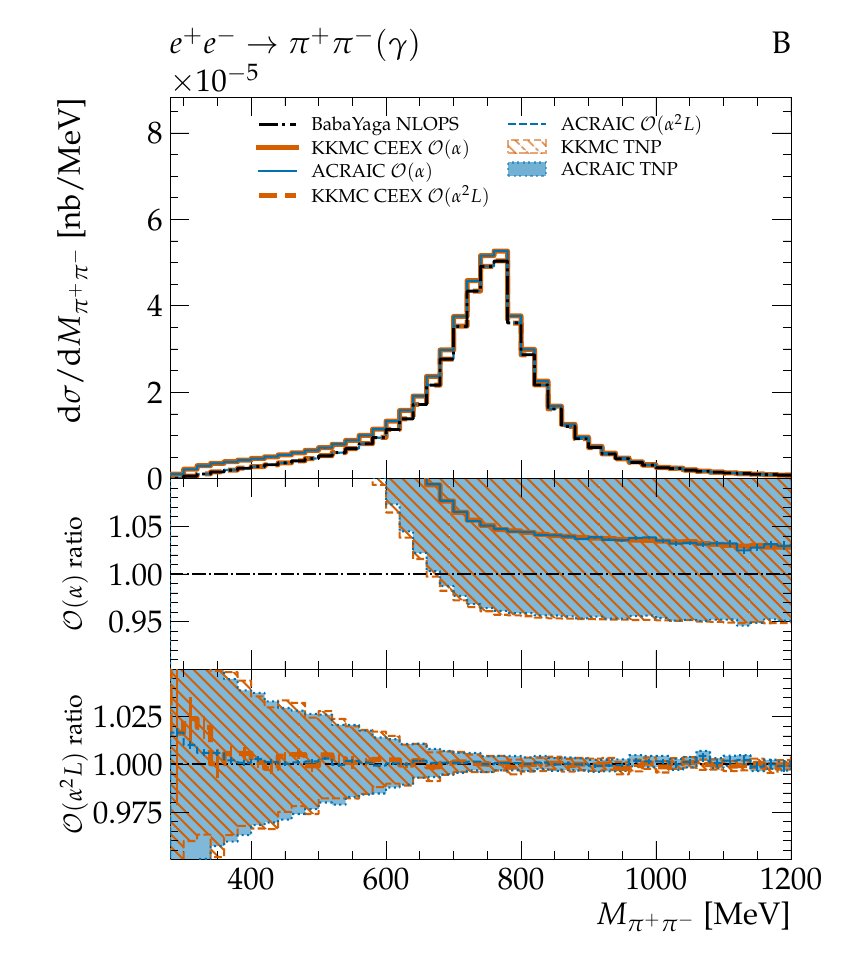}}
  \caption{The pion-pair mass $M_{\pi^+\pi^-}$ in the \BES{} (a) and \Bscen{} (b)
    scenarios. Curves and bands as in \Cref{fig:bby_angles}, with the
    next-to-leading-order $\pi^+\pi^-\gamma$ prediction of \ACRAIC (solid, purple).}
  \label{fig:bby_bes3_b}
\end{figure}

\begin{figure}[htb]
  \centering
  \subfloat[]{\includegraphics[width=0.45\textwidth]{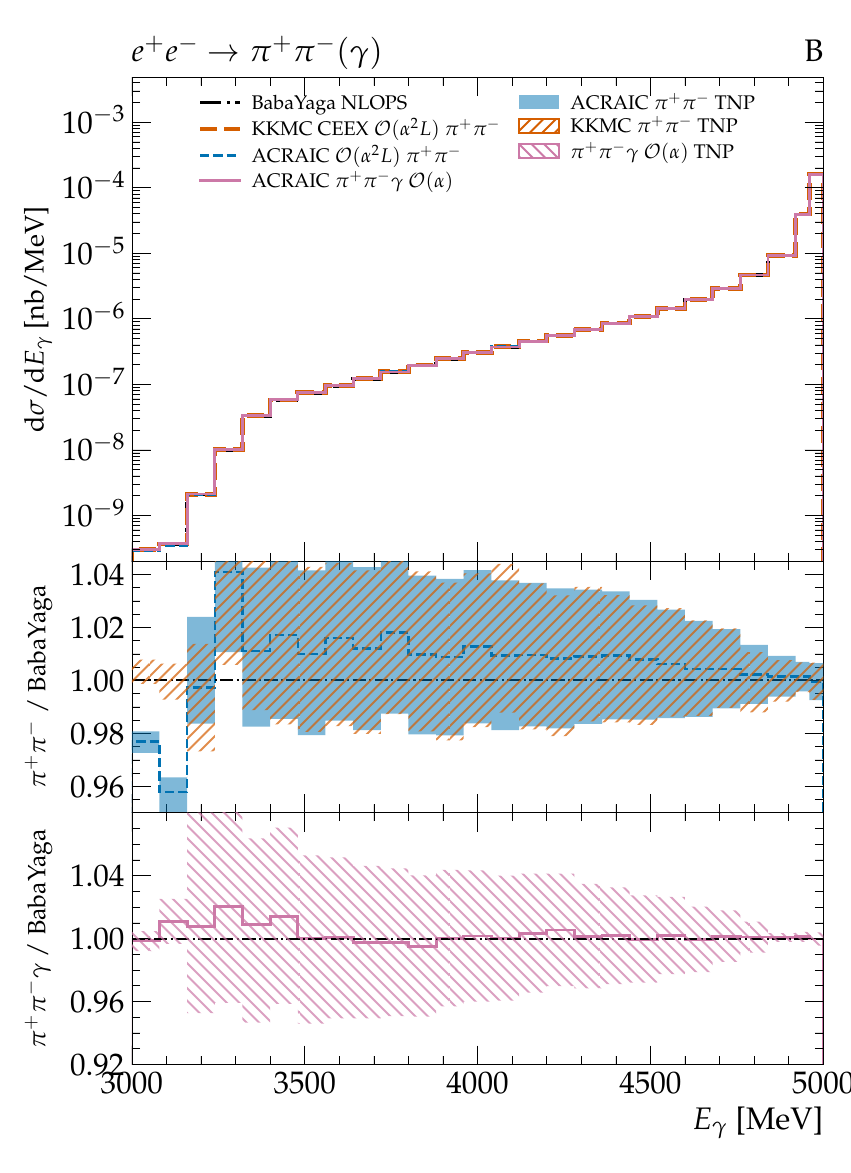}}\hfill
  \subfloat[]{\includegraphics[width=0.45\textwidth]{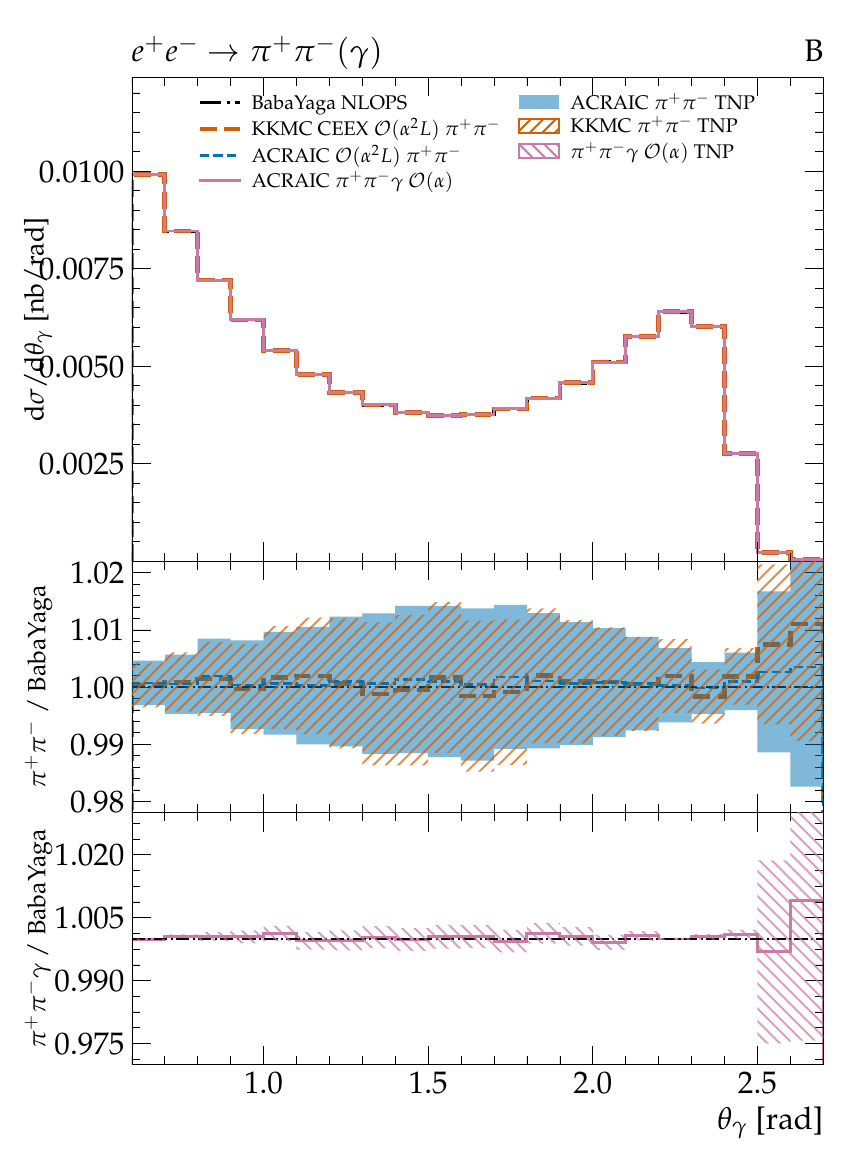}}\\
  \subfloat[]{\includegraphics[width=0.45\textwidth]{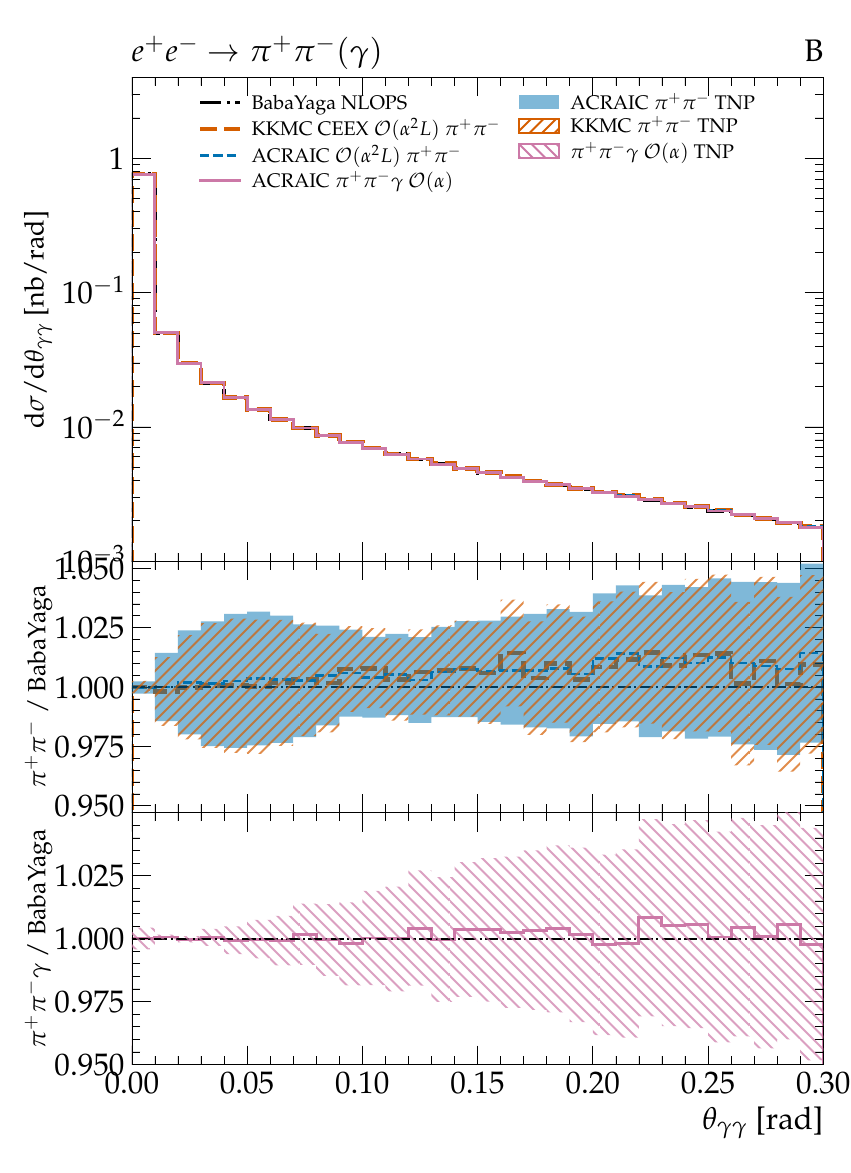}}
  \caption{The energy (a) and the polar angle (b) of the hardest photon, and its
      angle to the direction opposite the pion pair (c), at \Bscen{}, as in
      \Cref{fig:b_pipig_b} but as ratios to \BabaYaga.}
  \label{fig:bby_b_photon}
\end{figure}

\paragraph{\BES{} and \Bscen{}.} In the pion-pair mass
(\cref{fig:bby_bes3_b}) the $\pi^+\pi^-\gamma$ prediction of \ACRAIC agrees
with \BabaYaga to $\sim0.1\%$ over the whole range. At $\mathcal{O}(\alpha^2L)$
the $\pi^+\pi^-$ predictions of both generators agree with it to a few tenths
of a per cent, and to $\lesssim1\%$ in the lowest-mass bins at \Bscen{}. At $\mathcal{O}(\alpha)$ they lie far above it,
outside their bands, below $M_{\pi^+\pi^-}\sim600$~MeV and the photon-energy
peak, by tens of per cent at \BES{} and up to several times at
\Bscen{}. These events need a second hard photon, present in the parton shower and at
$\mathcal{O}(\alpha^2L)$ but only eikonally at $\mathcal{O}(\alpha)$
(\cref{sec:tnp_breakdown}). In
the photon spectra at \Bscen{} (\cref{fig:bby_b_photon}) the
$\pi^+\pi^-\gamma$ prediction agrees with \BabaYaga to $\sim0.1\%$ in both
angles and in the energy above the lowest bins. The $\pi^+\pi^-$ predictions
agree to $\sim0.1\%$ in the polar angle and lie up to $\sim1\%$ above below
the photon-energy peak and at larger angles to the direction opposite
the pion pair, within their bands.

\subsection*{Fixed order}
\begin{figure}[htb]
  \centering
  \subfloat[]{\includegraphics[width=0.45\textwidth]{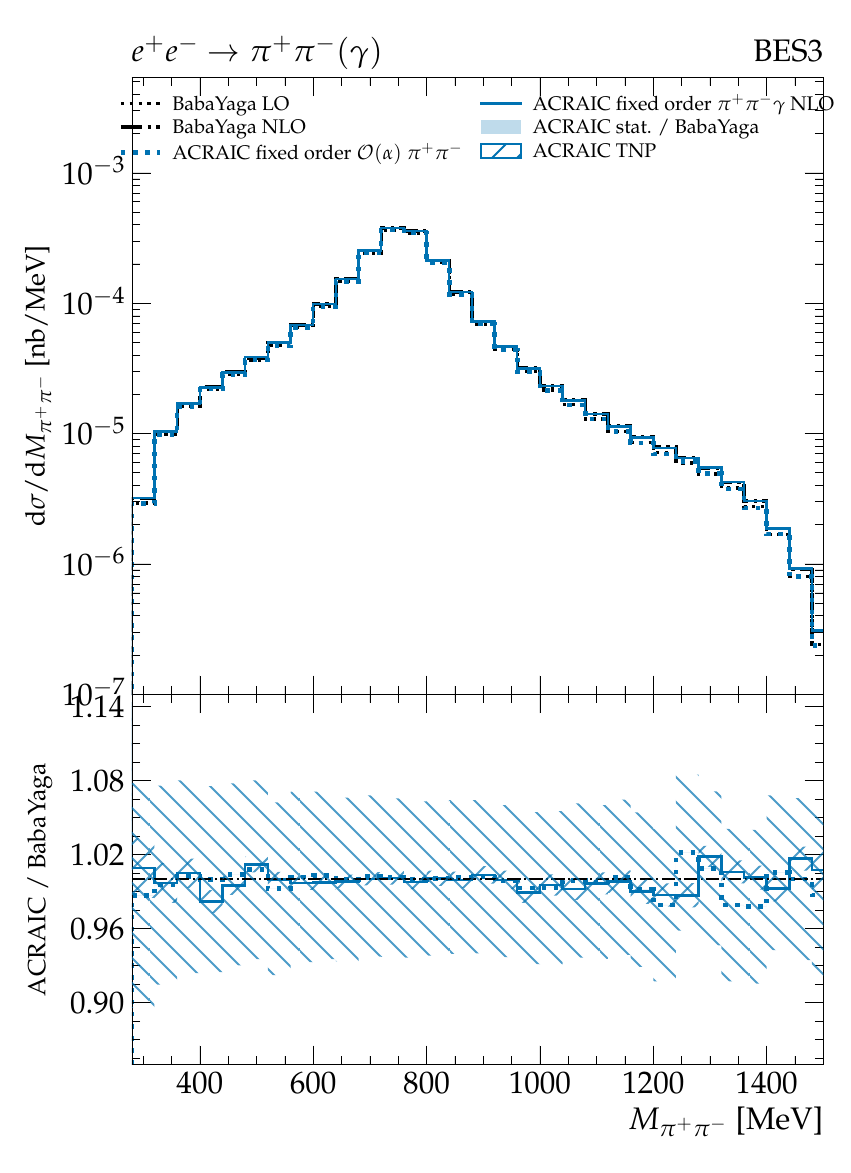}}\hfill
  \subfloat[]{\includegraphics[width=0.45\textwidth]{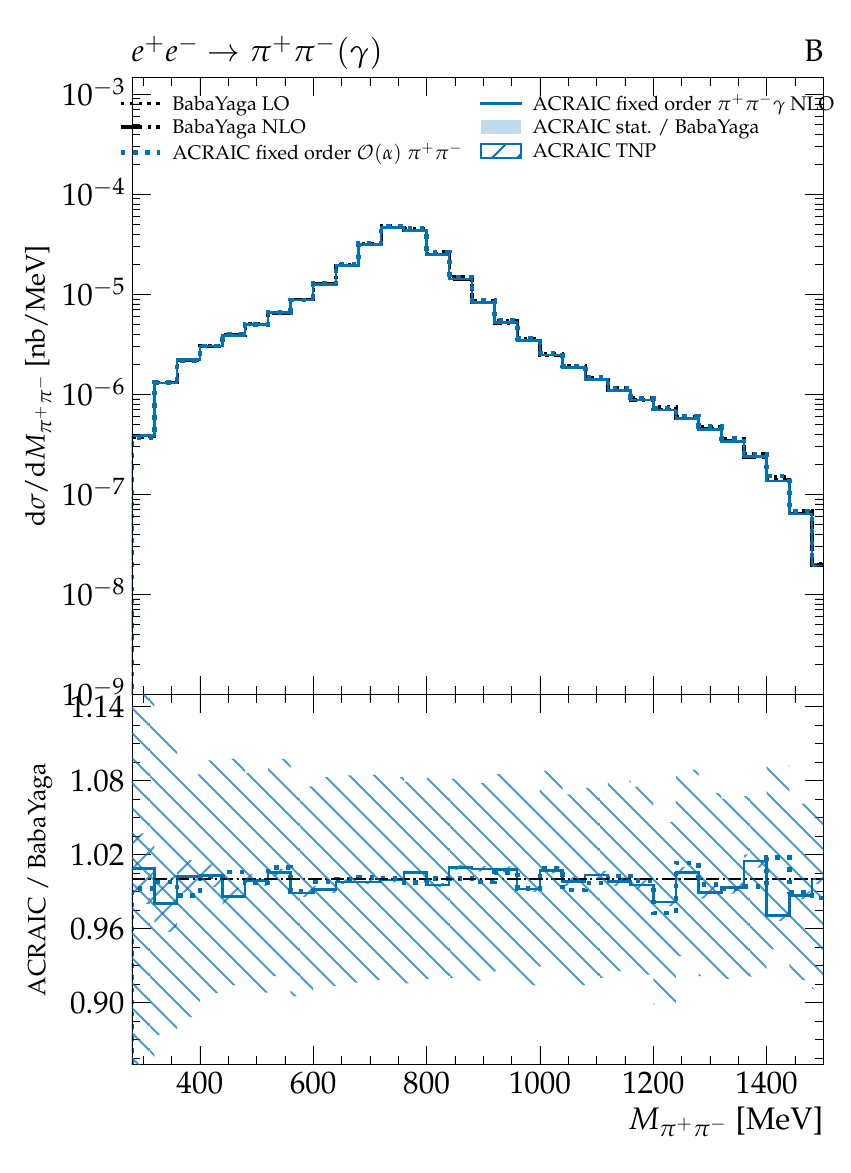}}
  \caption{The pion-pair mass $M_{\pi^+\pi^-}$ in the \BES{} (a) and \Bscen{} (b) fiducial regions at
    fixed order. The upper panels show the exact LO (thin, dotted) and NLO (dash-dotted) of \BabaYaga
    and the fixed-order $\pi^+\pi^-$ (thick, dotted) and $\pi^+\pi^-\gamma$ (solid) calculations of
    \ACRAIC. The lower panels show the ratios to \BabaYaga with the
    statistical uncertainty of \ACRAIC (shaded) and the fixed-order theory-nuisance-parameter bands
    (hatched, wide at leading and narrow at next-to-leading order), which size the missing next order.}
  \label{fig:fo_bby}
\end{figure}

\BabaYaga also provides the exact $\mathcal{O}(\alpha)$ corrections
without resummation. The strict fixed order of \ACRAIC is the next-to-leading
order of $\pi^+\pi^-$ at \CMD{}, where it agrees with \BabaYaga within the
$\sim0.01\%$ statistical precision, and in the tagged-photon scenarios the
$\pi^+\pi^-$ and $\pi^+\pi^-\gamma$ calculations at $\mathcal{O}(\alpha)$ for
the leading and next-to-leading order (\cref{fig:fo_bby}). There the cross
sections agree to $\sim0.1\%$, within the statistical precision of \ACRAIC,
except at leading order at \KLOEI{}, where \ACRAIC lies a few tenths of a per
cent, several standard deviations, above \BabaYaga for reasons not yet
identified. The distributions agree to a few per cent per bin away from the
tails.

\clearpage
\bibliographystyle{JHEP}
\bibliography{main}

\providecommand{\href}[2]{#2}\begingroup\raggedright\begin{thebibliography}{100}

\bibitem{Aoyama:2020ynm}
T.~Aoyama et~al., \emph{{The anomalous magnetic moment of the muon in the
  Standard Model}},
  \href{https://doi.org/10.1016/j.physrep.2020.07.006}{\emph{Phys. Rept.}
  {\bfseries 887} (2020) 1} [\href{https://arxiv.org/abs/2006.04822}{{\ttfamily
  2006.04822}}].

\bibitem{Muong-2:2021ojo}
{\scshape Muon g-2} collaboration, \emph{{Measurement of the Positive Muon
  Anomalous Magnetic Moment to 0.46 ppm}},
  \href{https://doi.org/10.1103/PhysRevLett.126.141801}{\emph{Phys. Rev. Lett.}
  {\bfseries 126} (2021) 141801}
  [\href{https://arxiv.org/abs/2104.03281}{{\ttfamily 2104.03281}}].

\bibitem{Muong-2:2023cdq}
{\scshape Muon g-2} collaboration, \emph{{Measurement of the Positive Muon
  Anomalous Magnetic Moment to 0.20~ppm}},
  \href{https://doi.org/10.1103/PhysRevLett.131.161802}{\emph{Phys. Rev. Lett.}
  {\bfseries 131} (2023) 161802}
  [\href{https://arxiv.org/abs/2308.06230}{{\ttfamily 2308.06230}}].

\bibitem{Muong-2:2025xyk}
{\scshape Muon g-2} collaboration, \emph{{Measurement of the Positive Muon
  Anomalous Magnetic Moment to 127~ppb}},
  \href{https://doi.org/10.1103/7clf-sm2v}{\emph{Phys. Rev. Lett.} {\bfseries
  135} (2025) 101802} [\href{https://arxiv.org/abs/2506.03069}{{\ttfamily
  2506.03069}}].

\bibitem{Jegerlehner:2009ry}
F.~Jegerlehner and A.~Nyffeler, \emph{{The Muon g-2}},
  \href{https://doi.org/10.1016/j.physrep.2009.04.003}{\emph{Phys. Rept.}
  {\bfseries 477} (2009) 1} [\href{https://arxiv.org/abs/0902.3360}{{\ttfamily
  0902.3360}}].

\bibitem{Athron:2025ets}
P.~Athron, K.~M{\"o}hling, D.~St{\"o}ckinger and H.~St{\"o}ckinger-Kim,
  \emph{{The muon magnetic moment and physics beyond the standard model}},
  \href{https://doi.org/10.1016/j.ppnp.2025.104225}{\emph{Prog. Part. Nucl.
  Phys.} {\bfseries 148} (2026) 104225}
  [\href{https://arxiv.org/abs/2507.09289}{{\ttfamily 2507.09289}}].

\bibitem{Muong-2:2015xgu}
{\scshape Muon g-2} collaboration, \emph{{Muon (g-2) Technical Design Report}},
   \href{https://arxiv.org/abs/1501.06858}{{\ttfamily 1501.06858}}.

\bibitem{Muong-2:2006rrc}
{\scshape Muon g-2} collaboration, \emph{{Final Report of the Muon E821
  Anomalous Magnetic Moment Measurement at BNL}},
  \href{https://doi.org/10.1103/PhysRevD.73.072003}{\emph{Phys. Rev. D}
  {\bfseries 73} (2006) 072003}
  [\href{https://arxiv.org/abs/hep-ex/0602035}{{\ttfamily hep-ex/0602035}}].

\bibitem{Muong-2:2026qnz}
{\scshape Muon g{\ensuremath{-}}2} collaboration, \emph{{Final Report on the
  Measurement of the Positive Muon Anomalous Magnetic Moment at Fermilab to 127
  ppb}},  \href{https://arxiv.org/abs/2606.17323}{{\ttfamily 2606.17323}}.

\bibitem{Aliberti:2025beg}
R.~Aliberti et~al., \emph{{The anomalous magnetic moment of the muon in the
  Standard Model: an update}},
  \href{https://doi.org/10.1016/j.physrep.2025.08.002}{\emph{Phys. Rept.}
  {\bfseries 1143} (2025) 1}
  [\href{https://arxiv.org/abs/2505.21476}{{\ttfamily 2505.21476}}].

\bibitem{Borsanyi:2020mff}
S.~Borsanyi et~al., \emph{{Leading hadronic contribution to the muon magnetic
  moment from lattice QCD}},
  \href{https://doi.org/10.1038/s41586-021-03418-1}{\emph{Nature} {\bfseries
  593} (2021) 51} [\href{https://arxiv.org/abs/2002.12347}{{\ttfamily
  2002.12347}}].

\bibitem{Boccaletti:2024guq}
A.~Boccaletti et~al., \emph{{Hybrid calculation of hadronic vacuum polarization
  in muon g {\ensuremath{-}} 2 to 0.48{\%}}},
  \href{https://doi.org/10.1038/s41586-026-10449-z}{\emph{Nature} {\bfseries
  653} (2026) 373} [\href{https://arxiv.org/abs/2407.10913}{{\ttfamily
  2407.10913}}].

\bibitem{Djukanovic:2024cmq}
D.~Djukanovic, G.~von Hippel, S.~Kuberski, H.B.~Meyer, N.~Miller, K.~Ottnad
  et~al., \emph{{The hadronic vacuum polarization contribution to the muon g
  {\ensuremath{-}} 2 at long distances}},
  \href{https://doi.org/10.1007/JHEP04(2025)098}{\emph{JHEP} {\bfseries 04}
  (2025) 098} [\href{https://arxiv.org/abs/2411.07969}{{\ttfamily
  2411.07969}}].

\bibitem{RBC:2024fic}
{\scshape RBC, UKQCD} collaboration, \emph{{Long-Distance Window of the
  Hadronic Vacuum Polarization for the Muon g-2}},
  \href{https://doi.org/10.1103/PhysRevLett.134.201901}{\emph{Phys. Rev. Lett.}
  {\bfseries 134} (2025) 201901}
  [\href{https://arxiv.org/abs/2410.20590}{{\ttfamily 2410.20590}}].

\bibitem{FermilabLatticeHPQCD:2024ppc}
{\scshape Fermilab Lattice, HPQCD,, MILC} collaboration, \emph{{Hadronic Vacuum
  Polarization for the Muon g-2 from Lattice QCD: Long-Distance and Full
  Light-Quark Connected Contribution}},
  \href{https://doi.org/10.1103/d583-yhfs}{\emph{Phys. Rev. Lett.} {\bfseries
  135} (2025) 011901} [\href{https://arxiv.org/abs/2412.18491}{{\ttfamily
  2412.18491}}].

\bibitem{Davier:2023fpl}
M.~Davier, A.~Hoecker, A.-M.~Lutz, B.~Malaescu and Z.~Zhang, \emph{{Tensions in
  $e^+e^-\rightarrow \pi ^+\pi ^-(\gamma )$ measurements: the new landscape of
  data-driven hadronic vacuum polarization predictions for the muon $g - 2$}},
  \href{https://doi.org/10.1140/epjc/s10052-024-12964-7}{\emph{Eur. Phys. J. C}
  {\bfseries 84} (2024) 721}
  [\href{https://arxiv.org/abs/2312.02053}{{\ttfamily 2312.02053}}].

\bibitem{CMD-3:2023alj}
{\scshape CMD-3} collaboration, \emph{{Measurement of the
  e+e-{\textrightarrow}{\ensuremath{\pi}}+{\ensuremath{\pi}}- cross section
  from threshold to 1.2~GeV with the CMD-3 detector}},
  \href{https://doi.org/10.1103/PhysRevD.109.112002}{\emph{Phys. Rev. D}
  {\bfseries 109} (2024) 112002}
  [\href{https://arxiv.org/abs/2302.08834}{{\ttfamily 2302.08834}}].

\bibitem{CMD-3:2023rfe}
{\scshape CMD-3} collaboration, \emph{{Measurement of the Pion Form Factor with
  CMD-3 Detector and its Implication to the Hadronic Contribution to Muon
  (g-2)}}, \href{https://doi.org/10.1103/PhysRevLett.132.231903}{\emph{Phys.
  Rev. Lett.} {\bfseries 132} (2024) 231903}
  [\href{https://arxiv.org/abs/2309.12910}{{\ttfamily 2309.12910}}].

\bibitem{KLOE:2008fmq}
{\scshape KLOE} collaboration, \emph{{Measurement of $\sigma(e^+ e^- \to \pi^+
  \pi^- \gamma(\gamma)$ and the dipion contribution to the muon anomaly with
  the KLOE detector}},
  \href{https://doi.org/10.1016/j.physletb.2008.10.060}{\emph{Phys. Lett. B}
  {\bfseries 670} (2009) 285}
  [\href{https://arxiv.org/abs/0809.3950}{{\ttfamily 0809.3950}}].

\bibitem{KLOE:2010qei}
{\scshape KLOE} collaboration, \emph{{Measurement of $\sigma(e^+ e^- \to \pi^+
  \pi^-)$ from threshold to 0.85 GeV$^2$ using Initial State Radiation with the
  KLOE detector}},
  \href{https://doi.org/10.1016/j.physletb.2011.04.055}{\emph{Phys. Lett. B}
  {\bfseries 700} (2011) 102}
  [\href{https://arxiv.org/abs/1006.5313}{{\ttfamily 1006.5313}}].

\bibitem{KLOE:2012anl}
{\scshape KLOE} collaboration, \emph{{Precision measurement of
  $\sigma(e^+e^-\rightarrow \pi^+\pi^-\gamma)/ \sigma(e^+e^-\rightarrow
  \mu^+\mu^-\gamma)$ and determination of the $\pi^+\pi^-$ contribution to the
  muon anomaly with the KLOE detector}},
  \href{https://doi.org/10.1016/j.physletb.2013.02.029}{\emph{Phys. Lett. B}
  {\bfseries 720} (2013) 336}
  [\href{https://arxiv.org/abs/1212.4524}{{\ttfamily 1212.4524}}].

\bibitem{KLOE-2:2017fda}
{\scshape KLOE-2} collaboration, \emph{{Combination of KLOE
  $\sigma\big(e^+e^-\rightarrow\pi^+\pi^-\gamma(\gamma)\big)$ measurements and
  determination of $a_{\mu}^{\pi^+\pi^-}$ in the energy range $0.10 < s < 0.95$
  GeV$^2$}}, \href{https://doi.org/10.1007/JHEP03(2018)173}{\emph{JHEP}
  {\bfseries 03} (2018) 173}
  [\href{https://arxiv.org/abs/1711.03085}{{\ttfamily 1711.03085}}].

\bibitem{BaBar:2012bdw}
{\scshape BaBar} collaboration, \emph{{Precise Measurement of the $e^+ e^- \to
  \pi^+\pi^- (\gamma)$ Cross Section with the Initial-State Radiation Method at
  BABAR}}, \href{https://doi.org/10.1103/PhysRevD.86.032013}{\emph{Phys. Rev.
  D} {\bfseries 86} (2012) 032013}
  [\href{https://arxiv.org/abs/1205.2228}{{\ttfamily 1205.2228}}].

\bibitem{BaBar:2026pic}
{\scshape BaBar} collaboration, \emph{{A New Precise Measurement of the
  $e^+e^-\to\pi ^+\pi ^-(\gamma)$ Cross Section with the Initial-State
  Radiation Method at BaBar}},
  \href{https://arxiv.org/abs/2611.04621}{{\ttfamily 2611.04621}}.

\bibitem{BaBar:2026jzp}
{\scshape BaBar} collaboration, \emph{{A New Precise Measurement of the
  $e^+e^-\to\pi ^+\pi ^-(\gamma)$ Cross Section with the Initial-State
  Radiation Method at BaBar}},
  \href{https://arxiv.org/abs/2611.04596}{{\ttfamily 2611.04596}}.

\bibitem{Jegerlehner:2026zwk}
F.~Jegerlehner, \emph{{Lepton Magnetic Moments: What They Tell Us}},
  \href{https://doi.org/10.5506/APhysPolBSupp.19.2-A1}{\emph{Acta Phys. Polon.
  Supp.} {\bfseries 19} (2026) 2}
  [\href{https://arxiv.org/abs/2601.08101}{{\ttfamily 2601.08101}}].

\bibitem{Hertzog:2025ssc}
D.W.~Hertzog and M.~Hoferichter, \emph{{The anomalous magnetic moment of the
  muon: status and perspectives}},
  \href{https://doi.org/10.1146/annurev-nucl-102422-040841}{\emph{Ann. Rev.
  Nucl. Part. Sci.} {\bfseries 76} (2026) 343}
  [\href{https://arxiv.org/abs/2512.16980}{{\ttfamily 2512.16980}}].

\bibitem{Gluza:2024sop}
J.~Gluza and B.~Karmakar, \emph{{Precision Physics at High-energy Colliders and
  Low-energy Connections}},
  \href{https://doi.org/10.5506/aphyspolbsupp.17.5-a3}{\emph{Acta Phys. Polon.
  Supp.} {\bfseries 17} (2024) 5}.

\bibitem{FCC:2025lpp}
{\scshape FCC} collaboration, \emph{{Future Circular Collider Feasibility Study
  Report: Volume 1, Physics, Experiments, Detectors}},
  \href{https://doi.org/10.1140/epjc/s10052-025-15077-x}{\emph{Eur. Phys. J. C}
  {\bfseries 85} (2025) 1468}
  [\href{https://arxiv.org/abs/2505.00272}{{\ttfamily 2505.00272}}].

\bibitem{Arbuzov:1998te}
A.B.~Arbuzov, E.A.~Kuraev, N.P.~Merenkov and L.~Trentadue, \emph{{Hadronic
  cross-sections in electron - positron annihilation with tagged photon}},
  \href{https://doi.org/10.1088/1126-6708/1998/12/009}{\emph{JHEP} {\bfseries
  12} (1998) 009} [\href{https://arxiv.org/abs/hep-ph/9804430}{{\ttfamily
  hep-ph/9804430}}].

\bibitem{Binner:1999bt}
S.~Binner, J.H.~Kuhn and K.~Melnikov, \emph{{Measuring sigma(e+ e-
  ---{\ensuremath{>}} hadrons) using tagged photon}},
  \href{https://doi.org/10.1016/S0370-2693(99)00658-9}{\emph{Phys. Lett. B}
  {\bfseries 459} (1999) 279}
  [\href{https://arxiv.org/abs/hep-ph/9902399}{{\ttfamily hep-ph/9902399}}].

\bibitem{Kuraev:1985hb}
E.A.~Kuraev and V.S.~Fadin, \emph{{On Radiative Corrections to e+ e- Single
  Photon Annihilation at High-Energy}}, {\emph{Sov. J. Nucl. Phys.} {\bfseries
  41} (1985) 466}.

\bibitem{Berends:1987ab}
F.A.~Berends, W.L.~van Neerven and G.J.H.~Burgers, \emph{{Higher Order
  Radiative Corrections at LEP Energies}},
  \href{https://doi.org/10.1016/0550-3213(88)90313-6}{\emph{Nucl. Phys. B}
  {\bfseries 297} (1988) 429}.

\bibitem{Nicrosini:1987ch}
O.~Nicrosini and L.~Trentadue, \emph{{Soft Photons and Second Order Radiative
  Corrections in $e^+e^-$ Annihilation}},
  \href{https://doi.org/10.1016/0370-2693(87)90815-6}{\emph{Phys. Lett. B}
  {\bfseries 196} (1987) 551}.

\bibitem{Nicrosini:1988hi}
O.~Nicrosini and L.~Trentadue, \emph{{Second Order Electromagnetic Radiative
  Corrections to $e^+e^- \to \mu^+ \mu^-$}},
  \href{https://doi.org/10.1007/BF01555133}{\emph{Z. Phys. C} {\bfseries 39}
  (1988) 479}.

\bibitem{Berends:1973jb}
F.A.~Berends, K.J.F.~Gaemers and R.~Gastmans, \emph{{Hard photon corrections
  for Bhabha scattering}},
  \href{https://doi.org/10.1016/0550-3213(74)90329-0}{\emph{Nucl. Phys. B}
  {\bfseries 68} (1974) 541}.

\bibitem{Berends:1983mi}
F.A.~Berends, R.~Kleiss and S.~Jadach, \emph{{Monte Carlo simulation of
  radiative corrections to the processes $e^+e^- \to \mu^+\mu^-$ and $e^+e^-
  \to q\bar{q}$ in the $Z^0$ region}},
  \href{https://doi.org/10.1016/0010-4655(83)90073-5}{\emph{Comput. Phys.
  Commun.} {\bfseries 29} (1983) 185}.

\bibitem{Jadach:1984hwn}
S.~Jadach and Z.~W\k{a}s, \emph{{QED $\mathcal{O}(\alpha^3)$ radiative
  corrections to the reaction $e^+e^- \to \tau^+\tau^-$ including spin and mass
  effects}}, {\emph{Acta Phys. Polon. B} {\bfseries 15} (1984) 1151}.

\bibitem{Jadach:1988ec}
S.~Jadach and B.F.L.~Ward, \emph{{Multi - Photon Monte Carlo for Bhabha
  Scattering at Low Angles}},
  \href{https://doi.org/10.1103/PhysRevD.40.3582}{\emph{Phys. Rev. D}
  {\bfseries 40} (1989) 3582}.

\bibitem{Jadach:1996is}
S.~Jadach, E.~Richter-Was, B.F.L.~Ward and Z.~Was, \emph{{BHLUMI 4.04: A Monte
  Carlo event generator for LEP luminosity determination}},
  \href{https://doi.org/10.1016/S0010-4655(96)00155-2}{\emph{Comput. Phys.
  Commun.} {\bfseries 102} (1997) 269}.

\bibitem{Jadach:1993hs}
S.~Jadach, B.F.L.~Ward and Z.~Was, \emph{{The Monte Carlo program KORALZ,
  version 4.0, for the lepton or quark pair production in $e^+e^-$
  collisions}},
  \href{https://doi.org/10.1016/0010-4655(94)90088-4}{\emph{Comput. Phys.
  Commun.} {\bfseries 79} (1994) 503}.

\bibitem{Yennie:1961ad}
D.R.~Yennie, S.C.~Frautschi and H.~Suura, \emph{{The infrared divergence
  phenomena and high-energy processes}},
  \href{https://doi.org/10.1016/0003-4916(61)90151-8}{\emph{Annals Phys.}
  {\bfseries 13} (1961) 379}.

\bibitem{Baier:1965jz}
V.N.~Baier and V.A.~Khoze, \emph{{Radiation accompanying two particle
  annihilation of an electron - positron pair}}, {\emph{Sov. Phys. JETP}
  {\bfseries 21} (1965) 1145}.

\bibitem{Baier:1966jf}
V.N.~Baier, V.S.~Fadin and V.A.~Khoze, \emph{{Photon bremsstrahlung in
  collisions of high-energy electrons}}, {\emph{Sov. Phys. JETP} {\bfseries 24}
  (1966) 760}.

\bibitem{Kuraev:1977rp}
E.A.~Kuraev and G.V.~Meledin, \emph{{QED Distributions for Hard Photon Emission
  in $e^+e^- \to \mu^+ \mu^- \gamma$}},
  \href{https://doi.org/10.1016/0550-3213(77)90141-9}{\emph{Nucl. Phys. B}
  {\bfseries 122} (1977) 485}.

\bibitem{Baier:1980kx}
V.N.~Baier, E.A.~Kuraev, V.S.~Fadin and V.A.~Khoze, \emph{{Inelastic Processes
  in Quantum Electrodynamics at High-Energies}},
  \href{https://doi.org/10.1016/0370-1573(81)90140-X}{\emph{Phys. Rept.}
  {\bfseries 78} (1981) 293}.

\bibitem{Berends:1981rb}
F.A.~Berends, R.~Kleiss, P.~De~Causmaecker, R.~Gastmans and T.T.~Wu,
  \emph{{Single Bremsstrahlung Processes in Gauge Theories}},
  \href{https://doi.org/10.1016/0370-2693(81)90685-7}{\emph{Phys. Lett. B}
  {\bfseries 103} (1981) 124}.

\bibitem{Chen:1974wv}
M.-S.~Chen and P.M.~Zerwas, \emph{{Secondary Reactions in electron - Positron
  (electron) Collisions}},
  \href{https://doi.org/10.1103/PhysRevD.11.58}{\emph{Phys. Rev. D} {\bfseries
  11} (1975) 58}.

\bibitem{Budnev:1975poe}
V.M.~Budnev, I.F.~Ginzburg, G.V.~Meledin and V.G.~Serbo, \emph{{The two photon
  particle production mechanism. Physical problems. Applications. Equivalent
  photon approximation}},
  \href{https://doi.org/10.1016/0370-1573(75)90009-5}{\emph{Phys. Rept.}
  {\bfseries 15} (1975) 181}.

\bibitem{Passarino:1979jh}
G.~Passarino and M.J.G.~Veltman, \emph{{One-loop corrections for $e^+ e^-$
  annihilation into $\mu^+ \mu^-$ in the Weinberg model}},
  \href{https://doi.org/10.1016/0550-3213(79)90234-7}{\emph{Nucl. Phys. B}
  {\bfseries 160} (1979) 151}.

\bibitem{Bardin:1989tq}
D.Y.~Bardin, M.S.~Bilenky, M.~Chizhov, A.~Sankar, O.~Dzhibuti, O.M.~Fedorenko
  et~al., \emph{{ZFITTER: An Analytical program for fermion pair production in
  $e^+e^-$ annihilation}}, \href{https://doi.org/10.1007/BF01548202}{\emph{Z.
  Phys. C} {\bfseries 44} (1989) 491}.

\bibitem{Bardin:1999ak}
D.Y.~Bardin, P.~Christova, L.~Kalinovskaya, A.~Olshevski, G.~Passarino,
  T.~Riemann et~al., \emph{{ZFITTER v.6.21: A Semi-analytical program for
  fermion pair production in e+ e- annihilation}},
  \href{https://doi.org/10.1016/S0010-4655(00)00152-1}{\emph{Comput. Phys.
  Commun.} {\bfseries 133} (2001) 229}.

\bibitem{Rodrigo:2001kf}
G.~Rodrigo, H.~Czyz, J.H.~Kuhn and M.~Szopa, \emph{{Radiative return at NLO and
  the measurement of the hadronic cross-section in electron positron
  annihilation}}, \href{https://doi.org/10.1007/s100520200912}{\emph{Eur. Phys.
  J. C} {\bfseries 24} (2002) 71}
  [\href{https://arxiv.org/abs/hep-ph/0112184}{{\ttfamily hep-ph/0112184}}].

\bibitem{Czyz:2000wh}
H.~Czyz and J.H.~Kuhn, \emph{{Four pion final states with tagged photons at
  electron positron colliders}},
  \href{https://doi.org/10.1007/s100520000553}{\emph{Eur. Phys. J. C}
  {\bfseries 18} (2001) 497}
  [\href{https://arxiv.org/abs/hep-ph/0008262}{{\ttfamily hep-ph/0008262}}].

\bibitem{Czyz:2002np}
H.~Czyz, A.~Grzelinska, J.H.~Kuhn and G.~Rodrigo, \emph{{The Radiative return
  at phi and B factories: Small angle photon emission at next-to-leading
  order}}, \href{https://doi.org/10.1140/epjc/s2003-01134-4}{\emph{Eur. Phys.
  J. C} {\bfseries 27} (2003) 563}
  [\href{https://arxiv.org/abs/hep-ph/0212225}{{\ttfamily hep-ph/0212225}}].

\bibitem{Czyz:2003ue}
H.~Czyz, A.~Grzelinska, J.H.~Kuhn and G.~Rodrigo, \emph{{The Radiative return
  at Phi and B factories: FSR at next-to-leading order}},
  \href{https://doi.org/10.1140/epjc/s2004-01605-0}{\emph{Eur. Phys. J. C}
  {\bfseries 33} (2004) 333}
  [\href{https://arxiv.org/abs/hep-ph/0308312}{{\ttfamily hep-ph/0308312}}].

\bibitem{Czyz:2004ua}
H.~Czyz, J.H.~Kuhn, E.~Nowak and G.~Rodrigo, \emph{{Nucleon form-factors, B
  meson factories and the radiative return}},
  \href{https://doi.org/10.1140/epjc/s2004-01864-7}{\emph{Eur. Phys. J. C}
  {\bfseries 35} (2004) 527}
  [\href{https://arxiv.org/abs/hep-ph/0403062}{{\ttfamily hep-ph/0403062}}].

\bibitem{Czyz:2005as}
H.~Czyz, A.~Grzelinska, J.H.~Kuhn and G.~Rodrigo, \emph{{Electron-positron
  annihilation into three pions and the radiative return}},
  \href{https://doi.org/10.1140/epjc/s2006-02614-7}{\emph{Eur. Phys. J. C}
  {\bfseries 47} (2006) 617}
  [\href{https://arxiv.org/abs/hep-ph/0512180}{{\ttfamily hep-ph/0512180}}].

\bibitem{Czyz:2007wi}
H.~Czyz, A.~Grzelinska and J.H.~Kuhn, \emph{{Spin asymmetries and correlations
  in lambda-pair production through the radiative return method}},
  \href{https://doi.org/10.1103/PhysRevD.75.074026}{\emph{Phys. Rev. D}
  {\bfseries 75} (2007) 074026}
  [\href{https://arxiv.org/abs/hep-ph/0702122}{{\ttfamily hep-ph/0702122}}].

\bibitem{Czyz:2010hj}
H.~Czyz, A.~Grzelinska and J.H.~Kuhn, \emph{{Narrow resonances studies with the
  radiative return method}},
  \href{https://doi.org/10.1103/PhysRevD.81.094014}{\emph{Phys. Rev. D}
  {\bfseries 81} (2010) 094014}
  [\href{https://arxiv.org/abs/1002.0279}{{\ttfamily 1002.0279}}].

\bibitem{Campanario:2013uea}
F.~Campanario, H.~Czy{\.z}, J.~Gluza, M.~Gunia, T.~Riemann, G.~Rodrigo et~al.,
  \emph{{Complete QED NLO contributions to the reaction $e^+e^- \to
  \mu^+\mu^-\gamma$ and their implementation in the event generator PHOKHARA}},
  \href{https://doi.org/10.1007/JHEP02(2014)114}{\emph{JHEP} {\bfseries 02}
  (2014) 114} [\href{https://arxiv.org/abs/1312.3610}{{\ttfamily 1312.3610}}].

\bibitem{Campanario:2019mjh}
F.~Campanario, H.~Czy{\.z}, J.~Gluza, T.~Jeli{\'n}ski, G.~Rodrigo, S.~Tracz
  et~al., \emph{{Standard model radiative corrections in the pion form factor
  measurements do not explain the $a_\mu$ anomaly}},
  \href{https://doi.org/10.1103/PhysRevD.100.076004}{\emph{Phys. Rev. D}
  {\bfseries 100} (2019) 076004}
  [\href{https://arxiv.org/abs/1903.10197}{{\ttfamily 1903.10197}}].

\bibitem{WorkingGrouponRadiativeCorrections:2010bjp}
{\scshape Working Group on Radiative Corrections, Monte Carlo Generators for
  Low Energies} collaboration, \emph{{Quest for precision in hadronic cross
  sections at low energy: Monte Carlo tools vs. experimental data}},
  \href{https://doi.org/10.1140/epjc/s10052-010-1251-4}{\emph{Eur. Phys. J. C}
  {\bfseries 66} (2010) 585} [\href{https://arxiv.org/abs/0912.0749}{{\ttfamily
  0912.0749}}].

\bibitem{BaBar:2023xiy}
{\scshape BaBar} collaboration, \emph{{Measurement of additional radiation in
  the initial-state-radiation processes
  e+e-$\to${\ensuremath{\mu}}+{\ensuremath{\mu}}-{\ensuremath{\gamma}} and
  e+e-$\to${\ensuremath{\pi}}+{\ensuremath{\pi}}-{\ensuremath{\gamma}} at
  BABAR}}, \href{https://doi.org/10.1103/PhysRevD.108.L111103}{\emph{Phys. Rev.
  D} {\bfseries 108} (2023) L111103}
  [\href{https://arxiv.org/abs/2308.05233}{{\ttfamily 2308.05233}}].

\bibitem{Balossini:2006wc}
G.~Balossini, C.M.~Carloni~Calame, G.~Montagna, O.~Nicrosini and F.~Piccinini,
  \emph{{Matching perturbative and parton shower corrections to Bhabha process
  at flavour factories}},
  \href{https://doi.org/10.1016/j.nuclphysb.2006.09.022}{\emph{Nucl. Phys. B}
  {\bfseries 758} (2006) 227}
  [\href{https://arxiv.org/abs/hep-ph/0607181}{{\ttfamily hep-ph/0607181}}].

\bibitem{Balossini:2008xr}
G.~Balossini, C.~Bignamini, C.M.C.~Calame, G.~Montagna, O.~Nicrosini and
  F.~Piccinini, \emph{{Photon pair production at flavour factories with per
  mille accuracy}},
  \href{https://doi.org/10.1016/j.physletb.2008.04.007}{\emph{Phys. Lett. B}
  {\bfseries 663} (2008) 209}
  [\href{https://arxiv.org/abs/0801.3360}{{\ttfamily 0801.3360}}].

\bibitem{Budassi:2024whw}
E.~Budassi, C.M.~Carloni~Calame, M.~Ghilardi, A.~Gurgone, G.~Montagna,
  M.~Moretti et~al., \emph{{Pion pair production in $e^{+}$e$^{-}$ annihilation
  at next-to-leading order matched to Parton Shower}},
  \href{https://doi.org/10.1007/JHEP05(2025)196}{\emph{JHEP} {\bfseries 05}
  (2025) 196} [\href{https://arxiv.org/abs/2409.03469}{{\ttfamily
  2409.03469}}].

\bibitem{Budassi:2026lmr}
E.~Budassi, C.M.~Carloni~Calame, M.~Ghilardi, A.~Gurgone, G.~Montagna,
  M.~Moretti et~al., \emph{{Radiative return at NLOPS accuracy}},
  \href{https://doi.org/10.1007/JHEP05(2026)221}{\emph{JHEP} {\bfseries 05}
  (2026) 221} [\href{https://arxiv.org/abs/2601.19530}{{\ttfamily
  2601.19530}}].

\bibitem{Banerjee:2020rww}
P.~Banerjee, T.~Engel, A.~Signer and Y.~Ulrich, \emph{{QED at NNLO with
  McMule}}, \href{https://doi.org/10.21468/SciPostPhys.9.2.027}{\emph{SciPost
  Phys.} {\bfseries 9} (2020) 027}
  [\href{https://arxiv.org/abs/2007.01654}{{\ttfamily 2007.01654}}].

\bibitem{Fang:2025mhn}
Y.~Fang, S.~Kollatzsch, M.~Rocco, A.~Signer, Y.~Ulrich and M.~Zoller,
  \emph{{Disperon QED}},
  \href{https://doi.org/10.21468/SciPostPhys.20.4.116}{\emph{SciPost Phys.}
  {\bfseries 20} (2026) 116}
  [\href{https://arxiv.org/abs/2512.10709}{{\ttfamily 2512.10709}}].

\bibitem{Aliberti:2024fpq}
R.~Aliberti et~al., \emph{{Radiative corrections and Monte Carlo tools for
  low-energy hadronic cross sections in $e^+ e^-$ collisions}},
  \href{https://doi.org/10.21468/SciPostPhysCommRep.9}{\emph{SciPost Phys.
  Comm. Rep.} (2025) 9} [\href{https://arxiv.org/abs/2410.22882}{{\ttfamily
  2410.22882}}].

\bibitem{CarloniCalame:2026vfc}
C.M.~Carloni~Calame, M.~Ghilardi, A.~Gurgone, G.~Montagna, M.~Moretti,
  O.~Nicrosini et~al., \emph{{Structure-dependent radiative corrections to
  \(\left. e^{+}e^{-}\rightarrow\pi^{+}\pi^{-}\gamma \right.\) in the GVMD
  approach}}, \href{https://doi.org/10.1016/j.physletb.2026.140666}{\emph{Phys.
  Lett. B} {\bfseries 879} (2026) 140666}
  [\href{https://arxiv.org/abs/2603.28621}{{\ttfamily 2603.28621}}].

\bibitem{CarloniCalame:2026gdm}
C.M.~Carloni~Calame, M.~Ghilardi, A.~Gurgone, G.~Montagna, M.~Moretti,
  O.~Nicrosini et~al., \emph{{Next-to-leading order corrections to radiative
  pion pair production via dispersion relations}},
  \href{https://doi.org/10.1103/8zd9-sytq}{\emph{Phys. Rev. D} {\bfseries 114}
  (2026) 076022} [\href{https://arxiv.org/abs/2607.24642}{{\ttfamily
  2607.24642}}].

\bibitem{Dave:2026pvq}
T.~Dave, J.~Paltrinieri, P.~Petit~Ros{\`a}s and W.J.~Torres~Bobadilla,
  \emph{{Tensor decomposition of
  e+e{\ensuremath{-}}{\textrightarrow}{\ensuremath{\pi}}+{\ensuremath{\pi}}{\ensuremath{-}}{\ensuremath{\gamma}}
  to higher orders in the dimensional regulator}},
  \href{https://doi.org/10.1016/j.physletb.2026.140952}{\emph{Phys. Lett. B}
  {\bfseries 881} (2026) 140952}
  [\href{https://arxiv.org/abs/2604.16251}{{\ttfamily 2604.16251}}].

\bibitem{PetitRosas:2026iuq}
P.~Petit~Ros{\`a}s, O.~Shekhovtsova and W.J.~Torres~Bobadilla, \emph{{Radiative
  return meets GVMD}},
  \href{https://doi.org/10.1016/j.physletb.2026.140864}{\emph{Phys. Lett. B}
  {\bfseries 881} (2026) 140864}
  [\href{https://arxiv.org/abs/2603.13171}{{\ttfamily 2603.13171}}].

\bibitem{Pozzoli:2026eiu}
M.~Pozzoli and W.J.~Torres~Bobadilla, \emph{{First look at the evaluation of
  two-loop Feynman integrals for radiative return processes, }},
  \href{https://arxiv.org/abs/2607.03343}{{\ttfamily 2607.03343}}.

\bibitem{Fadin:2023phc}
V.S.~Fadin and R.N.~Lee, \emph{{Two-loop radiative corrections to
  e$^{+}$e$^{-}$ $\to$ {\ensuremath{\gamma}}{\ensuremath{\gamma}}$^{*}$ cross
  section}}, \href{https://doi.org/10.1007/JHEP11(2023)148}{\emph{JHEP}
  {\bfseries 11} (2023) 148}
  [\href{https://arxiv.org/abs/2308.09479}{{\ttfamily 2308.09479}}].

\bibitem{Badger:2023xtl}
S.~Badger, J.~Kry{\'s}, R.~Moodie and S.~Zoia, \emph{{Lepton-pair scattering
  with an off-shell and an on-shell photon at two loops in massless QED}},
  \href{https://doi.org/10.1007/JHEP11(2023)041}{\emph{JHEP} {\bfseries 11}
  (2023) 041} [\href{https://arxiv.org/abs/2307.03098}{{\ttfamily
  2307.03098}}].

\bibitem{PetitRosas:2025xhm}
P.~Petit~Ros{\`a}s and W.J.~Torres~Bobadilla, \emph{{Fast evaluation of Feynman
  integrals for Monte Carlo generators}},
  \href{https://doi.org/10.1007/JHEP09(2025)210}{\emph{JHEP} {\bfseries 09}
  (2025) 210} [\href{https://arxiv.org/abs/2507.12548}{{\ttfamily
  2507.12548}}].

\bibitem{Sherpa:2024mfk}
{\scshape Sherpa} collaboration, \emph{{Event generation with Sherpa 3}},
  \href{https://doi.org/10.1007/JHEP12(2024)156}{\emph{JHEP} {\bfseries 12}
  (2024) 156} [\href{https://arxiv.org/abs/2410.22148}{{\ttfamily
  2410.22148}}].

\bibitem{Krauss:2022ajk}
F.~Krauss, A.~Price and M.~Sch{\"o}nherr, \emph{{YFS Resummation for Future
  Lepton-Lepton Colliders in SHERPA}},
  \href{https://doi.org/10.21468/SciPostPhys.13.2.026}{\emph{SciPost Phys.}
  {\bfseries 13} (2022) 026}
  [\href{https://arxiv.org/abs/2203.10948}{{\ttfamily 2203.10948}}].

\bibitem{Price:2025fiu}
A.~Price and F.~Krauss, \emph{{Toward a fully automated differential NNLOEW
  generator for lepton colliders}},
  \href{https://doi.org/10.1103/rt7d-7jf9}{\emph{Phys. Rev. D} {\bfseries 113}
  (2026) 073008} [\href{https://arxiv.org/abs/2512.04959}{{\ttfamily
  2512.04959}}].

\bibitem{Jadach:1999vf}
S.~Jadach, B.F.L.~Ward and Z.~Was, \emph{{The Precision Monte Carlo event
  generator K K for two fermion final states in e+ e- collisions}},
  \href{https://doi.org/10.1016/S0010-4655(00)00048-5}{\emph{Comput. Phys.
  Commun.} {\bfseries 130} (2000) 260}
  [\href{https://arxiv.org/abs/hep-ph/9912214}{{\ttfamily hep-ph/9912214}}].

\bibitem{Jadach:2022mbe}
S.~Jadach, B.F.L.~Ward, Z.~W\k{a}s, S.A.~Yost and A.~Siodmok,
  \emph{{Multi-photon Monte Carlo event generator KKMCee for lepton and quark
  pair production in lepton colliders}},
  \href{https://doi.org/10.1016/j.cpc.2022.108556}{\emph{Comput. Phys. Commun.}
  {\bfseries 283} (2023) 108556}
  [\href{https://arxiv.org/abs/2204.11949}{{\ttfamily 2204.11949}}].

\bibitem{Jadach:2000ir}
S.~Jadach, B.F.L.~Ward and Z.~Was, \emph{{Coherent exclusive exponentiation for
  precision Monte Carlo calculations}},
  \href{https://doi.org/10.1103/PhysRevD.63.113009}{\emph{Phys. Rev. D}
  {\bfseries 63} (2001) 113009}
  [\href{https://arxiv.org/abs/hep-ph/0006359}{{\ttfamily hep-ph/0006359}}].

\bibitem{Price:2026qop}
A.~Price, \emph{{Automation of amplitude-level QED resummation with ACRAIC}},
  \href{https://arxiv.org/abs/2610.28283}{{\ttfamily 2610.28283}}.

\bibitem{Tackmann:2024kci}
F.J.~Tackmann, \emph{{Beyond scale variations: perturbative theory
  uncertainties from nuisance parameters}},
  \href{https://doi.org/10.1007/JHEP08(2025)098}{\emph{JHEP} {\bfseries 08}
  (2025) 098} [\href{https://arxiv.org/abs/2411.18606}{{\ttfamily
  2411.18606}}].

\bibitem{Lim:2024nsk}
M.A.~Lim and R.~Poncelet, \emph{{Robust estimates of theoretical uncertainties
  at fixed-order in perturbation theory}},
  \href{https://doi.org/10.1103/7g5k-4y3v}{\emph{Phys. Rev. D} {\bfseries 112}
  (2025) L111901} [\href{https://arxiv.org/abs/2412.14910}{{\ttfamily
  2412.14910}}].

\bibitem{Schonherr:2008av}
M.~Schonherr and F.~Krauss, \emph{{Soft Photon Radiation in Particle Decays in
  SHERPA}}, \href{https://doi.org/10.1088/1126-6708/2008/12/018}{\emph{JHEP}
  {\bfseries 12} (2008) 018} [\href{https://arxiv.org/abs/0810.5071}{{\ttfamily
  0810.5071}}].

\bibitem{PhysRev.124.953}
M.~Gell-Mann and F.~Zachariasen, \emph{Form factors and vector mesons},
  \href{https://doi.org/10.1103/PhysRev.124.953}{\emph{Phys. Rev.} {\bfseries
  124} (1961) 953}.

\bibitem{RevModPhys.50.261}
T.H.~Bauer, R.D.~Spital, D.R.~Yennie and F.M.~Pipkin, \emph{The hadronic
  properties of the photon in high-energy interactions},
  \href{https://doi.org/10.1103/RevModPhys.50.261}{\emph{Rev. Mod. Phys.}
  {\bfseries 50} (1978) 261}.

\bibitem{Gounaris:1968mw}
G.J.~Gounaris and J.J.~Sakurai, \emph{{Finite width corrections to the vector
  meson dominance prediction for $\rho \to e^+ e^-$}},
  \href{https://doi.org/10.1103/PhysRevLett.21.244}{\emph{Phys. Rev. Lett.}
  {\bfseries 21} (1968) 244}.

\bibitem{Hoefer:2001mx}
A.~Hoefer, J.~Gluza and F.~Jegerlehner, \emph{{Pion pair production with higher
  order radiative corrections in low energy e+ e- collisions}},
  \href{https://doi.org/10.1007/s100520200916}{\emph{Eur. Phys. J. C}
  {\bfseries 24} (2002) 51}
  [\href{https://arxiv.org/abs/hep-ph/0107154}{{\ttfamily hep-ph/0107154}}].

\bibitem{Bystritskiy:2005ib}
Y.M.~Bystritskiy, E.A.~Kuraev, G.V.~Fedotovich and F.V.~Ignatov, \emph{{The
  Cross sections of the muons and charged pions pairs production at
  electron-positron annihilation near the threshold}},
  \href{https://doi.org/10.1103/PhysRevD.72.114019}{\emph{Phys. Rev. D}
  {\bfseries 72} (2005) 114019}
  [\href{https://arxiv.org/abs/hep-ph/0505236}{{\ttfamily hep-ph/0505236}}].

\bibitem{Arbuzov:2020foj}
A.B.~Arbuzov, T.V.~Kopylova and G.A.~Seilkhanova,
  \emph{{Forward{\textendash}backward asymmetry in
  electron{\textendash}positron annihilation into pion or kaon pairs
  revisited}}, \href{https://doi.org/10.1142/S0217732320502107}{\emph{Mod.
  Phys. Lett. A} {\bfseries 35} (2020) 2050210}
  [\href{https://arxiv.org/abs/2003.14054}{{\ttfamily 2003.14054}}].

\bibitem{Sakurai:1972wk}
J.J.~Sakurai and D.~Schildknecht, \emph{{Generalized vector dominance and
  inelastic electron - proton scattering}},
  \href{https://doi.org/10.1016/0370-2693(72)90300-0}{\emph{Phys. Lett. B}
  {\bfseries 40} (1972) 121}.

\bibitem{Ignatov:2022iou}
F.~Ignatov and R.N.~Lee, \emph{{Charge asymmetry in e+e{\ensuremath{-}}
  {\textrightarrow} {\ensuremath{\pi}}+{\ensuremath{\pi}}{\ensuremath{-}}
  process}}, \href{https://doi.org/10.1016/j.physletb.2022.137283}{\emph{Phys.
  Lett. B} {\bfseries 833} (2022) 137283}
  [\href{https://arxiv.org/abs/2204.12235}{{\ttfamily 2204.12235}}].

\bibitem{Colangelo:2014dfa}
G.~Colangelo, M.~Hoferichter, M.~Procura and P.~Stoffer, \emph{{Dispersive
  approach to hadronic light-by-light scattering}},
  \href{https://doi.org/10.1007/JHEP09(2014)091}{\emph{JHEP} {\bfseries 09}
  (2014) 091} [\href{https://arxiv.org/abs/1402.7081}{{\ttfamily 1402.7081}}].

\bibitem{Colangelo:2022lzg}
G.~Colangelo, M.~Hoferichter, J.~Monnard and J.~Ruiz~de Elvira,
  \emph{{Radiative corrections to the forward-backward asymmetry in
  e$^{+}$e$^{−}$ {\textrightarrow}
  {\ensuremath{\pi}}$^{+}${\ensuremath{\pi}}$^{−}$}},
  \href{https://doi.org/10.1007/JHEP08(2022)295}{\emph{JHEP} {\bfseries 08}
  (2022) 295} [\href{https://arxiv.org/abs/2207.03495}{{\ttfamily
  2207.03495}}].

\bibitem{Gleisberg:2008fv}
T.~Gleisberg and S.~Hoeche, \emph{{Comix, a new matrix element generator}},
  \href{https://doi.org/10.1088/1126-6708/2008/12/039}{\emph{JHEP} {\bfseries
  12} (2008) 039} [\href{https://arxiv.org/abs/0808.3674}{{\ttfamily
  0808.3674}}].

\bibitem{Jadach:1998wp}
S.~Jadach, B.F.L.~Ward and Z.~Was, \emph{{Global positioning of spin GPS scheme
  for half spin massive spinors}},
  \href{https://doi.org/10.1007/s100520100818}{\emph{Eur. Phys. J. C}
  {\bfseries 22} (2001) 423}
  [\href{https://arxiv.org/abs/hep-ph/9905452}{{\ttfamily hep-ph/9905452}}].

\bibitem{Buccioni:2019sur}
F.~Buccioni, J.-N.~Lang, J.M.~Lindert, P.~Maierh{\"o}fer, S.~Pozzorini,
  H.~Zhang et~al., \emph{{OpenLoops 2}},
  \href{https://doi.org/10.1140/epjc/s10052-019-7306-2}{\emph{Eur. Phys. J. C}
  {\bfseries 79} (2019) 866}
  [\href{https://arxiv.org/abs/1907.13071}{{\ttfamily 1907.13071}}].

\bibitem{Hahn:2000kx}
T.~Hahn, \emph{{Generating Feynman diagrams and amplitudes with FeynArts 3}},
  \href{https://doi.org/10.1016/S0010-4655(01)00290-9}{\emph{Comput. Phys.
  Commun.} {\bfseries 140} (2001) 418}
  [\href{https://arxiv.org/abs/hep-ph/0012260}{{\ttfamily hep-ph/0012260}}].

\bibitem{Mertig:1990an}
R.~Mertig, M.~B{\"o}hm and A.~Denner, \emph{{FEYN CALC: Computer algebraic
  calculation of Feynman amplitudes}},
  \href{https://doi.org/10.1016/0010-4655(91)90130-D}{\emph{Comput. Phys.
  Commun.} {\bfseries 64} (1991) 345}.

\bibitem{Shtabovenko:2023idz}
V.~Shtabovenko, R.~Mertig and F.~Orellana, \emph{{FeynCalc 10: Do multiloop
  integrals dream of computer codes?}},
  \href{https://doi.org/10.1016/j.cpc.2024.109357}{\emph{Comput. Phys. Commun.}
  {\bfseries 306} (2025) 109357}
  [\href{https://arxiv.org/abs/2312.14089}{{\ttfamily 2312.14089}}].

\bibitem{Passarino:1978jh}
G.~Passarino and M.J.G.~Veltman, \emph{{One Loop Corrections for e+ e-
  Annihilation Into mu+ mu- in the Weinberg Model}},
  \href{https://doi.org/10.1016/0550-3213(79)90234-7}{\emph{Nucl. Phys. B}
  {\bfseries 160} (1979) 151}.

\bibitem{Denner:1991kt}
A.~Denner, \emph{{Techniques for calculation of electroweak radiative
  corrections at the one loop level and results for W physics at LEP-200}},
  \href{https://doi.org/10.1002/prop.2190410402}{\emph{Fortsch. Phys.}
  {\bfseries 41} (1993) 307} [\href{https://arxiv.org/abs/0709.1075}{{\ttfamily
  0709.1075}}].

\bibitem{Hahn:1998yk}
T.~Hahn and M.~Perez-Victoria, \emph{{Automatized one loop calculations in
  four-dimensions and D-dimensions}},
  \href{https://doi.org/10.1016/S0010-4655(98)00173-8}{\emph{Comput. Phys.
  Commun.} {\bfseries 118} (1999) 153}
  [\href{https://arxiv.org/abs/hep-ph/9807565}{{\ttfamily hep-ph/9807565}}].

\bibitem{Patel:2015tea}
H.H.~Patel, \emph{{Package-X: A Mathematica package for the analytic
  calculation of one-loop integrals}},
  \href{https://doi.org/10.1016/j.cpc.2015.08.017}{\emph{Comput. Phys. Commun.}
  {\bfseries 197} (2015) 276}
  [\href{https://arxiv.org/abs/1503.01469}{{\ttfamily 1503.01469}}].

\bibitem{Patel:2016fam}
H.H.~Patel, \emph{{Package-X 2.0: A Mathematica package for the analytic
  calculation of one-loop integrals}},
  \href{https://doi.org/10.1016/j.cpc.2017.04.015}{\emph{Comput. Phys. Commun.}
  {\bfseries 218} (2017) 66}
  [\href{https://arxiv.org/abs/1612.00009}{{\ttfamily 1612.00009}}].

\bibitem{Shtabovenko:2016whf}
V.~Shtabovenko, \emph{{FeynHelpers: Connecting FeynCalc to FIRE and
  Package-X}}, \href{https://doi.org/10.1016/j.cpc.2017.04.014}{\emph{Comput.
  Phys. Commun.} {\bfseries 218} (2017) 48}
  [\href{https://arxiv.org/abs/1611.06793}{{\ttfamily 1611.06793}}].

\bibitem{Rodrigo:2001jr}
G.~Rodrigo, A.~Gehrmann-De~Ridder, M.~Guilleaume and J.H.~Kuhn, \emph{{NLO QED
  corrections to ISR in e+ e- annihilation and the measurement of sigma(e+ e-
  ---{\ensuremath{>}} hadrons) using tagged photons}},
  \href{https://doi.org/10.1007/s100520100784}{\emph{Eur. Phys. J. C}
  {\bfseries 22} (2001) 81}
  [\href{https://arxiv.org/abs/hep-ph/0106132}{{\ttfamily hep-ph/0106132}}].

\bibitem{Kuhn:2002xg}
J.H.~Kuhn and G.~Rodrigo, \emph{{The Radiative return at small angles: Virtual
  corrections}}, \href{https://doi.org/10.1007/s10052-002-0999-6}{\emph{Eur.
  Phys. J. C} {\bfseries 25} (2002) 215}
  [\href{https://arxiv.org/abs/hep-ph/0204283}{{\ttfamily hep-ph/0204283}}].

\bibitem{Denner:2016kdg}
A.~Denner, S.~Dittmaier and L.~Hofer, \emph{{Collier: a fortran-based Complex
  One-Loop LIbrary in Extended Regularizations}},
  \href{https://doi.org/10.1016/j.cpc.2016.10.013}{\emph{Comput. Phys. Commun.}
  {\bfseries 212} (2017) 220}
  [\href{https://arxiv.org/abs/1604.06792}{{\ttfamily 1604.06792}}].

\bibitem{Fang:2026naz}
Y.~Fang, S.~Kollatzsch, A.~Signer and M.~Zoller, \emph{{Radiative pion-pair
  production in disperon QED}},
  \href{https://arxiv.org/abs/2609.00912}{{\ttfamily 2609.00912}}.

\bibitem{Jadach:2001jx}
S.~Jadach, M.~Melles, B.F.L.~Ward and S.A.~Yost, \emph{{Exact differential
  O(alpha**2) results for hard bremsstrahlung in e+ e- annihilation to 2f at
  and beyond LEP-2 energies}},
  \href{https://doi.org/10.1103/PhysRevD.65.073030}{\emph{Phys. Rev. D}
  {\bfseries 65} (2002) 073030}
  [\href{https://arxiv.org/abs/hep-ph/0109279}{{\ttfamily hep-ph/0109279}}].

\bibitem{radiomontecarlow2-results}
{RadioMonteCarLow 2 Working Group}, ``{Monte Carlo results: scenario
  definitions}.''
  \url{https://gitlab.com/radiomontecarlow2/monte-carlo-results}.

\bibitem{BESIII:2015equ}
{\scshape BESIII} collaboration, \emph{{Measurement of the $e^+ e^- \to \pi^+
  \pi^-$ cross section between 600 and 900 MeV using initial state radiation}},
  \href{https://doi.org/10.1016/j.physletb.2015.11.043}{\emph{Phys. Lett. B}
  {\bfseries 753} (2016) 629}
  [\href{https://arxiv.org/abs/1507.08188}{{\ttfamily 1507.08188}}].

\bibitem{Penin:2005eh}
A.A.~Penin, \emph{{Two-loop photonic corrections to massive Bhabha
  scattering}},
  \href{https://doi.org/10.1016/j.nuclphysb.2005.11.016}{\emph{Nucl. Phys. B}
  {\bfseries 734} (2006) 185}
  [\href{https://arxiv.org/abs/hep-ph/0508127}{{\ttfamily hep-ph/0508127}}].

\end{thebibliography}\endgroup

\end{document}